 \documentclass[final,1p,times]{elsarticle}

\usepackage{amsmath}
\usepackage{amssymb}
\usepackage{amstext}
\usepackage{mathrsfs}
\usepackage{amsxtra}
\usepackage{subfigure}
\usepackage{color}
\usepackage{feynmf}
\usepackage{eucal}
\usepackage{soul}
\usepackage{latexsym}
\usepackage{amsthm}
\usepackage{simplewick}

\usepackage[symbol]{footmisc}

\usepackage{tikz}
\usepackage[compat=1.1.0]{tikz-feynman}
\usetikzlibrary{arrows,shapes}
\usetikzlibrary{trees}
\usetikzlibrary{matrix,arrows} 				
\usetikzlibrary{positioning}				
\usetikzlibrary{calc,through}				
\usetikzlibrary{decorations.pathreplacing}  
\usepackage{pgffor}							

\usetikzlibrary{decorations.pathmorphing}	
\usetikzlibrary{decorations.markings}
\tikzset{
    vector/.style={decorate, decoration={snake}, draw},
	provector/.style={decorate, decoration={snake,amplitude=2.5pt}, draw},
	antivector/.style={decorate, decoration={snake,amplitude=-2.5pt}, draw},
    fermion/.style={draw=black, postaction={decorate},
        decoration={markings,mark=at position .55 with {\arrow[draw=black]{>}}}},
    fermionbar/.style={draw=black, postaction={decorate},
        decoration={markings,mark=at position .55 with {\arrow[draw=black]{<}}}},
    fermionnoarrow/.style={draw=black},
    gluon/.style={decorate, draw=black,
        decoration={coil,amplitude=4pt, segment length=5pt}},
    scalar/.style={dashed,draw=black, postaction={decorate},
        decoration={markings,mark=at position .55 with {\arrow[draw=black]{>}}}},
    scalarbar/.style={dashed,draw=black, postaction={decorate},
        decoration={markings,mark=at position .55 with {\arrow[draw=black]{<}}}},
    scalarnoarrow/.style={dashed,draw=black},
    electron/.style={draw=black, postaction={decorate},
        decoration={markings,mark=at position .55 with {\arrow[draw=black]{>}}}},
	bigvector/.style={decorate, decoration={snake,amplitude=4pt}, draw},
}
\tikzstyle{block} = [draw, rectangle, 
    minimum height=3em, minimum width=6em]

\begin{document}

\begin{frontmatter}



\title{Spin-wave theory in a randomly disordered  lattice: A Heisenberg ferromagnet
}

\author{Z. J. Weiss\footnote[2]{Deceased.}  and A. R. Massih}
\address{Quantum Technologies AB, Uppsala Science Park, SE-751 83 Uppsala, Sweden\\}

\begin{abstract}
Starting from the hamiltonian for the Heisenberg ferromagnet which comprise randomly distributed nonmagnetic ions as impurities in a Bravais lattice, we  express the spin operators by means of the Dyson-Maleev transformation in terms of the Bose operators of the second quantization. Then by using methods of quantum statistical field theory, we derive the partition function and the free energy for the system. We adopt  the Matsubara thermal perturbation method to a portion of the hamiltonian which describes the interaction between magnons and the stationary field of nonmagnetic ions. Upon  averaging over all possible distributions of impurities, we express the free energy of the system as a function of the mean impurity concentration. Subsequently, we set up the double-time single particle Green function at temperature $T$ in the momentum space in terms of magnon operators  and derive the equation of motion for the Green function through the Heisenberg equation of motion and then solve the resulting equation. From this, we calculate the self-energy and then the spectral density function for the system. We apply the formalism to the case of the simple cubic lattice and compute the density of states, the spectral density function and the lifetime of the magnons as a function of energy  for several values of the mean concentration of nonmagnetic ions in the ferromagnetic lattice. We calculate the magnon energy spectrum as a function of the average  impurity  concentration fraction $c$, which shows that for low lying states, the excitation energy increases continuously with $c$ in the studied range $0.1 \le c \le 0.7$. We also use the spectral density function to compute some thermodynamical quantities through the magnon occupation number. We have obtained closed form expressions for the configurationally averaged physical quantities of interest  in a unified fashion as functions of the mean concentration  of nonmagnetic impurities $c$ to any order of $c$  applicable below a critical percolation concentration $c_p$. The quantities of interest comprise the thermodynamic potential (free energy), the spin-wave self-energy and the spectral density function from which other quantities can be derived.
\end{abstract}

\end{frontmatter}

\section{Introduction}
\label{sec:intro}
There has been  a resurgence of interest on the effects of impurities in spin-1/2 Heisenberg chains  recently due to experimental realizations in solid state systems and in particular in ultracold gases  \cite{bohrdt2018dynamic}. The impurities effectively can display as missing sites or couplings, which give rise to isolated finite chain segments, that  attain characteristic boundary correlation functions, which would lead to impurity-induced changes in the Knight shift, the susceptibility, the static structure factor, and the ordering temperature \cite{bohrdt2018dynamic}.
 
Similarly, the discovery of a ferromagnetic transition at temperatures above 100 K in the diluted III-V magnetic semiconductors,  actualized by doping a semiconducting host material with low concentrations of magnetic impurities, has generated a great deal of interest from both the experimental and theoretical vantage point due to their potential in spintronics applications \cite{ohno1998making,vzutic2004spintronics,macdonald2005ferromagnetic}. The recent advancements of the field of magnonics in general, which address the use of spin waves (magnons) to transmit, store and process information, and associated computing have been subject of recent reviews \cite{barman2021,chumak2022roadmap}.

Actually, the effect of disorder on magnetism is an old issue in diluted spin systems \cite{Wolfram_Callaway_1963,takeno1963spin,izyumov1966spin,izyumov1966theory,murray1966low,murray1966determination, Hone_et_al_1966,takeno1968low,kaneyoshi1969contribution,edwards1971green,foo1972coherent,tahir1972spatially,matsubara1973application,tahir1977effects,dvey1978magnetic}, albeit only during the last decade or so disorder effects in magnetic semiconductors have   been considered. Most of the past experimental findings have been on antiferromagnetic materials \cite{cowley1972properties,coombs1975magnetic} with some in  certain ferromagnets \cite{korenblit1978ferromagnetism}. However, magnetic excitations, e.g. in  the dilute two-dimensional ferromagnetic Heisenberg system K$_2$Cu$_{1-x}$Zn$_{x}$F$_4$,  in which spin-1/2 for the Cu ions have been observed \cite{wagner1978spin,wagner1979spin}.  Another compound, the mixed spinal ferrite Mg$_{1-x}$Zn$_x$Fe$_2$O$_4$ can for instance serve as a pertinent example, in which Zn ions, distributed randomly on the octahedral sites, give rise to many interesting phenomena depending on the value of $x$ \cite{daSilva2010effect,saqib2019structural}. 

In this paper, we will revisit an old problem of spin wave theory in randomly disordered magnetic systems, where nonmagnetic impurities are randomly distributed in a ferromagnetic lattice. In the first part of the paper, we utilize the Matsubara imaginary-time formulation of finite-temperature many body physics \cite{Matsubara_1955,AGD_1963,Coleman_2015} to calculate through the partition function the free energy of the system under consideration, from which other thermodynamic quantities can be derived. In the second part of the paper, the same physical problem will be tackled by means of time-dependent Green functions \cite{Bogolyubov_1959,Bonch-Bruevich_1962,Zubarev_1960,rudoy2011bogoliubov}. This technique has been used in various branches of quantum statistical physics \cite{Kubo_1957,AGD_1963,Coleman_2015} and has also been turned out to be very useful in theory of magnetism \cite{Oguchi_Honma_1963,Callen_1963,Tahir-Kheli_1964,Tyablikov_1967,nolting2009quantum}. The main advantage of the thermodynamic or temperature  Green function technique (as is sometimes called) is its physical interpretation of spin waves  in terms of the quasiparticle concept.  Before proceeding with the present method, a brief account of the development of theoretical approaches  would be useful to put the context of the present paper among such diverse approaches developed over the years.

\subsection{Development of theoretical approaches}\label{S:survey}
There has been a good amount theoretical work on dilute ferromagnets in the past, where the host material, usually nonmagnetic, is doped with a small concentration of magnetic ions. Brout in 1959 \cite{brout1959statistical}  developed a statistical mechanical model of a random ferromagnetic system in which paramagnetic impurities were exchange-coupled in a nonmagnetic substrate. Using the methods of the cluster expansion of the partition function and semi-invariants, Brout calculated the free energy averaged over random sites. The Curie temperature $T_c$ for the system was calculated as a function of nonmagnetic ion concentration in the weak dilute regime. It was argued that in the limit of the long-range exchange interaction, compared to interatomic spacing,  $T_c$ increases  linearly with the concentration, whereas for the short-range  interaction this increase is highly nonlinear at very low concentrations.

Following Brout's work, the problem attracted a fair deal of attention. Elliott \cite{elliott1960some} examined the problem within the constant-coupling approximation. Smart \cite{smart1960behavior} and subsequently Charap \cite{charap1962critical} generalized the Bethe-Peierls-Weiss cluster procedure to evaluate the concentration dependence of the $T_c$.  Elliott and Heap \cite{elliott1962theory} examined the behavior of the paramagnetic susceptibility of the system as a power series in the magnetic concentration.  Their calculations confirmed Brout's conjecture, namely at concentrations near the pure limit, the dependence of $T_c$ on concentration was by and large linear, but the extrapolation to the critical concentration limit was prone to a large uncertainty.

Wolfram and Callaway \cite{Wolfram_Callaway_1963}  and independently Takeno \cite{takeno1963spin} considered the effect of a single substituted impurity ion on the spin-wave spectrum of an insulating ferromagnet.  They supposed that the spin of the impurity ion and the effective exchange interaction are different from those of the host atoms.  They noted that by using the Heisenberg spin chain model on a simple cubic lattice with a single magnetic ion impurity, the wave function for a spin wave associated with a defect must transform according to one of the irreducible representations of the cubic point group. In particular, there are three types of magnon impurity modes, namely s-like, p-like and d-like modes or waves corresponding to the representations of the cubic point group $O_h$.  Among these modes,  the s-like mode is of particular interest, which directly is associated with the motion of the impurity spin.

 Izyumov and Medvedev \cite{izyumov1965impurity,izyumov1965properties,izyumov1966peculiarities,izyumov1967incoherent,izyumov1973magnetically}, Izyumov \cite{izyumov1966spin,izyumov1966theory} treated  an analogous problem as Wolfram \& Callaway \cite{Wolfram_Callaway_1963} and Takeno \cite{takeno1963spin}, namely a Heisenberg ferromagnet in a cubic lattice containing a low concentration of impurity atoms with different spin and exchange integrals. They calculated the Green function of one-magnon excitations by series expansion in powers of the perturbation introduced by the impurity and then averaging the terms of the series over all possible configurations of the impurities. From this, they calculated the density of states for the spin waves in the low frequency limit. Subsequently,  Izyumov and Medvedev treated the case where the impurity is nonmagnetic \cite{izyumov1967incoherent,izyumov1973magnetically}.

 Subsequently, Murray \cite{murray1966low,murray1966determination}  studied a ferromagnetic Heisenberg system with a random arrangement of two different types of atom with spins $S_1$ and $S_2$ and different exchange couplings. She treated the low-frequency, long-wavelength limit ($ka \ll 1$, $k$ the wave vector magnitude, $a$ the lattice constant) by combining a perturbation scheme with a variational calculation.  This lead to spin-wave energies or a dispersion relation at low concentration of magnetic ions in the form $E(k) =2SJ(ka)^2\phi(p)$, where $S$ is the magnitude of the spin, $J$ is the exchange integral and  $\phi(p)$ is called a "stiffness coefficient", which only depends on a fraction   $p$  of magnetic atoms and the lattice structure \cite{murray1966low,murray1966determination}, with $\phi(1)=1$.   Murray's approximative calculations indicated that there is a region above the percolation threshold $p_c$ where ferromagnetism is unstable.\footnote{The critical concentration $p_c$ of classical percolation theory is the concentration at which infinitely extended clusters of magnetic ions open up, a phenomenon  associated with the onset of ferromagnetism.} However, by correcting the errors in Murray's calculations,  Last \cite{last1972percolation} has shown that Murray's  method cannot give an accurate estimate for $p_c$. Last's arguments indicate that $\phi(p) \to 0$ as $p \to p_c$. On the same issue, Kumar and Harris \cite{kumar1973spin} noted that one should distinguish between $p_c$ and the critical concentration $x_c \equiv p^\ast$  for the occurrence of long-range order in the zero-temperature limit. In particular, they argued that $p_c$ is close, if not exactly equal to $x_c$. Furthermore, they noted that Murray's model, (i) allows $\phi(p)$ to be negative, although the system becomes ferromagnetic, and (ii) has a finite bound for $\phi(p)$  when  $p<p_c$ despite that there is no long-wavelength spin waves in this domain.
 
 Hone, Callen and Walker \cite{Hone_et_al_1966}, about the same time as Murray, explored the thermodynamic properties of a single substituted impurity ion on the spin-wave spectrum of an insulating ferromagnet. In particular, they set up the equations of motions of the  temperature-dependent, double-time, retarded Green functions for the spins in the distorted lattice and calculated the magnetization of the impurity ion, and the energy and weight of the s-state localized mode, as a function of temperature in the simple cubic lattice.
 
 In a later paper, Takeno and Homma \cite{takeno1968low} investigated the low-energy magnon spectrum and some physical properties at low temperatures of an impure Heisenberg ferromagnetic lattice in which each impurity spin is weakly exchange-coupled with the rest of the lattice \cite{takeno1968low}. They obtained analytic expressions for low-energy magnon resonant modes, which do not seem to depend sensitively on a specific lattice and impurity models. Moreover, they calculated the effect of the low-lying magnon impurity mode on the low-temperature magnon specific heat. Their  main result  was that an impurity spin, which is weakly exchange-coupled with its surroundings in a Heisenberg ferromagnetic cubic lattice, is very likely to produce a sharp low-lying  s-like resonant magnon mode. All the aforementioned investigators were able to show the existence of a spin wave mode localized on the magnetic ion impurity, in addition to the presence of modified spin wave band of the nonmagnetic host lattice.

Kaneyoshi \cite{kaneyoshi1970contribution} studied a spin-wave theory of a dilute Heisenberg ferromagnet, i.e. magnetic atoms randomly embedded in nonmagnetic lattice,  by using  a Green function technique \cite{kaneyoshi1969contribution}. The Green function for one-magnon excitations was calculated in two ways, a decoupling method and a diagram method. An expression for $\phi(p)$ was approximately computed for the simple cubic lattice, which was used to determine the critical impurity concentration for which the ferromagnetic ground state gets unstable with respect to the formation of long-wavelength spin waves.  In particular, Kaneyoshi obtained  $p_c=0.329$  \cite{kaneyoshi1969contribution}.  In a subsequent paper, Kaneyoshi  \cite{kaneyoshi1970contribution} generalized  his method to finite temperature range, and studied the dilute Heisenberg ferromagnet in two ways, the molecular field approximation and the Tyablikov approximation (equivalent to random-phase approximation or RPA). In the molecular field approximation, he obtained  an expression for the averaged spin moment at any lattice site by a diagram method. For the simple cubic crystal, numerical computations of the averaged moment with a spin-1/2 host were presented and the Curie temperature depending on the concentration of magnetic atoms was calculated.

 Edwards and Jones \cite{edwards1971green},  using a method similar to that developed earlier by Kaneyoshi  \cite{kaneyoshi1969contribution},  studied the  behavior of spin waves in a dilute ferromagnetic system where the spins occupy random positions on a cubic lattice \cite{edwards1971green}. They treated the low-frequency long-wavelength limit  and calculated the Green function from its equations of motion using the Tyablikov decoupling approximation. Upon the impurity averaging the Green function,  they evaluated the self-energy for the system. The averaged Green function provides  the renormalization and damping of a spin wave mode due to disorder.  They showed that the vacancy (or nonmagnetic ion) concentration $c\equiv(1-p)$ on the lattice offers an appropriate expansion parameter in the perturbation (Born) series for the self-energy. More specifically, Edwards and Jones  by a diagrammatic approach treated the scattering of spin waves with impurities and calculated the real part of the self-energy in the frequency domain $\omega$, $\mathrm{Re}\Sigma(k,\omega)$ up to order $c^2\equiv(1-p)^2$ in the long-wavelength limit. The spin-wave energy $E(k)\sim\mathrm{Re}\Sigma(k,\omega)=2SJ(ak)^2\phi(c)$ calculated as the value of $\omega$ for which the Green function exhibits a sharp peak,  vanishes at some critical concentration $c=c_c$ where $\phi(c_c)=0$.  Edwards and Jones \cite{edwards1971green} derived several approximate expressions for $\phi(c)$, which for the simple cubic lattice yields $c_c$-values in the range of 0.609 to 0.653, which are lower than the value computed by Kaneyoshi \cite{kaneyoshi1969contribution}, i.e.  $c_c=0.674$. The critical concentration $c_c$ can be compared with the percolation concentration $c_p=0.693$ for the simple cubic lattice \cite{sykes1964critical}. The Edwards-Jones treatment \cite{edwards1971green} is confined to low energy excitations of a magnetically ordered lattice with a small concentration of nonmagnetic impurities, $c<0.5$.

In the aforementioned studies \cite{izyumov1966spin,murray1966low,murray1966determination} including those of Kaneyoshi  \cite{kaneyoshi1969contribution} and Edwards and Jones \cite{edwards1971green}, the higher terms in the impurity concentration were ignored in the calculation of the Green function. Thereby, the validity of accuracy of the calculated concentration dependence of the stiffness coefficient $\phi(p)$ remained irresolute. Furthermore, these authors derived the critical concentration $p_c$ from the condition that the long-wavelength spin waves become unstable, which may not be the same concentration at which the critical (Curie) temperature $T_c$ vanishes. Matsubara \cite{matsubara1973application} generalized the foregoing calculations of spin waves in random spin system to higher concentrations by employing a coherent  potential approximation (CPA) in a random lattice setting. He derived an energy  dispersion relation  and a self-energy  $\Sigma(k,\omega)$, which he calculated for cubic lattices, expressed in terms of  a $3 \times 3$-matrix function. He further showed that in the long-wavelength or infrared (IR) limit, $ka \ll 1$,  $\Sigma(k,\omega) =\phi(p)(ka)^2$ with $\phi(p)$ expressed in terms of a matrix series, which can be truncated by a function with good approximation. The obtained formulae should be valid to higher concentrations of spin, however, no explicit computational results are shown in \cite{matsubara1973application}. Moreover, Matsubara in \cite{matsubara1973application} derived general formulae for the Green function, the spectral density function and density of states, but without any numerical computations. He, nevertheless, outlines a  computation scheme for these quantities.

 In another approach, Tahir-Kheli \cite{tahir1972spatially} proposed a model of dilute Heisenberg ferromagnet in which bonds (exchange interactions) are removed between pairs of sites at random. His method is equivalent to CPA and gives  a spin-wave Green function, $G(\vec{k},\omega)$, as a function wave vector and frequency in appropriate form through which he calculated the spectral density function and subsequently the density of states as a function of magnetic ion concentration. In addition, he found that the spin-wave stiffness coefficient varies linearly with the concentration of missing bonds ($ \equiv c$) at all concentrations. If the concentration of bonds is interpreted as the concentration of magnetic sites, the spin-wave curve is in good agreement with the Monte Carlo simulations reported in  \cite{Kirkpatrick1973nature,kirkpatrick1973percolation}, but this is not sufficient for a theoretical justification. Apparently this bond model yields an incorrect dilute ferromagnet limit, because it neglects the correlations between the bonds removed around a missing magnetic site. Nonetheless, Tahir-Kheli's approximation appears to be quantitatively fair at higher magnetic ion concentrations where all CPA treatments have impediments.
Moreover,  Tahir-Kheli in  \cite{tahir1972spatially} has treated an approximation to a "bond" problem in percolation theory rather a "site" problem, which is suitable to a dilute magnet. The critical concentration differs  in these two problems \cite{shante1971introduction}.

The CPA, in its original form, could only be used if the impurity terms in the hamiltonian are site-diagonal.  Takeno \cite{takeno1968solution}  derived an expression to remove this impediment, however, no parameter for impurity-impurity coupling appears in his results. The literature on theories and properties of randomly disordered crystals and related physical systems including CPA and its applications up to 1974 has been reviewed in \cite{elliott1974theory,Rickayzen_1984}. A detailed review of the fundamental aspects  of CPA is given in  \cite{yonezawa1973coherent}; see also \cite{nolting2008fundamentals}.  Elliott and Pepper \cite{elliott1973excitations} studied spin waves in randomly dilute Heisenberg ferromagnets and antiferromagnets adapting  Takeno's  CPA approach \cite{takeno1968solution} for the special case where the defect creates a site-diagonal perturbation. In the dilute ferromagnet problem, because a nonmagnetic impurity affects also neighboring sites,  Elliott and Pepper interpreted the CPA self-consistency equation as a matrix equation in the space of the vacancy and its nearest neighbors, which is also an effective medium CPA model. They considered spin waves in a cubic crystal as in \cite{Wolfram_Callaway_1963} with three types of local (magnetic) oscillations: s--, p--, and d--waves noted earlier. They derived a set of equations for the self-energies for the respective modes, which were then computed numerically.  They computed the self-energies as a function of energy (or frequency $\omega$) and the nonmagnetic impurity concentration $c$. They further computed the line shapes for neutron scattering through the imaginary part of the Green function,  $\propto \mathrm{Im}G(\vec{k},\omega)$,  and  finally determined  the dispersion relation for the spin waves in the dilute ferromagnet crystal plus the density of states.

Numerical computations based on the Elliott-Pepper CPA indicate that results are satisfactory at high energies, i.e. the spin-wave peaks in $\mathrm{Im}G(\vec{k},\omega)$ behave sensibly. But, at low energies the  utilized CPA method fails to keep the s-wave resonance at $\omega = 0$; implying that the method cannot be used to obtain a meaningful result for the critical concentration  \cite{pepper1972}. In particular, Pepper's numerics \cite{pepper1972} show that at $c=0.5$ the spin-wave energy calculated by the CPA  becomes zero and  for  $c>0.5$ the energies are negative for small $k$.   In fact, any peak in $\mathrm{Im}G(\vec{k},\omega)$ for small  $k$
must appear for some $\omega< 0$, i.e.,  no peak should be present at $\omega> 0$. As a result, the obtained dispersion relation  goes to negative energy at some finite $k$ for all $c>0.5$, and the density of states comprised some states  for  $\omega < 0$ to remain finite into this region.  This kind of behavior is obviously unphysical. It is a consequence of the incorrect resonance in the self-energy of s-modes according to Pepper's analysis \cite{pepper1972}.

 Elliott and Pepper numerical computations of spin-wave density of states as a function energy or frequency $\omega$ for several values of nonmagnetic ion concentrations ($0.01\le c\le 0.6$) exhibit low-energy resonances for $c=0.1$ and $c=0.3$, which have peaks around $\omega \approx 0$ \cite{elliott1973excitations}. Pepper  \cite{pepper1972} did attempt to remove this low-frequency resonances by various stratagems, but to no avail.  Nevertheless, the work of Pepper and Elliott paved the way for a more satisfactory CPA treatment of the site problem for the dilute magnet as briefed next.

 A more satisfactory CPA treatment building on the Elliott-Pepper approach  \cite{ elliott1973excitations,pepper1972} and calculations in \cite{kumar1973spin}, was developed by Harris et al. \cite{harris1974excitations}  to describe spin waves in a dilute Heisenberg ferromagnet at $T = 0$. Harris et al. calculated the full scattering matrix (T-matrix) of an isolated vacancy in an effective medium and obtained the self-energy in self-consistent form. They  added an extra term to the Heisenberg hamiltonian, referred to as pseudopotential, to remove the spurious degrees of freedom associated with the fictitious spins on the vacancy sites, which showed up in the  Elliott-Pepper formulation. Harris et al. \cite{harris1974excitations} computed the spectral functions and density of states for such pseudopotentials (with different interaction constants)  numerically as a function of spin-wave frequency for various values of the nonmagnetic ion concentration $c$, and compared the results with some specific solutions, namely  Pad\'e approximants \cite{nickel1974method}, the effective exchange model of Tahir-Kheli \cite{tahir1972spatially}, and the  CPA results of Elliott and Pepper \cite{elliott1973excitations}.  Harris et al's  CPA \cite{harris1974excitations} at intermediate concentration $c \lesssim 0.4$  provides a satisfactory treatment of the spectral density function or $\mathrm{Im}G(k,\omega)$ and related properties in good agreement with the Pad\'e approximant results. However, near the critical percolation concentration $c_p \approx 0.7$,  none of the chosen pseudopotential constants led to satisfactory agreement with the Pad\'e approximant approach, which is considered as a benchmark.

In order to ameliorate the aforementioned calculations, Theumann and Tahir-Kheli \cite{theumann1975excitations} presented yet another CPA approach to the problem of the randomly diluted ferromagnet. The main difference between the Theumann--Tahir-Kheli approach and the work of Harris et al. \cite{harris1974excitations} is that the former authors avoid using the bosonic representation of the Heisenberg hamiltonian. Instead, they define an effective medium by means of nonlocal perturbation potentials generated by vacancies or nonmagnetic ions. Theumann and Tahir-Kheli  developed an alternative CPA formalism that is not based on the multiple scattering theory, but instead is based on the generalization of the so-called  path method \cite{brouers1975theories}  applied to the problem of localization of magnon states. They computed the spectral density function, the density of states and other properties and compared their numerical results \cite{theumann1975excitations} with those of Harris et al. \cite{harris1974excitations} and the Pad\'e procedure \cite{nickel1974method}. The accuracy of the Theumann--Tahir-Kheli results is fully comparable to that of \cite{harris1974excitations}, in the small- and intermediate-vacancy-concentration regimes ($c \lesssim 0.3$). For higher vacancy concentrations,  $0.4 \lesssim  c \lesssim 0.6$,  their results appear to be of better quality to those given in  \cite{harris1974excitations}.

The CPA results of Theumann and Tahir-Kheli \cite{theumann1975excitations} regarding the density of states and the dynamic structure factor  $S(\vec{k},\omega)\propto \mathrm{Im}G(\vec{k},\omega)$ for dilute ferromagnet have been compared with the results of direct numerical simulations with good agreement \cite{alben1977spin}. In a direct numerical simulation of the Heisenberg ferromagnet, Alben et al. \cite{alben1977spin} solved $N$ (= 10 000) simultaneous differential equations of motion for the Green function in cubic lattices through which they computed the density of states and  $S(\vec{k},\omega)$. They compared their  $S(\vec{k},\omega)$, as a function $\omega$ for $c=0.5$, with those of Harris et al.  \cite{harris1974excitations}, Theumann and Tahir-Kheli \cite{theumann1975excitations}  and the Pad\'e approximants \cite{nickel1974method}.  The agreement between the latter two works were excellent but the former deviated in the $\omega$-region where   $S(\vec{k},\omega)$ peaks.

The magnetic behavior of a disordered alloy A$_x$B$_{1-x}$ in which A (magnetic) and B (nonmagnetic) atoms are randomly  distributed on a cubic lattice has been investigated by means of a Heisenberg hamiltonian and a Green function method in \cite{dvey1978magnetic}. The authors in \cite{dvey1978magnetic} derived, within the linear spin-wave approximation, the Dyson equation for the configuration averaged Green function. This equation was decoupled, by a quadratic approximation, through which they calculated the energy dispersion relation and the density of states for the system. In addition, the critical temperature and the critical concentration $x_c$, below which no bulk ferromagnetism exists were calculated. As an example, for the case of the nearest-neighbor interactions between the spins on the simple cubic lattice, they obtained $x_c=0.296$ compared with the critical percolation value $p_c=0.307$  \cite{sykes1964critical}.

Salzberg et al.  \cite{salzberg1976spin} employed a Heisenberg hamiltonian for dilute ferromagnet using the cluster-Bethe-lattice (CBL) approximation \cite{katsura1974bethe} to compute the Green function and the density of spin-wave states as a function of spin-wave frequency for several values of $p$ in the range of 0.25 to 1.0 in body-centered cubic  and simple-cubic lattices with clusters of various sizes.  The utilized CBL method gives a plausible description of localized magnetic excitations but with $\delta$-function type anomalies. These anomalies are attributed (i) to isolated magnetic clusters, which result in localized modes comprising one zero-frequency mode, and to (ii) local excitations within the "bulk" ferromagnet which do not propagate beyond a few atoms due to fluctuations in the local configurations  \cite{salzberg1976spin}.  Salzberg et al.   showed that $T_c\to 0$ as $p \to p_c$.

 An extension of the conventional CPA  or generalized CPA \cite{blackman1971generalized} to account for the presence of disorder in spin chains both in off-diagonal and inhomogeneous terms that appears in the Dyson type equation for the spin-spin correlation (Green) function was considered in \cite{theumann1974generalized,lage1977generalized}. In particular, the diagrammatic series for this equation was expressed in terms of two quantities: the self-energy $\Sigma(\vec{k})$, arising solely from the intrachain interaction energy and a term referred to as the end correction  $\Delta(\vec{k})$, which accounts for the effect of disorder in the inhomogeneous term of the Dyson equation; see ref.  \cite{lage1977generalized} for details. The obtained equation for the averaged Green function is exactly equivalent to that obtained by Harris  et al.'s  more approximate treatment  \cite{harris1974excitations} if the hard-core potential constant,  introduced by Harris  et al. to remove the zero-frequency response from a vacancy, is set equal to 1. However, one cannot show that the expression for the the end correction  $\Delta(\vec{k})$ in \cite{lage1977generalized} is equal to the corresponding term in \cite{harris1974excitations}. A detailed treatment of the  critical properties of dilute Heisenberg magnet (bond- and site-diluted with concentration $p$) is given in \cite{stinchcombe1979critical}.  A scaling theory of critical phenomena has been used  to obtain the critical exponents. In particular, the critical curve (Curie temperature versus concentration $p$) of the three-dimensional diluted Heisenberg model  on simple cubic lattice is calculated.

The  specific heat and the dynamic structure factor  $S(\vec{k},\omega)$ for dilute one-dimensional  Heisenberg chain have been calculated analytically  at low temperatures \cite{mcgurn1983spin}, where the spin excitations are treated as free bosons. The latter quantity was determined from the Fermi Golden Rule for the scattering of a segment of $n$ spins, then  summed over $N$ segments to obtain the bulk properties   \cite{mcgurn1983spin}. The results for $S(\vec{k},\omega)$ versus $\omega$ for several values of $k$ and concentration $c$ were compared with  direct numerical calculations  \cite{alben1977spin}. As the magnetic ion concentration is decreased from $p=1$, McGurn and Thorpe \cite{mcgurn1983spin} found that the long-range periodicity breaks down. Accordingly,  not only does the $p=1$ spin wave peak broaden but new peaks do appear in $S(\vec{k},\omega)$.

The thermodynamic of the dilute one-dimensional Heisenberg ferromagnet in an external field was further investigated in \cite{rettori1988effect}; wherein the free energy of a chain segment was calculated for both the quantum case at low temperatures and the classical case, for a single segment with $n$-spins. Then the free energy of the dilute system was evaluated by summing over all the segments $N$, i.e. the total number of sites. Analytical expressions for the specific heat and the magnetic susceptibility as a function of temperature and  concentration were derived \cite{rettori1988effect}.

Hilbert and Nolting have studied the effect of substitutional disorder on the magnetic properties of diluted Heisenberg spin systems applicable to ferromagnetic diluted semiconductors \cite{hilbert2004disorder}. In such materials, a small fraction of the nonmagnetic host-semiconductor ions is replaced by ions, which have a localized magnetic moment or spins \cite{reed2001room,park2002group}. These magnetic ions are usually randomly distributed over the lattice sites. Hilbert and Nolting solved the equation of motion for the magnon Green function  numerically using the Tyablikov decoupling approximation for finite systems in a cubic lattice. They computed the spectral density as a function of magnon energy and through which estimated the magnetization and Curie temperature in the thermodynamic limit. The results of their computations indicate that, for short-range interaction, no ferromagnetic magnetic order exists below the critical percolation concentration, but  for  long-range interaction, the Curie temperature  increases linearly with the concentration of spins.

In a follow-up article, Tang and Nolting \cite{tang2007effects} studied the effects of both dilution and disorder on the magnetic behavior of diluted Heisenberg spin systems applicable to diluted magnetic semiconductors. They combined the methods of supercells \cite{martin2016interacting} and  augmented space formalism \cite{mookerjee1973averaged,mookerjee1973new} to  evaluate and appraise  the impact of position disorder of magnetic ions on magnetization and the Curie temperature in these systems. The size of the supercell determines the concentration of magnetic ions in the host materials. The computed spectral density was used to calculate the temperature dependence of magnetization and the Curie temperature of  the system. The method used by Tang and Nolting is applicable to the case of the finite size systems, but  it includes the long-range exchange integrals and treats the spins quantum mechanically.

Bouzerar and Bruno \cite{bouzerar2002rpa} developed a comprehensive model, based on the Green function formalism, for evaluating the magnetic attributes of disordered Heisenberg ferromagnets with long-range interactions.  The considered system was a binary alloy A$_{1-c}$B$_c$ where A and B can be either magnetic or nonmagnetic ions. They used the standard Tyablikov approximation  ($\equiv$ RPA) to decouple the statistically averaged Green functions in the equation of motion. Furthermore, the Green function equations were expressed in terms of $2 \times 2$ matrix by means of a generalized CPA which accounts for the presence of off-diagonal disorder \cite{blackman1971generalized,theumann1974generalized,lage1977generalized,whitelaw1981single}.  Bouzerar and Bruno treated simultaneously and self-consistently the RPA-CPA equations. A cumulant expansion method \cite{yonezawa1973coherent,Yonezawa_1968} was used  to express the averaged Green function in the momentum space in terms of the self-energy $\Sigma(\vec{k},\omega)$ and  the end correction  $\Delta(\vec{k},\omega)$, which accounts the effect of disorder. Thereafter, they evaluated $\Sigma(\vec{k},\omega)$ and  $\Delta(\vec{k},\omega)$ in power series expansion. The Bouzerar-Bruno model \cite{bouzerar2002rpa} provides a method  to compute the Curie temperature, the spectral functions, and the temperature dependence of the magnetization of each constituent as a function of concentration of impurity.
 Moreover,  they have proposed a simplified treatment of the p--, d--, f--scattering contributions of the self-energy which is difficult to treat analytically in the case of long-range interactions (exchange integrals).

 In the aforementioned generalized CPA approach of Theumann \cite{theumann1974generalized}  to the disordered Heisenberg ferromagnet of a binary alloy, the equation of motion for the Green function basically involves three kinds of disorder:  A diagonal disorder that depends explicitly on the site spin, an off-diagonal disorder depending on the exchange parameters, and  an environmental disorder, i.e. a term given by the static field induced at one site by the presence of neighboring spins. The former two types of disorder were properly treated in the Bouzerar-Bruno model, while for the latter one,  Theumann \cite{theumann1974generalized} utilized the virtual crystal approximation, where  fluctuations in the spin configuration surrounding a given A or B ion are neglected  and  the local static field is replaced by its average value. This is a kind of mean-field approximation, which causes discrepancy in the $n$th moment of the density of states for $n \ge 2$ for the dilute ferromagnet \cite{theumann1973dilute}.

Tang and Nolting \cite{tang2006magnetic} combined the methods of Theumann \cite{theumann1974generalized}  and the Bouzerar-Bruno RPA-CPA  for calculating the temperature dependence of magnetization and the Curie temperature of disordered Heisenberg binary (A$_{1-c}$B$_c$) spin system with long-range interaction. The  long-range exchange integrals used  comprised  a power-law decaying and an oscillating Ruderman-Kittel-Kasuya-Yosida (RKKY) exchange interaction. They calculated the magnon spectral density function self-consistently to obtain the magnetization and the Curie temperature on the simple  cubic  lattice for various values of $c$. The results indicate a strong influence of ferromagnetic long-range exchange integrals on  magnetization and the Curie temperature.

More recently, Buczek et al. \cite{buczek2016magnons} have studied spin excitation spectra of one-, two-, and three-dimensional magnets containing nonmagnetic impurities in a wide range of impurity concentrations starting from the Heisenberg model. They carried out their investigation by both direct numerical simulations in large supercells and using a  CPA method.  In their model, magnetic and nonmagnetic ions are randomly distributed on the sites of a crystal lattice. In their direct numerical simulations, the magnetic susceptibility was computed for  different configurations (typically between 100 and 1000) and subsequently averaged. They sampled the configurations by a Monte Carlo method, where the averaging was done and the standard deviation of the mean was computed. Buezek et al.'s CPA  is based  on the Matsubara -Yonezawa formulation \cite{matsubara1973application}, but  generalized  to  the case of complex crystals with multiple sites in the primitive cell, optional number of atomic species forming the disordered crystal, and  arbitrary dimensionality  \cite{buczek2016magnons}. It also takes into account the off-diagonal disorder as in \cite{lage1977generalized}. Buczek et al. through their CPA calculated the generalized average susceptibility in the momentum-frequency domain, viz.  $\bar\chi(\vec{k}, \omega)$. We note that the dynamic structure factor, determined in an inelastic neutron scattering experiment, is $S(\vec{k},\omega)\propto \mathrm{Im}\bar\chi(\vec{k},\omega)$ and the magnetization $M(\vec{k}, \omega)=\bar\chi(\vec{k}, \omega) B(\vec{k}, \omega)$ with  $B(\vec{k}, \omega)$ being the modulated magnetic field of wave-vector $\vec{k}$ and frequency $\omega$.

Buczek et al. \cite{buczek2016magnons} discussed the way a realistic electronic structure can affect the properties of imperfect magnets, by considering  Fe$_{1-x}$Al$_x$ compound (Fe magnetic, Al nonmagnetic) at several values of $x$ in three and two dimensions at zero kelvin. They computed   $\mathrm{Im}\bar\chi(\mathbf{k},E)$  versus $E$ (in meV) both by a Monte Carlo technique and a CPA for the three-dimensional Fe$_{0.7}$Al$_{0.3}$ at the wave-vectors (0.125,0,0)$2\pi/a$ and (0.375,0,0)$2\pi/a$ with remarkable agreement. However, to our knowledge,  no such measured data (usually obtained by neutron scattering) on this compound have been reported in the literature so that one could compare or verify these computational data.

The method of Buczek et al.   \cite{buczek2016magnons,buczek2018spin} has recently been applied by Paischer et al. \cite{paischer2021spin} to study the effect of temperature and disorder on the magnetic behavior  of the crystal Fe$_{1-x}$Co$_x$, a bi-magnetic alloy with respective Curie  temperatures of 1043 K (Fe) and 1388 K (Co). The CPA method of \cite{buczek2016magnons} is augmented with an RPA method \cite{bouzerar2002rpa,Callen_1963} to include the influence of temperature. The results of the $T_c$ computations using this CPA-RPA method as a function of $x$ in the range of $x=0.1$ to  $x=0.5$, assuming a bcc lattice for the compound, exhibit overestimations of the measured values. This is partly attributed to by not accounting  a structural phase transition of the Fe-Co system at elevated temperatures,  which is expected to affect the Curie temperature.

Paischer et al. \cite{paischer2021spin}  also computed  $\mathrm{Im}\bar\chi(\vec{k},E)$ (equivalent to the spectral function) and fitted the data to a Lorentzian function from which the full width at half maximum or the inverse lifetime of magnon $\Gamma(\vec{k})^{-1}$ is determined. The results are presented in terms of  $\Gamma(\vec{k})^{-1}$ for certain $k$ values (points in the Brillouin zone) (i) as a function Co concentration at zero kelvin and (ii) as a function of temperature in the ferromagnetic domain at $x=0.2$. The computational data show that for different modes with different $k$, the density of available finite states will vary during a temperature rise.  As the temperature was raised, the normalized widths increased for low-energy acoustic magnons, but decreased for magnons at the top of the acoustic branch and in the optical branch  \cite{paischer2021spin}. The magnon width varied with $x$ in different ways depending on the selected $k$ values. The results of these computations have not yet been compared with experimental data.

\subsection{Method and outline of the present paper}\label{S:outline}
 In the present paper, we start from the Heisenberg hamiltonian  of ferromagnetism and represent  the spins  in terms of  Boson creation and annihilation operators. In the first part of the paper, we use Matsubara's perturbation method \cite{Matsubara_1955} to expand the partition function and therefrom calculate the free energy in terms of the mean concentration of nonmagnetic ions. For the sake of mathematical simplicity, we limit our investigation to the case of one-component lattice, however in principle, there is no technical impediment  in extending our formalism to multicomponent lattices, e.g. the spinel crystal structures. The basic assumptions of our model are:
\begin{itemize}
\item The nonmagnetic impurities are assumed to be quenched  at their lattice positions, i.e.  their positions are fixed.
\item The magnon-magnon interaction, which gives rise to short-wavelength spin waves, is neglected, i.e. the system is in the low temperature limit \cite{Tyablikov_1967,kittel1987quantum}.
\item The Heisenberg model is constructed in a Bravais lattice with nearest-neighbor interaction in the presence of an external magnetic field.
\item Our method is applicable to a regime where the nonmagnetic impurity concentration is below the critical percolation concentration $c_p$ , i.e. $0 \le c <c_p$, where for the simple cubic lattice $c_p\approx 0.7$.
\end{itemize}
As a consequence of the first item, the concentration field of lattice impurities (nonmagnetic ions) is stationary and here is treated as a c-number.

In the second part of the paper, we set up the double-time single particle Green function at temperature $T$ in momentum (wave-vector) space in terms of magnon operators. We derive the equation of motion for the Green function through the Heisenberg equation of motion and solve the equation. From that, we calculate the self-energy function and subsequently the spectral density function for the system. Next, we perform averaging over impurity concentration in a  scheme and express all the quantities of interest in terms of the nonmagnetic ion or impurity mean concentration. The Green function technique utilized is not limited to low temperature domains, hence the aforementioned second item  can in principle be avoided at the expense of some additional calculations.
 
 The principal merit of our method perhaps is that we obtain closed form expressions for the configurationally averaged physical quantities of interest  in a unified fashion as functions of the mean concentration (fraction) of nonmagnetic impurities $c$ to any order of $c$  applicable below a critical percolation concentration $c_p$. The quantities of interest comprise the thermodynamic potential (free energy), the spin-wave self-energy and the spectral density function from which other quantities can be derived. For example,  from the self-energy expression, we have calculated the magnon lifetime for a range of impurity concentrations as a function of frequency in the simple cubic lattice.

In Section \ref{sec:model}, we describe the model hamiltonian for the system under consideration, where it is expressed in terms of Bose operators in the wave vector representation. In Section \ref{sec:matsubara}, we set up the partition function and employ the Matsubara method to calculate the density matrix of the canonical ensemble and the associated S--matrix. The partition function is then calculated from the ensemble average of the S--matrix. The thermodynamic potential or the  free energy (the interaction part) is the logarithm of  the ensemble averaged S--matrix, expressed as the sum of closed loop connected diagrams in a perturbation series. Next, we average the free energy over the impurity distributions and express the mean free energy as a function of the mean impurity (nonmagnetic ion) concentration.

In Section \ref{sec:spinwave}, we present the spin wave formalism for the model. We set up the equation of motion for the Green function, and by solving it, we derive the relations for the self-energy, the magnon lifetime,  and the spectral density function. The averaging  of these quantities over impurity distribution is also done in this section. In Section \ref{sec:result}, we present the main results of our calculations by applying the obtained formulae to the case of the simple cubic lattice. We compute the density of states, the spectral density function and the lifetime of the magnons as a function of energy (frequency) for several values of impurity concentrations. We discuss the results in Section \ref{sec:discuss} and conclude the paper with  some remarks in Section \ref{sec:conclude}. Some technical details are relegated to six appendices.

\section{The Model}
\label{sec:model}
We consider a Heisenberg model for ferromagnet on a Bravais lattice of $N$ sites containing $n$ nonmagnetic impurities. The impurity  in the system is characterized by a random variable $c_j$, with $c_j=1$, if the site $j$ is occupied by an impurity  and $c_j=0$, otherwise.  Alternatively, we may consider that the site $j$ is occupied by a spin (magnetic ion) with a probability $p_j$ and unoccupied (nonmagnetic ion impurity) with probability $1-p_j$, independently of other sites, so $p_j \equiv (1-c_j)$. The Heisenberg hamiltonian for the system is
\begin{equation}
\label{eqn:hh1}
\mathcal{H} = -\frac{1}{2}\sum_{i,j}J(|\mathbf{r}_i-\mathbf{r}_j|)p_i p_j \mathbf{S}_i\cdot\mathbf{S}_j-B\sum_{j}p_jS_j^z\; ,
\end{equation}
\noindent
where $\mathbf{S}_j$ is the spin vector operator (magnetic moment) at site $j$, $J(|\mathbf{r}_i-\mathbf{r}_j|)>0$ is the exchange interaction integrals, depending only on the distance between the sites,  $B \equiv \mu_B H/2$, $\mu_B$ is the Bohr magneton, and $H$ is an external magnetic field. Furthermore, the three spin components at any given site obey the SU(2) (angular momentum) commutator algebra
\begin{equation}
\label{eqn:spin-commute}
[S_i^\mu, S_j^\nu] = i\hbar\delta_{ij}\varepsilon^{\mu\nu\sigma}S_k^\sigma,
\end{equation}
\noindent
where $\mu,\nu,\sigma$ stand for the Cartesian indices, $\varepsilon^{\mu\nu\sigma}$ is the antisymmetric Levi-Civita  tensor and $\hbar$ is the reduced Planck constant which we set $\hbar=1$, otherwise noted.

We express now spin operators for $S \ge 1/2$, $\mathbf{S}_j=(S_j^x,S_j^y,S_j^z)$ and $S_j^\pm=S_j^x \pm iS_j^y$ by the Dyson-Maleev representation \cite{Dyson_1956a,Maleev_1958,Tyablikov_1967,nolting2009quantum} in the form
\begin{equation}
\label{eqn:spin-compx}
S_j^+  = \sqrt{2S}\Big(1-\frac{n_j}{2S}\Big)b_j;\quad
S_j^-  =  \sqrt{2S}b_j^\dagger;\quad
S_j^z  = S-n_j,
\end{equation}
where   $n_j=b_j^\dagger\,b_j$ and $b_j^\dagger$ ,  $b_j$ are Bose  creation,  annihilation operators  satisfying the commutator algebra: $[b_i,b^\dagger_j] = \delta_{i,j}$, $[b_i,b_j] = [b^\dagger_i,b^\dagger_j] = 0$. We also set $\tilde{c}_j=c_j-c$, where $c=n/N$ is the mean concentration of nonmagnetic ions (impurities) per lattice site, $n$ is the total number of nonmagnetic ions in the system, $\sum_j \tilde{c}_j = 0$, and  $J=\frac{1}{N}\sum_{i,j}J(|\mathbf{r}_i-\mathbf{r}_j|)$.

We first write Eq. \eqref{eqn:hh1} in the form
\begin{equation}
\label{eqn:hh2}
\mathcal{H} = -B\sum_j p_jS_j^z -\frac{1}{2}\sum_{i,j}p_ip_jJ_{ij}\Big[\frac{1}{2}\big(S_i^+ S_j^-+S_i^-S_j^+\big)+S_i^zS_j^z\Big],
\end{equation}
\noindent
where  $J_{ij}\equiv J(|\mathbf{r}_i-\mathbf{r}_j|)$. Next, inserting Eq. \eqref{eqn:spin-compx} into Eq. \eqref{eqn:hh2} and writing  the hamiltonian as: $\mathcal{H} = E_0 +\mathcal{H}_2 + \mathcal{H}_2^\prime +  \mathcal{H}_4^\prime$, where
\begin{subequations}
\begin{eqnarray}
\label{eqn:E0}
E_0 &=& -BN(1-c)S-\frac{1}{2}N(1-c)^2S^2J,\\
\label{eqn:h2}
\mathcal{H}_2 &=& B(1-c)\sum_{j}b_j^\dagger b_j + (1-c)^2 S \sum_{i,j}J_{ij}(b_i^\dagger b_i - b_i^\dagger b_j),\\
\label{eqn:h2p}
\mathcal{H}_2^\prime &=& -B\sum_{j}b_j^\dagger b_j \tilde{c}_j - 2(1-c) S \sum_{i,j}J_{ij}(b_i^\dagger b_i - b_i^\dagger b_j)\tilde{c}_i,\\
\label{eqn:h4p}
\mathcal{H}_4^\prime &=& S\sum_{i,j}J_{ij}(b_i^\dagger b_i - b_i^\dagger b_j)\tilde{c}_i \tilde{c}_j.
\end{eqnarray}
\end{subequations}
Here, we have neglected the contribution of higher order Bose operators, $O(b_j^3)$, etc. to the hamiltonian \cite{Weiss_1964a}. Fourier transforming now $b_j$, $\tilde{c}_j$  and $J_{ij}$  from the  lattice  spatial coordinates to the lattice momentum space $\mathbf{k} \equiv \vec{k}$ according to
\begin{eqnarray}
\label{eqn:b-k}
b_j &=& N^{-1/2}\sum_\mathbf{k} e^{i \mathbf{k} \cdot \mathbf{r}_j} b_\mathbf{k}; \quad b_j^\dagger = N^{-1/2}\sum_\mathbf{k} e^{-i \mathbf{k} \cdot \mathbf{r}_j} b_\mathbf{k}^\dagger,\\
\tilde{c}_j &=& N^{-1/2}\sum_{\mathbf{k}}e^{i\mathbf{k}\cdot \mathbf{r}_j}c_\mathbf{k}; \quad J_{ij} = N^{-1}\sum_\mathbf{k} J_\mathbf{k} e^{i \mathbf{k} \cdot (\mathbf{r}_i-\mathbf{r}_j)},
\label{eqn:J-k}
\end{eqnarray}
we obtain
\begin{subequations}
\begin{eqnarray}
\label{eqn:ft-h2}
\mathcal{H}_{2} &=& \sum_\mathbf{k}\epsilon_\mathbf{k}\,b_\mathbf{k}^\dagger b_\mathbf{k},\\
\label{eqn:ft-h2p}
\mathcal{H}_2^\prime &=& -\frac{1}{\sqrt{N}} \sum_{\mathbf{k},\mathbf{q}} \gamma_{\mathbf{k}} b_{\mathbf{k}+\mathbf{q}}^\dagger b_{\mathbf{k}}
c_{\mathbf{q}},\\
\label{eqn:ft-h4p}
\mathcal{H}_4^\prime &=& \frac{1}{N} \sum_{\mathbf{k},\mathbf{q}_1,\mathbf{q}_2}\Delta_{\mathbf{k},\mathbf{q}_2}
b_{\mathbf{k}+\mathbf{q}_1+\mathbf{q}_2}^\dagger b_{\mathbf{k}}c_{\mathbf{q}_1}c_{\mathbf{q}_2},
\end{eqnarray}
\label{eqn:ft-h}
\end{subequations}
where $\epsilon_\mathbf{k} \equiv \alpha +(1-c)^2\varepsilon_{\mathbf{k}}$, $\alpha \equiv 2(1-c)B$, $\varepsilon_\mathbf{k}=SJ_{0\mathbf{k}}$,  $J_{0\mathbf{k}} \equiv J_0-J_\mathbf{k}$, $\gamma_{\mathbf{k}} \equiv 2[B+(1-c)\varepsilon_\mathbf{k}]$, $J_0 \equiv J_{\mathbf{k}=0}$, $J_\mathbf{k} = \sum_n J(\boldsymbol{a}_n) e^{-i\mathbf{k} \cdot \boldsymbol{a}_n}$,  $\boldsymbol{a}_j$ is a vector directed from one site to a nearest neighbor site $j$, and $\Delta_{\mathbf{k},\mathbf{q}_2} \equiv 2S(J_{\mathbf{q}_2}-J_{\mathbf{k}+\mathbf{q}_2})$ with $\Delta_{\mathbf{k},0}=\varepsilon_\mathbf{k}$. Here, $c_\mathbf{q}$ describes the field of nonmagnetic impurity  in the momentum  (wave vector) space $\mathbf{q}$, defined as $c_\mathbf{q} =  N^{-1/2}\sum_{j}  \tilde{c}_j e^{-i \mathbf{q} \cdot \mathbf{r}_j}$
where  $r_j$ is the position at site $j$  with nonmagnetic impurity concentration $c_j$  subtracted from its mean value $\bar c_j =c$ before Fourier transformation; $c_\mathbf{q}$ is a c-number. The  operators $b_\mathbf{k}^{\dagger}$ and $b_\mathbf{k}$ obey the usual Boson commutation rules: $[b_\mathbf{k},b_\mathbf{k^\prime}^{\dagger}] = \delta_{\mathbf{k},\mathbf{k^\prime}}$, $[b_\mathbf{k},b_\mathbf{k^\prime}] = [b_\mathbf{k}^{\dagger},b_\mathbf{k^\prime}^{\dagger}]=0$.

\begin{figure}[htbp]
   \centering
\begin{tikzpicture}[line width=1.5 pt, scale=1]
	\draw [fermion](2,0) -- (3.5,0);
	\draw[fill=black] (3.5,0) circle (.075cm);
	\draw[vector] (3.5,0)--(3.5,1.25);
	\node[red] at (3.5,1.25) {\LARGE${\times}$};
	\draw [fermion](3.5,0) -- (5,0);	
	\node at  (2.75,0.3) {\small$\mathbf{k}$};
	\node at  (3.25,0.65) {\small$\mathbf{q}$};
	\node at  (4.25,0.3) {\small$\mathbf{k}+\mathbf{q}$};
	\node at (3.5,-0.75) {$(a)$};
\begin{scope}[shift={(9,0)}]
	\draw[fermionbar] (0:1.75)--(0,0);
	\draw[fill=black] (0,0) circle (.075cm);
	\draw[vector] (100:1.45)--(0,0);
	\draw[vector] (80:1.45)--(0,0);
	\draw[fermion] (180:1.75)--(0,0);
	\node at (120:1) {\small$\mathbf{q}_1$};
	\node at (60:1) {\small$\mathbf{q}_2$};
	\node at  (-1,0.35) {\small$\mathbf{k}$};
	\node at (15:1.2) {\small$\mathbf{k}+\mathbf{q}_1+\mathbf{q}_2$};
	\node[red,line width=5 pt] at (0,1.45) {\huge{$\boldsymbol{\times}$}};	
	\node at (0,-0.75) {$(b)$};
\end{scope}
\end{tikzpicture}
\caption{\small{(a) $\mathcal{H}_2^\prime \sim \sum_{\mathbf{k,q}}\gamma_\mathbf{k} b_\mathbf{k+q}^{\dagger}b_\mathbf{k}c_\mathbf{q}$: The incoming magnon $b_\mathbf{k}$ with momentum \textbf{k} collides with the field of an stationary impurity $c_\mathbf{q}$ (red cross) through momentum \textbf{q} (wavy-line), then an outgoing recoiled magnon $b_\mathbf{k+q}^{\dagger}$  with momentum \textbf{k}+\textbf{q} is emitted. The black dot denotes the vertex of the interaction. (b) $\mathcal{H}_4^\prime\sim \sum_{\mathbf{k},\mathbf{q}_1,\mathbf{q}_2}\Delta_{\mathbf{k},\mathbf{q}_2}
b_{\mathbf{k}+\mathbf{q}_1+\mathbf{q}_2}^\dagger b_{\mathbf{k}}c_{\mathbf{q}_1}c_{\mathbf{q}_2}$: Simultaneous absorption and emission of magnon of momentum \textbf{k} interacting with the field of  an impurity through  $\mathbf{q}_1$ and  $\mathbf{q}_2$, then emerging with momentum $\mathbf{k}+\mathbf{q}_1+\mathbf{q}_2$.}}
\label{fig:mag-defect-scat-1}
\end{figure}
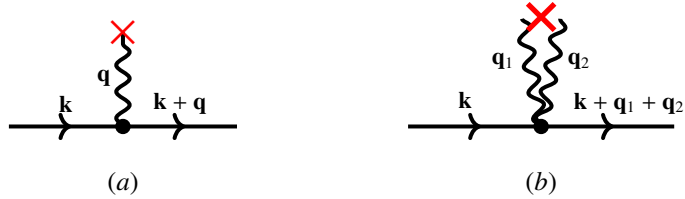

The first two terms in the hamiltonian, $E_0$ and $\mathcal{H}_{2}$  represent the free magnon field and $\mathcal{H}_2^\prime$, is the interaction of magnons with the field of nonmagnetic impurities.  The contribution of the term $\mathcal{H}_4^\prime$ to the partition function is shown to be identically zero  \cite{Weiss_1964a}, and thereby can be neglected at the outset.  The magnon-impurity interaction term in Eq. \eqref{eqn:ft-h2p} can be represented by a scattering process, in which a magnon (spin-wave) with momentum (wave-vector) $\mathbf{k}$ collides with a nonmagnetic impurity, then recoiling with momentum $\mathbf{k}+\mathbf{q}$ as shown by the diagram in Fig. \ref{fig:mag-defect-scat-1}(a).  $\mathcal{H}_4^\prime$ represents a process of simultaneous absorption and emission of magnon  interacting with the field of a nonmagnetic impurity, Fig. \ref{fig:mag-defect-scat-1}(b). However,  this is basically the same process as  the preceding one, i.e., if  we put   $\mathbf{k}=\mathbf{q}_1+\mathbf{q}_2$ in  $\mathcal{H}_2^\prime$ and perform the summation over the remaining degrees of freedom. Indeed, if  $\sum_{\mathbf{q}_2}\Delta_{\mathbf{k},\mathbf{q}_2}=0$, then 
\[\sum_{\mathbf{k},\mathbf{q}_1+\mathbf{q}_2,\mathbf{q}_2}\!\!\!\!\negthickspace \Delta_{\mathbf{k},\mathbf{q}_2} b_{\mathbf{k}+\mathbf{q}_1+\mathbf{q}_2}^\dagger b_{\mathbf{k}} c_{\mathbf{q}_1+\mathbf{q}_2}=0\]
But, with $2S=1$, we can write
\begin{align}
\sum_{\mathbf{q}_2}\Delta_{\mathbf{k},\mathbf{q}_2} &=\sum_{\mathbf{q}_2}J_{\mathbf{q}_2}-\sum_{\mathbf{q}_2}J_{\mathbf{k}+\mathbf{q}_2}=\sum_{\mathbf{q}_2}\sum_{\mathbf{R}_n}\Big(e^{-i\mathbf{q}_2\cdot \mathbf{R}_n} - e^{-i(\mathbf{k}+\mathbf{q}_2)\cdot \mathbf{R}_n}\Big)J(\mathbf{R}_n)\nonumber\\
&= N \sum_{\mathbf{R}_n}J(\mathbf{R}_n)\Big( 1 -  e^{-i\mathbf{k} \cdot \mathbf{R}_n}\Big) \delta (\mathbf{R}_n) =0.\nonumber
\end{align}

Thus, the Heisenberg hamiltonian can  be  expressed in terms of the operators $b_\mathbf{k}^\dagger$ and $b_\mathbf{k}$, describing the interaction of magnons with the Fourier transformed field of impurities:  
\begin{equation}
\label{eqn:heisenberg}
\mathcal{H} = E_0+\sum_{\mathbf{k}} \epsilon_\mathbf{k} b^\dagger_\mathbf{k} b_\mathbf{k}
 -\frac{1}{\sqrt{N}}\sum_{\mathbf{k,q}}\gamma_\mathbf{k} b_\mathbf{k+q}^{\dagger}b_\mathbf{k}c_\mathbf{q}.
\end{equation}
\noindent
Equation \eqref{eqn:heisenberg} will  serve as a starting point for our computations. 

\section{Thermodynamic potential}
\label{sec:matsubara}
\subsection{The partition function}
\label{sec:zustands}

The partition function $ \mathcal{Z}$  can be expressed as a functional of the nonmagnetic impurity field $c_j$  through the trace of the density matrix:
\begin{equation}
\label{eqn:zust_1}
\mathcal{Z} = \mathrm{Tr} [\rho] = \mathrm{Tr} [\exp(-\beta\mathcal{H})]
\end{equation}
\noindent
where  $\beta \equiv (k_BT)^{-1}$ and  $\rho = \exp(-\beta\mathcal{H}) $  is the density matrix of a canonical ensemble satisfying the Bloch equation \cite{Matsubara_1955,Bloch_1932}
\begin{equation}
\label{eqn:bloch-eq}
\frac{\partial \rho}{\partial \beta} = -\mathcal{H}e^{-\beta \mathcal{H}}.
\end{equation}
\noindent
In the standard Matsubara approach \cite{Matsubara_1955}, one puts
\begin{eqnarray}
\label{eqn:h-tot}
\mathcal{H} &=&  \mathcal{H}_0 + \mathcal{H}_\mathrm{I},\\
\label{eqn:matsu-rho}
\rho &=& e^{-\beta \mathcal{H}_0} \mathcal{S}(\beta) \equiv  \rho_0 \mathcal{S}(\beta),\\
\label{eqn:matsu-s}
\mathcal{S}(\beta) &=& e^{\beta \mathcal{H}_0} e^{-\beta \mathcal{H}},\\
\label{eqn:matsu-h0}
\mathcal{H}_0 &=&  E_0 + \sum_{\mathbf{k}} \epsilon_\mathbf{k} \hat n_\mathbf{k},\\
\label{eqn:matsu-hI}
\mathcal{H}_I &=& -\frac{1}{\sqrt{N}} \sum_{\mathbf{k},\mathbf{q}} \gamma_{\mathbf{k}} b_{\mathbf{k}+\mathbf{q}}^\dagger b_{\mathbf{k}}
c_{\mathbf{q}},\\
\label{eqn:no-opertaor}
\hat n_\mathbf{k} &=& b^\dagger_\mathbf{k} b_\mathbf{k}.
\end{eqnarray}
The equation for $\mathcal{S}(\beta)$, from Eqs.  (\ref{eqn:bloch-eq})-(\ref{eqn:matsu-s}), becomes
\begin{align}
\label{eqn:matsu-eq}
\frac{\partial \mathcal{S}(\beta)}{\partial \beta} &= -\widetilde{\mathcal{H}}_\mathrm{I} \mathcal{S}(\beta),\\
\label{eqn:matsu-htilde}
\widetilde{\mathcal{H}}_\mathrm{I}(\beta) &= e^{\beta \mathcal{H}_0} \mathcal{H}_\mathrm{I} e^{-\beta \mathcal{H}_0}.
\end{align}
\noindent
The formal solution of Eq. \eqref{eqn:matsu-eq} with the initial condition $\mathcal{S}(0)=1$ is
\begin{eqnarray}
\mathcal{S}(\beta) &=& \mathrm{T}\exp\Big[-\!\!\int_0^\beta \! \! d\tau \, \widetilde{\mathcal{H}}_\mathrm{I}(\tau)\Big] \\
 &=& \sum_{n=0}^\infty \frac{(-1)^n}{n!}\int_0^\beta d\tau_1 \int_0^{\beta} \dots \int_0^{\beta}d\tau_n
 \mathrm{T} \big[\widetilde{\mathcal{H}}_{\mathrm{I}}(\tau_1) \dots \widetilde{\mathcal{H}}_{\mathrm{I}}(\tau_n)\big],
\label{eqn:matsu-sol}
\end{eqnarray}
where $\mathrm{T}[\bullet]$ denotes the usual chronological ordering of the factors (T-product), i.e. with the time increasing from right to left, $0 <\tau <\beta$ is the Matsubara time domain, $\tau \to it$ ($t=-i\beta$) being a fictitious imaginary time,  and $\mathcal{S}(\beta)$ is analogous to the S--matrix of field theory. Finally, we can write the partition function in the form
\begin{equation}
\label{eqn:matsu-zust}
\mathcal{Z} =  \mathcal{Z}_0 \langle \mathcal{S}(\beta) \rangle_\beta,
\end{equation}
\noindent
where $\mathcal{Z}_0= \mathrm{Tr}[\exp{(-\beta \mathcal{H}_0)}]$ and $\langle \mathcal{S}(\beta) \rangle_\beta=\text{Tr}[\rho_0\,\mathcal{S(\beta)}]/\text{Tr} [\rho_0]$ or 
\begin{equation}
\langle \mathcal{S}(\beta) \rangle_\beta  = \sum_{n=0}^\infty \frac{(-1)^n}{n!}\int_0^\beta d\tau_1  \dots \int_0^{\beta}\!\!\!\!d\tau_n\;
 \langle \mathrm{T} \big[\widetilde{\mathcal{H}}_{\mathrm{I}}(\tau_1) \dots \widetilde{\mathcal{H}}_{\mathrm{I}}(\tau_n)\big]\rangle_\beta.
\label{eqn:matsu-sav}
\end{equation}
We should bear in mind that $\langle \mathcal{S}(\beta) \rangle_\beta$ is a functional of $c_j$, so upon its computation it must be once more averaged over all possible $c_j$. Note also that $\mathcal{Z}_0$ is merely a function of the average defect concentration $c$ and hence does not require additional averaging. Now we express the Helmholtz potential or free energy using  Eq. (\ref{eqn:matsu-zust})
\begin{equation}
F  =  -\frac{1}{\beta} \log \mathcal{Z} = -\beta^{-1} \log \mathcal{Z}_0-\beta^{-1} \log \langle \mathcal{S}(\beta) \rangle_\beta
\label{eqn:helmholtz}
\end{equation}
The first term on the right-side of Eq. \eqref{eqn:helmholtz}, using Eq. \eqref{eqn:matsu-h0},  can readily be evaluated
\begin{equation}
\mathcal{Z}_0 = \sum_{n_{\mathbf{k}}}\langle n_{\mathbf{k}}|e^{-\beta\mathcal{H}_0}|n_{\mathbf{k}}\rangle =\prod_\mathbf{k}\frac{e^{\beta E_0}}{1-e^{-\beta \epsilon_\mathbf{k}}}.
\label{eqn:zust0}
\end{equation}
The corresponding noninteracting Helmholtz potential, $F_0\equiv -\beta^{-1}\log \mathcal{Z}_0$, is expressed as
\begin{equation}
F_0 = E_0+\frac{\mathcal{V}}{\beta}\int \frac{d^3k}{(2\pi)^3} \log\big[1-e^{-\beta\epsilon_\mathbf{k}}\big],
\label{eqn:helm0}
\end{equation}
where  we took the continuum momentum limit $\sum_{\mathbf{k}} \to \mathcal{V}/(2\pi)^3 \int d\mathbf{k}$ and $\mathcal{V}$ is the volume of the system. The last term in  Eq. (\ref{eqn:helmholtz}) is designated as  $F_{\mathrm{I}} \equiv -\beta^{-1}\log \langle \mathcal{S}(\beta) \rangle_\beta $.

The key element in the formalism is the calculation of the expectation value of the S--matrix. Following the treatment and notation in  \S 15 of  Ref. \cite{AGD_1963}, we write Eq.  \eqref{eqn:matsu-sav} in the form
\begin{equation}
\langle \mathcal{S} \rangle_\mathrm{con} \equiv 1+ \ln \langle \mathcal{S}(\beta) \rangle_\beta = 1 +  \sum_{j=1}^\infty \Xi_j\; ,
\label{eqn:agd-sav}
\end{equation}
where $\sum_j\Xi_j$ is the sum of all connected closed loop (vacuum) graphs with a special arrangement of vertices at $\tau_1, \tau_2,\dots,\tau_n$,  see \ref{sec:appA0} for an example.
Thus, the (magnetic-nonmagnetic) interacting part of the Helmholtz potential is
\begin{equation}
F_{\mathrm{I}}= -\beta ^{-1} \Big(\langle \mathcal{S} \rangle_\mathrm{con} -1\Big) = -\frac{1}{\beta}\sum_{j=1}^\infty \Xi_j.
\label{eqn:helmI}
\end{equation}
One can show $\Xi_1=0$ and the second term in this series is (\ref{sec:appA0})
\begin{equation}
\Xi_2 = \frac{\beta}{2N} \sum_{\mathbf{k},\mathbf{q}} \frac{\gamma_{\mathbf{k}}\gamma_{\mathbf{k+q}}}{\epsilon_\mathbf{k+q}-\epsilon_\mathbf{k}}\big(n_\mathbf{k}-n_\mathbf{k+q}\big)c_\mathbf{q}c_\mathbf{-q},
\label{eqn:Gamma-2}
\end{equation}
or by changing the order of summation and simplifying
\begin{equation}
\Xi_2 = \frac{\beta}{N} \sum_{\mathbf{k},\mathbf{q}}  \gamma_{\mathbf{k}}\gamma_{\mathbf{k+q}}\Bigg( \frac{n_\mathbf{k}}{\epsilon_\mathbf{k+q}-\epsilon_\mathbf{k}}\Bigg) c_\mathbf{q}c_\mathbf{-q},
\label{eqn:Gamma-2s}
\end{equation}
where $n_\mathbf{k} \equiv \langle \hat n_\mathbf{k}\rangle_\beta$ is the thermal average number of magnons with a vector $\mathbf{k}$:
\begin{equation}
n_\mathbf{k} = \frac{1}{e^{\beta\epsilon_\mathbf{k}}-1}.
\label{eqn:amn}
\end{equation}
Furthermore, the $n$-th term in the series in  Eq. \eqref{eqn:helmI} is written as
\begin{equation}\begin{split}
\Xi_n & = \frac{\beta}{2N^{n/2}} \sum_{\substack{\mathbf{k}_1\dots\mathbf{k}_n \\ \mathbf{q}_1\dots\mathbf{q}_n}}
 \gamma_{\mathbf{k}_1}\delta(\mathbf{k}_n+\mathbf{q}_n-\mathbf{k}_1)\\
 &\times \prod_{i=2}^{n}\frac{\gamma_{\mathbf{k}_i}}{\varepsilon_{\mathbf{k}_1}-\varepsilon_{\mathbf{k}_i}}
 [(n_{\mathbf{k}_1}+1)n_{\mathbf{k}_i}+(-1)^{n-1}n_{\mathbf{k}_2}(n_{\mathbf{k}_i}+1)]\\
 &\times \delta(\mathbf{k}_{i-1}+\mathbf{q}_{i-1}-\mathbf{k}_i)c_{\mathbf{q}_1}c_{\mathbf{q}_i}.
\label{eqn:Gamma-n}
\end{split}\end{equation}


\subsection{Impurity averaging the free energy}
\label{sec:mean-field}
In order to evaluate the interaction part of the free energy in Eq. (\ref{eqn:helmI}), which accounts the change in free energy due to the impurities, we assume that the products of impurity fields $c_{\mathbf{q}_1}c_{\mathbf{q}_i}$ appearing in Eq.  \eqref{eqn:Gamma-n} deviate little from their mean values $\langle c_{\mathbf{q}_1}c_{\mathbf{q}_i}\rangle_c$, where the subscript $c$ signifies the average taken over all possible distribution of impurities  in the lattice. The number of possible distributions of $n$ impurities between $N$ lattice sites is simply
\begin{equation}
Z_c = \frac{N!}{n!(N-n)!}.
\label{eqn:zc1}
\end{equation}
Each distribution state can be described by the $N$-dimensional vector $|c_j\rangle$. Thus the distribution state has $Z_c$ dimensions.

Let us represent the $c_j$ numbers as commuting operators with the following properties
\begin{equation}
c_j |c_j\rangle =
\begin{cases}
0|c_j\rangle, & \mathrm{if} \quad  c_j=0\\
1|c_j\rangle, & \mathrm{if} \quad c_j=1
\end{cases}
\label{eqn:cj-op}
\end{equation}
Furthermore, we define a density matrix $\rho_c$ whose elements $|c_j\rangle\rho_c\langle c_j|$ give the probability distribution represented by the vector $|c_j\rangle$. The simplest choice for $\rho_c$ is
\begin{equation}
|c_j\rangle\rho_c\langle c_j| = Z_c^{-1},
\label{eqn:zc2}
\end{equation}
which corresponds to the entirely random distribution of impurities in the lattice. Now the mean value of any arbitrary function of $c_j$ operators, $f(c_j)$, will be
\begin{equation}
\label{eqn:mean-c}
\langle f(c_j)\rangle_c = \frac{\mathrm{Tr}[ \rho_c f(c_j)]}{\mathrm{Tr} \rho_c}.
\end{equation}
\noindent
The averaged interacting free energy is
\begin{equation}
\langle  F_{\mathrm{I}} \rangle_c = -\frac{1}{\beta}\sum_{j=1}^\infty \langle  \Xi_j \rangle_c.
\label{eqn:helmI-mf}
\end{equation}
With such a premise  the average of $c_{\mathbf{q}_i}$ and its  products can be evaluated;  see  \ref{sec:cumav}
\begin{align}
\label{eqn:cq1}
\langle c_{\mathbf{q}}\rangle_c  &= 0,\\
\label{eqn:cq2}
\langle c_{\mathbf{q}_1} c_{\mathbf{q}_2} \rangle_c  &= c(1-c)\delta(\mathbf{q}_1+\mathbf{q}_2),\\
\label{eqn:cq3}
\langle c_{\mathbf{q}_1} c_{\mathbf{q}_2} c_{\mathbf{q}_3}\rangle_c  &= \frac{c(1-3c+2c^2)}{\sqrt{N}}\delta(\mathbf{q}_1+\mathbf{q}_2+\mathbf{q}_3),\\
\vdots \notag\\
\label{eqn:cqn}
\langle c_{\mathbf{q}_1} c_{\mathbf{q}_2}\dots c_{\mathbf{q}_n}\rangle_c  &= \frac{[c(1-c)^n+(1-c)(-1)^n c^n]}{N^{n/2-1}}
\delta(\sum_{i=1}^{n}\mathbf{q}_i).
\end{align}

Recall now that $c\equiv n/N$ is the mean macroscopic concentration of impurities. So the mean field interacting free energy Eq. \eqref{eqn:helmI-mf}) is an infinite sum of Eq. \eqref{eqn:Gamma-n} averaged over the distribution of impurities according to Eq. \eqref{eqn:cqn}. Note that each $\langle  \Xi_j \rangle_c$ contributes to $\langle  F_{\mathrm{I}} \rangle_c$ with a term proportional to $c$. Therefore, the infinite sum \eqref{eqn:helmI-mf} cannot be truncated by its first couple of terms, or in any number of terms, to obtain the exact result in the dilute-impurity solution regime. Making use of $\delta$-functions, we rewrite Eq. \eqref{eqn:helmI-mf} in the form
\begin{align}
\label{eqn:helmI-mf1}
\langle  F_{\mathrm{I}} \rangle_c & = -\frac{N}{2} \sum_{n=2}^{\infty} \frac{p_n(c)}{N^n} \sum_{\mathbf{k}_1\dots\mathbf{k}_n}
 \prod_{i=2}^{n}\frac{\gamma_{\mathbf{k}_1}\gamma_{\mathbf{k}_i}}{\varepsilon_{\mathbf{k}_1}-\varepsilon_{\mathbf{k}_i}}
 \big[(n_{\mathbf{k}_1}+1)n_{\mathbf{k}_i}+(-1)^{n-1}n_{\mathbf{k}_1}(n_{\mathbf{k}_i}+1)\big],\\
p_n(c) & \equiv  c(1-c)^n + (-1)^n(1-c) c^n.
\label{eqn:pnc}
\end{align}

To make further simplifications, we now introduce two new functions:
\begin{subequations}
\begin{align}
\label{eqn:V-}
V^-(\mathbf{k}) &= \frac{1}{N}\sum_{\mathbf{k}_1} \frac{\gamma_{\mathbf{k}_1}}{\epsilon_{\mathbf{k}}-\epsilon_{\mathbf{k}_1}} n_{\mathbf{k}_1},\\
\label{eqn:V+}
V^+(\mathbf{k}) &= \frac{1}{N}\sum_{\mathbf{k}_1} \frac{\gamma_{\mathbf{k}_1}}{\epsilon_{\mathbf{k}}-\epsilon_{\mathbf{k}_1}} (n_{\mathbf{k}_1}+1).
\end{align}
\end{subequations}
With the aid of these two auxiliary functions and recalling that $2\epsilon_{\mathbf{k}}=\alpha+(1-c)\gamma_{\mathbf{k}}$, we can readily sum up the infinite series (\ref{eqn:helmI-mf1}) to obtain
\begin{equation}
\langle  F_{\mathrm{I}} \rangle_c  = - \frac{1}{2}\sum_{n=2}^{\infty} \sum_{\mathbf{k}} \gamma_{\mathbf{k}} \Big\{(n_{\mathbf{k}}+1)
  \big[V^-(\mathbf{k})\big]^{n-1}
  + (-1)^{n-1} n_{\mathbf{k}}  \big[V^+(\mathbf{k})\big]^{n-1}
 \Big\}p_n(c).
\label{eqn:helmI-mf2}
\end{equation}
 So, the second-order term ($n=2$) contribution to the interacting free energy reads
\begin{equation}
\langle  F_{\mathrm{I}}^{(2)} \rangle_c  =  \frac{c}{N} \sum_{\mathbf{k}} \sum_{\mathbf{k}_1} \frac{\gamma_{\mathbf{k}} \gamma_{\mathbf{k}_1}}{\gamma_{\mathbf{k}}-\gamma_{\mathbf{k}_1}} (n_\mathbf{k}-n_{\mathbf{k}_1}).
\label{eqn:helmI-mf2-1}
\end{equation}
Now, if $\forall$ $\mathbf{k}$, $|V^\pm(\mathbf{k})| < 1$,  we can perform the summation over the index $n$ to arrive at
\begin{equation}\begin{split}
\langle  F_{\mathrm{I}} \rangle_c & = -N\frac{c(1-c)\varv}{2} \int \frac{d^3k}{(2\pi)^3} \,\gamma_{\mathbf{k}} \Big\{(n_{\mathbf{k}}+1)
  \frac{V^-(\mathbf{k})}{\big[1-(1-c)V^-(\mathbf{k})\big] \big[1+c V^-(\mathbf{k})\big]}\\
 & - n_{\mathbf{k}} \frac{V^+(\mathbf{k})}{\big[1+(1-c)V^+(\mathbf{k})\big] \big[1-c V^+(\mathbf{k})\big]}
 \Big\}.
\label{eqn:helmI-mf3}
\end{split}\end{equation}
where we went to continuum  momentum with $\varv=\mathcal{V}/N$. Note that  $V^\pm(\mathbf{k})$ depend on the mean impurity concentration $c$ through $\gamma_{\mathbf{k}}$ and $\epsilon_{\mathbf{k}}$, namely
\begin{subequations}
\begin{align}
\label{eqn:V-c}
V^-(\mathbf{k}) &= \frac{2}{(1-c)^2N}\sum_{\mathbf{k}_1} \frac{B+(1-c)\varepsilon_{\mathbf{k}_1}}{\varepsilon_{\mathbf{k}}-\varepsilon_{\mathbf{k}_1}} n_{\mathbf{k}_1},\\
\label{eqn:V+c}
V^+(\mathbf{k}) &= \frac{2}{(1-c)^2N}\sum_{\mathbf{k}_1} \frac{B+(1-c)\varepsilon_{\mathbf{k}_1}}{\varepsilon_{\mathbf{k}}-\varepsilon_{\mathbf{k}_1}} (n_{\mathbf{k}_1}+1),
\end{align}
\end{subequations}
We can write  Eq. \eqref{eqn:helmI-mf3} in a more compact form by introducing  $U^\pm(\mathbf{k})$
\begin{equation}
\langle  F_{\mathrm{I}} \rangle_c  = -N c(1-c)\frac{\varv}{2}\int \frac{d^3k}{(2\pi)^3} \,\gamma_{\mathbf{k}} \Big[(n_{\mathbf{k}}+1)
  U^-(\mathbf{k}) - n_{\mathbf{k}}U^+(\mathbf{k})
 \Big],
\label{eqn:helmI-mf4}
\end{equation}
\begin{equation}
\text{where} \quad U^\pm(\mathbf{k}) = \frac{V^\pm(\mathbf{k})}{\big[1\pm(1-c)V^\pm(\mathbf{k})\big] \big[1\mp c V^\pm(\mathbf{k})\big]}.
\label{eqn:Upm}
\end{equation}
Finally, the total Helmholtz free energy per ion is obtained by adding  Eqs. \eqref{eqn:helm0} and  \eqref{eqn:helmI-mf4}, $F=F_0+\langle  F_{\mathrm{I}} \rangle_c $, and then dividing by $N$:
\begin{equation}\begin{split}
\frac{F}{N} & = -\big[(1-c)SB+\frac{1}{2}(1-c)^2S^2J\big] + \frac{\varv}{\beta}\int \frac{d^3k}{(2\pi)^3} \log\Big[1-e^{-\beta \big[2(1-c)B+(1-c)^2\varepsilon_\mathbf{k}\big]}\Big] \\
 & - c(1-c) \frac{\varv}{2} \int \frac{d^3k}{(2\pi)^3} \,\gamma_{\mathbf{k}} \big[(n_{\mathbf{k}}+1)
  U^-(\mathbf{k}) - n_{\mathbf{k}}U^+(\mathbf{k}) \big].
 \label{eqn:helm-tot}
\end{split}\end{equation}
 The second term on the right-hand side  of Eq. \eqref{eqn:helm-tot} represents the  Bloch free energy of an ideal Bose gas of magnons. Its expansion in powers of temperature gives a well-known relation  in the low temperature ($\beta \gg 1$) region  of the  ferromagnet \cite{Dyson_1956b,Vaks1968thermodynamics}; see section \ref{sec:dos-thermo}. The third term in Eq. \eqref{eqn:helm-tot} is the contribution of the magnon-impurity interaction to the free energy.
\section{Spin waves}
\label{sec:spinwave}
\subsection{Equation of motion}
\label{sec:Greens}
We start by writing the  double-time single-particle Green function at temperature  $T$  in $\mathbf{k}$-space, $G(\mathbf{k},\mathbf{k^\prime}; t,t^\prime)$, in   retarded $G^R$ and advanced  $G^A$ forms, in terms of the magnon operators
\begin{equation}
\label{eqn:rgreen-ft-pauli-op}
G^R(\mathbf{k},\mathbf{k^\prime};t,t^\prime) = -i\theta(t-t^\prime) \langle [b_\mathbf{k}(t),b^{\dagger}_\mathbf{k^\prime}(t^\prime)]\rangle.
\end{equation}
\noindent
\begin{equation}
\label{eqn:agreen-ft-pauli}
G^A(\mathbf{k},\mathbf{k^\prime};t,t^\prime) = i\theta(t^\prime-t) \langle [b_\mathbf{k}(t),b^{\dagger}_\mathbf{k^\prime}(t^\prime)]\rangle.
\end{equation}
\noindent
Here, $\theta(x)$ is the usual Heaviside step-function and the symbol $\langle \dots \rangle$ denotes thermal average over the grand canonical Gibbs ensemble, viz.
\begin{equation}
\label{eqn:therm-av}
\langle \dots \rangle = \mathcal{Z}^{-1} \mathrm{Tr} \big(e^{-\beta\mathcal{H}^\prime} \dots \big),
\end{equation}
\noindent
$\mathcal{Z} = \mathrm{Tr} \exp(-\beta\mathcal{H}^\prime)$, $\mathcal{H}^\prime \equiv \mathcal{H}-\mu_c N$, $\mu_c$ the chemical potential, $\beta=1/T$, where we set $k_B \equiv 1$.

The equation of motion for the retarded Green function is
\begin{equation}
\label{eqn:rgreen-ft-pauli}
i\partial_t G^R(\mathbf{k},\mathbf{k^\prime};t,t^\prime) = \delta(t-t^\prime)\delta_{\mathbf{k},\mathbf{k^\prime}} -i\theta(t-t^\prime)\langle [i\partial_t b_\mathbf{k}(t),b^{\dagger}_\mathbf{k^\prime}(t^\prime)]\rangle,
\end{equation}
\noindent
with the time derivative of $b_\mathbf{k}(t)$ in the Heisenberg picture, viz.
\begin{equation}
\label{eqn:pauli-heisenberg}
\partial_t b_\mathbf{k}(t) = i[\mathcal{H},b_\mathbf{k}].
\end{equation}
\noindent
Substituting for $\mathcal{H}$  from Eq. \eqref{eqn:heisenberg} and using  the aforementioned Boson commutation rules 
\begin{equation}
\label{eqn:pauli-heisenberg-2}
i\partial_t b_\mathbf{k}(t) = \epsilon_\mathbf{p} b_\mathbf{k}(t)
 -\frac{1}{\sqrt{N}}\sum_{\mathbf{p}}\gamma_\mathbf{p} c_{\mathbf{k}-\mathbf{p}}b_\mathbf{p}(t).
\end{equation}
\noindent
Inserting this result into Eq. \eqref{eqn:rgreen-ft-pauli} yields
\begin{equation}\begin{split}
\label{eqn:eom-rgreen-ff}
i\partial_t G^R(\mathbf{k},\mathbf{k^\prime};t,t^\prime) &= -\delta(t-t^\prime)\delta_{\mathbf{k},\mathbf{k^\prime}} + \epsilon_\mathbf{k}G^R(\mathbf{k},\mathbf{k^\prime};t,t^\prime)\\
& -\frac{1}{\sqrt{N}}\sum_{\mathbf{p}}\gamma_\mathbf{p} c_{\mathbf{k}-\mathbf{p}}G^R(\mathbf{p},\mathbf{k},t,t^\prime).
\end{split}\end{equation}
\noindent
Fourier transforming this equation from the time domain, assuming time-invariance,  to the frequency domain $\omega$, we write
\begin{equation}
\label{eqn:eom-rgreen-ft}
 (\omega-\epsilon_\mathbf{k}) G^R(\mathbf{k},\mathbf{k^\prime},\omega) + \frac{1}{\sqrt{N}}\sum_{\mathbf{p}}\gamma_\mathbf{p} c_{\mathbf{k}-\mathbf{p}}G^R(\mathbf{p},\mathbf{k},\omega)  = -\delta_{\mathbf{k},\mathbf{k^\prime}},
\end{equation}
\noindent
\begin{equation}
\label{eqn:ft-w-domain}
\text{with} \quad  G^R(\mathbf{k},\mathbf{k^\prime},\omega) = \int_{-\infty}^\infty G^R(\mathbf{k},\mathbf{k^\prime},t) e^{i\omega t} dt.
\end{equation}
\noindent
Similar calculations can be done to obtain the equation of motion for $G^A(\mathbf{k},\omega)$.

\subsection{Solving  equation of motion}
\label{sec:solve-eom}

In order to solve Eq. \eqref{eqn:eom-rgreen-ft} we make the ansatz
\begin{align}
\label{eqn:izyuov-ansatz}
G^R(\mathbf{p},\mathbf{k},\omega) &= G^R(\mathbf{k},\mathbf{k},\omega) K(\mathbf{p},\mathbf{k},\omega),\\
K(\mathbf{k},\mathbf{k},\omega) &= 1,
\end{align}
\noindent
Inserting Eq. \eqref{eqn:izyuov-ansatz} into  Eq. \eqref{eqn:eom-rgreen-ft}, for $\mathbf{k}=\mathbf{k^\prime}$, we obtain
\begin{equation}
\label{eqn:eom-green-sol1}
G^R(\mathbf{k},\mathbf{k},\omega) = \frac{1}{ \epsilon_\mathbf{k}-\omega-\Sigma(\mathbf{k},\omega)},
\end{equation}
\noindent
where as before $\epsilon_\mathbf{k} \equiv (1-c)^2\varepsilon_{\mathbf{k}}+2(1-c)B$ and $\Sigma(\mathbf{k},\omega)$ is the self-energy term given by
\begin{equation}
\label{eqn:self-energy-1}
\Sigma(\mathbf{k},\omega) = \frac{1}{\sqrt{N}} \sum_{\mathbf{p}}\gamma_\mathbf{p} c_{\mathbf{k}-\mathbf{p}}K(\mathbf{p},\mathbf{k},\omega).
\end{equation}
\noindent
On the other hand, for $\mathbf{k} \neq \mathbf{k}^\prime$ and $\forall$ $G(\mathbf{k},\mathbf{k},\omega)$, we have
\begin{equation}
\label{eqn:eom-Kgreen}
 (\omega-\epsilon_\mathbf{k}) K(\mathbf{k},\mathbf{k}^\prime,\omega) + \frac{1}{\sqrt{N}}\sum_{\mathbf{p}}\gamma_\mathbf{p} c_{\mathbf{k}-\mathbf{p}}K(\mathbf{p},\mathbf{k}^\prime,\omega)  = 0.
\end{equation}
\noindent
This relation also holds for  the advanced Green function, so we tacitly drop the superscripts $R$ and $A$ from $G$, and write Eq. \eqref{eqn:eom-green-sol1} as  a Dyson equation in the form
\begin{equation}
\label{eqn:dyson-eq}
G(\mathbf{k},\omega) =  G_0(\mathbf{k},\omega) + G_0(\mathbf{k},\omega)\Sigma(\mathbf{k},\omega)G(\mathbf{k},\omega),
\end{equation}
\noindent
where we put $G(\mathbf{k},\omega)\equiv G(\mathbf{k},\mathbf{k},\omega)$  and defined the free propagator as $G_0(\mathbf{k},\omega)=(\epsilon_\mathbf{k}-\omega)^{-1}$.

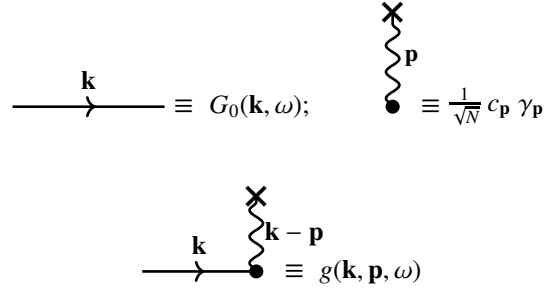
\begin{figure}
    \centering
\begin{tikzpicture}[line width=1.0 pt, scale=1]
	\draw[fermion] (-2,0)--(0,0);
	\node at (0.25,0) {\large{$\equiv$}};
	\node at (1.25,0) {$G_0(\mathbf{k},\omega$);};
	\node at (-1,0.35) {$\mathbf{k}$};
\begin{scope}[shift={(4.0,0)}]
	\draw[vector] (-1,0)--(-1,1.25);
	\draw[fill=black] (-1,0) circle (.075cm);
	\node at (-1,1.25) {\Large$\boldsymbol{\times}$};
	\node at (-0.5,0) {\large{$\equiv$}};
	\node at (0.35,0) {$\frac{1}{\sqrt{N}}\,c_\mathbf{p}\;\gamma_\mathbf{p}$};
	\node at (-0.75,0.65) {\large{\small{$\mathbf{p}$}}};
\end{scope}
\end{tikzpicture}
\vspace{1.5em}

\begin{tikzpicture}[line width=1.0 pt, scale=1]
\draw[fermion] (-2,0)--(-0.5,0);
\node at (0,0) {\large{$\equiv$}};
\node at (1,0) {$g(\mathbf{k},\mathbf{p},\omega)$};
\draw[fill=black] (-0.5,0) circle (.075cm);
\draw[vector] (-0.5,0)--(-0.5,1.);
\node at (-0.5,1.) {\Large$\boldsymbol{\times}$};
\node at (-1.25,0.35) {$\mathbf{k}$};
\node at (-0.0,0.50) {$\mathbf{k-p}$};
\end{tikzpicture}
\caption{\small{Convention for drawing the diagrams.}}
\label{fig:feynman-rule}
\end{figure}

 We now write the kernel function $K$ as
\begin{align}
\label{eqn:eom-Kgreen-sol}
K(\mathbf{k},\mathbf{k}^\prime,\omega) &= \sum_{\mathbf{p}} \mathfrak{g}(\mathbf{k},\mathbf{p},\omega) K(\mathbf{p},\mathbf{k}^\prime,\omega),\\
\label{eqn:B-fun}
\mathfrak{g}(\mathbf{k},\mathbf{p},\omega) &\equiv \frac{(1-c)^{-1}}{\sqrt{N}}\frac{\epsilon_\mathbf{p}}{\epsilon_\mathbf{k} - \omega} c_{\mathbf{k}-\mathbf{p}},
\end{align}
\noindent
Next, we separate Eq. \eqref{eqn:eom-Kgreen-sol}  as
\begin{equation}
\label{eqn:Kgreen-sol-sep}
K(\mathbf{k},\mathbf{k}^\prime,\omega) =\mathfrak{g}(\mathbf{k},\mathbf{k}^\prime,\omega) + \sum_{\mathbf{p} \neq \mathbf{k}^\prime} \mathfrak{g}(\mathbf{k},\mathbf{p},\omega) K(\mathbf{p},\mathbf{k}^\prime,\omega).
\end{equation}
\noindent
From this relation, $K$ may be solved by iteration, viz.
\begin{equation}\begin{split}
\label{eqn:Kgreen-sol-iter}
K(\mathbf{k},\mathbf{q},\omega) &= \mathfrak{g}(\mathbf{k},\mathbf{q},\omega) + \sum_{\mathbf{p} \neq \mathbf{q}} \mathfrak{g}(\mathbf{k},\mathbf{p},\omega) \mathfrak{g}(\mathbf{p},\mathbf{q},\omega)\\
& + \sum_{\substack{\mathbf{p}\neq\mathbf{q} \\ \mathbf{p}^\prime \neq \mathbf{q}}}\mathfrak{g}(\mathbf{k},\mathbf{p},\omega) \mathfrak{g}(\mathbf{p},\mathbf{p}^\prime,\omega)\mathfrak{g}(\mathbf{p}^\prime,\mathbf{q},\omega) + \dots
\end{split}\end{equation}
\noindent
Hence, Eq. \eqref{eqn:self-energy-1} is written as
\begin{equation}\begin{split}
\label{eqn:self-energy-g}
\Sigma(\mathbf{k},\omega)  &=  \frac{1}{\sqrt{N}}\sum_{\mathbf{p}} \Big[\mathfrak{g}(\mathbf{p},\mathbf{k},\omega) + \sum_{\mathbf{q} \neq \mathbf{k}} \mathfrak{g}(\mathbf{p},\mathbf{q},\omega) \mathfrak{g}(\mathbf{q},\mathbf{k},\omega)\\
& + \sum_{\substack{\mathbf{q}\neq\mathbf{k} \\ \mathbf{q}^\prime \neq \mathbf{k}}}\mathfrak{g}(\mathbf{p},\mathbf{q},\omega) \mathfrak{g}(\mathbf{q},\mathbf{q}^\prime,\omega)\mathfrak{g}(\mathbf{q}^\prime,\mathbf{k},\omega) + \dots \Big] \gamma_\mathbf{p} c_{\mathbf{k}-\mathbf{p}}
\end{split}\end{equation}
\noindent
Substituting for $\mathfrak{g}$ from  Eq. \eqref{eqn:B-fun}  results in
\begin{equation}
\label{eqn:self-energy-2}
\Sigma(\mathbf{q},\omega) = \sum_{n=1}^\infty  \sum_{\mathbf{k}_1\dots\mathbf{k}_{n+1}} \frac{\delta_{\mathbf{k}_{n+1},\mathbf{q}}\, \gamma_\mathbf{q} \, c_{\mathbf{q}-\mathbf{k}_1}}{N^{\frac{n+1}{2}}}\prod_{i=1}^n \frac{\gamma_{\mathbf{k}_i}c_{\mathbf{k}_i-\mathbf{k}_{i+1}}}{(\epsilon_{\mathbf{k}_i} - \omega)}.
\end{equation}
\noindent
Equations \eqref{eqn:Kgreen-sol-iter}, \eqref{eqn:self-energy-g} and the Dyson equation \eqref{eqn:dyson-eq} can be represented graphically as

\begin{equation}
\begin{tikzpicture}[line width=1.0 pt, scale=0.65]
	\node at (-4,0) {$K(\mathbf{k},\mathbf{q},\omega)$};
	\node at (-2.5,0) {\Large=};
	\draw [fermion](-2,0) -- (-0.5,0);
	\draw[fill=black] (-0.5,0) circle (.075cm);
	\draw[vector] (-0.5,0)--(-0.5,1.25);
	\node at (-0.5,1.25) {\Large$\boldsymbol{\times}$};
	\node at (0.25,0) {\Large+};
	\draw [fermion](1.0,0) -- (2.5,0);
	\draw[fill=black] (2.5,0) circle (.075cm);
	\draw[vector] (2.5,0)--(2.5,1.25);
	\node at (2.5,1.25) {\Large$\boldsymbol{\times}$};
	\draw [fermion](2.5,0) -- (3.5,0);
	\draw[fill=black] (3.5,0) circle (.075cm);
	\draw[vector] (3.5,0)--(3.5,1.25);
	\node at (3.5,1.25) {\Large$\boldsymbol{\times}$};
	\node at (4.25,0) {\Large+};
    \draw [fermion](5.0,0) -- (6.5,0);
    \draw[fill=black] (6.5,0) circle (.075cm);
    \draw[vector] (6.5,0)--(6.5,1.25);
    \node at (6.5,1.25) {\Large$\boldsymbol{\times}$};
    \draw [fermion](6.5,0) -- (7.5,0);
    \draw[fill=black] (7.5,0) circle (.075cm);
    \draw[vector] (7.5,0)--(7.5,1.25);
    \node at (7.5,1.25) {\Large$\boldsymbol{\times}$}; 
    \draw [fermion](7.5,0) -- (8.5,0);
    \draw[fill=black] (8.5,0) circle (.075cm);
    \draw[vector] (8.5,0)--(8.5,1.25);
    \node at (8.5,1.25) {\Large$\boldsymbol{\times}$}; 
    \node at (9.20,0) {\Large+};
     \node at (10.0,0) {\dots};
    
    \end{tikzpicture}
    \nonumber
 \end{equation}

 \begin{equation}
\begin{tikzpicture}[line width=1.0 pt, scale=0.65]
	\node at (-5.5,0) {$\Sigma(\mathbf{q},\omega)$};
	\node at (-4.5,0) {$\equiv$};
	\draw[fill=gray!30!white] (-3.5,0) circle (.5cm);
	\node at (-2.5,0) {\Large=};
	\draw [fermion](-2,0) -- (-0.5,0);
	\draw[fill=black] (-2,0) circle (.075cm);
	\draw[vector] (-2,0)--(-2,1.25);
	\node at (-2,1.25){\Large$\boldsymbol{\times}$};
	\draw[fill=black] (-0.5,0) circle (.075cm);
	\draw[vector] (-0.5,0)--(-0.5,1.25);
	\node at (-0.5,1.25) {\Large$\boldsymbol{\times}$};
	\node at (0.25,0) {\Large+};
	\draw[fill=black] (1.25,0) circle (.075cm);
	\draw[vector] (1.25,0)--(1.25,1.25);
	\node at (1.25,1.25) {\Large$\boldsymbol{\times}$};
	\draw [fermion](1.25,0) -- (2.25,0);
	\draw[fill=black] (2.25,0) circle (.075cm);
	\draw[vector] (2.25,0)--(2.25,1.25);
	\node at (2.25,1.25) {\Large$\boldsymbol{\times}$};
	\draw [fermion](2.25,0) -- (3.25,0);
	\draw[fill=black] (3.25,0) circle (.075cm);
	\draw[vector] (3.25,0)--(3.25,1.25);
	\node at (3.25,1.25) {\Large$\boldsymbol{\times}$};
	\node at (4.0,0) {\Large+};
	\draw[fill=black] (4.75,0) circle (.075cm);
    \draw [fermion](4.75,0) -- (5.75,0);
    \draw[vector] (4.75,0)--(4.75,1.25);
    \node at (4.75,1.25) {\Large$\boldsymbol{\times}$};
    \draw[fill=black] (5.75,0) circle (.075cm);
    \draw[vector] (5.75,0)--(5.75,1.25);
    \node at (5.75,1.25) {\Large$\boldsymbol{\times}$};
    \draw [fermion](5.75,0) -- (6.75,0);
    \draw[fill=black] (6.75,0) circle (.075cm);
    \draw[vector] (6.75,0)--(6.75,1.25);
    \node at (6.75,1.25) {\Large$\boldsymbol{\times}$}; 
    \draw [fermion](6.75,0) -- (7.75,0);
    \draw[fill=black] (7.75,0) circle (.075cm);
    \draw[vector] (7.75,0)--(7.75,1.25);
    \node at (7.75,1.25) {\Large$\boldsymbol{\times}$}; 
    \node at (8.35,0) {\Large+};
     \node at (9.0,0) {\dots};
    \end{tikzpicture}
      \nonumber
 \end{equation}
and
 \begin{equation}
\begin{tikzpicture}[line width=1.0 pt, scale=0.65]
	\node at (-5.5,0) {$G(\mathbf{q},\omega)$};
	\node at (-4.5,0) {$\equiv$};
	\draw[line  width=1.25mm] (-4.25,0)--(-2.80,0);
	\node at (-2.45,0) {\Large=};
	\draw [fermion](-2,0) -- (-0.5,0);
	\node at (0,0) {\Large+};
	\draw [fermion](0.5,0) -- (1.5,0);
	\draw[fill=gray!30!white] (1.95,0) circle (.4cm);
	\draw [fermion](2.35,0) -- (3.35,0);
	\node at (3.75,0) {\Large+};
	\draw [fermion](4.25,0) -- (5.25,0);
	\draw[fill=gray!30!white] (5.70,0) circle (.4cm);
    \draw [fermion](6.10,0) -- (7.10,0);
    \draw[fill=gray!30!white] (7.55,0) circle (.4cm);
    \draw [fermion](7.95,0) -- (8.95,0);
    \node at (9.25,0) {\Large+};
     \node at (10,0) {\Large{\dots}};
    \end{tikzpicture}
      \nonumber
 \end{equation}
 or the Dyson equation
 \begin{equation}
\begin{tikzpicture}[line width=1.0 pt, scale=0.65]
	\draw[line  width=1.25mm] (-4.25,0)--(-2.80,0);
	\node at (-2.45,0) {\Large=};
	\draw [fermion](-2,0) -- (-0.5,0);
	\node at (0,0) {\Large+};
	\draw [fermion](0.5,0) -- (1.5,0);
	\draw[fill=gray!30!white] (1.95,0) circle (.4cm);
	\draw[line  width=1.25mm] (2.35,0)--(3.35,0);
    \end{tikzpicture}
          \nonumber
 \end{equation}
 where the convention for drawing  the diagrams is shown in Fig. \ref{fig:feynman-rule}. We note that, in the present scheme, the single interaction (wavy line) diagram  does not appear in the expansion of $\Sigma(\mathbf{q},\omega)$.

A  quantity of interest is the one-particle spectral density function or SDF,  $A(\mathbf{k},\omega) $, where we make use  the standard device (e.g. \cite{Coleman_2015}) by replacing the real frequency $\omega$ with the complex frequency $\omega-i\delta$ where $\delta$ is a small positive  real number:
\begin{equation}
\label{eqn:sdf-def}
A(\mathbf{k},\omega) \equiv 2 \lim_{\delta \to 0} \big[ \, \mathrm{Im}\,G(\mathbf{k},\mathbf{k},\omega-i\delta)\big].
\end{equation}
\noindent
The self-energy has  both a real and an imaginary parts, using the notation in  \cite{Coleman_2015}, we write
\begin{equation}
\label{eqn:self-energy-split}
\Sigma(\mathbf{k},\omega-i\delta) = \Sigma^\prime(\mathbf{k},\omega) + i \,\Gamma(\mathbf{k},\omega).
\end{equation}
Hence, we obtain
\begin{equation}
\label{eqn:sdf-1}
A(\mathbf{k},\omega) =  \frac{2\Gamma(\mathbf{k},\omega)}{\big[\omega-\epsilon_\mathbf{k}+\Sigma^\prime (\mathbf{k},\omega)\big]^2+\big[\Gamma(\mathbf{k},\omega)\big]^2}.
\end{equation}
\noindent
 We note that if the self-energy is small,  $A(\mathbf{k},\omega)$ has a Lorentzian shape in the $\omega$-domain  of width $\Gamma$ centered around a renormalized energy $\epsilon^\ast_\mathbf{k}=\epsilon_\mathbf{k} + \Sigma^\prime(\mathbf{k},\epsilon^\ast_\mathbf{k} )$. By Taylor expanding $\Sigma^\prime(\mathbf{k},\omega)$ about this point, i.e. near $\omega \approx \epsilon^\ast_\mathbf{k}$, the  Green function can be expressed as \cite{Coleman_2015}
\begin{equation}
\label{eqn:gen-green}
G(\mathbf{k},\omega-i\delta) = \frac{Z_\mathbf{k}}{ \omega-\epsilon^\ast_\mathbf{k} - i \Gamma_\mathbf{k}^\ast},
\end{equation}
\noindent
where  $Z_\mathbf{k}$  is a renormalization factor: $Z_\mathbf{k}^{-1}=\big[1-\partial_\omega \Sigma^\prime (\mathbf{k},\omega)]_{\omega=\epsilon^\ast_\mathbf{k} }$.  For small $\Gamma_\mathbf{k}^\ast$, we have 
\begin{align}
\label{eqn:magnon-energy}
\epsilon^\ast_\mathbf{k}  &= \epsilon_\mathbf{k}+\Sigma^\prime(\mathbf{k},\epsilon^\ast_\mathbf{k} ),\\
\label{eqn:magnon-lifetime}
\Gamma_\mathbf{k}^\ast&= Z_\mathbf{k} \; \Gamma(\mathbf{k},\epsilon^\ast_\mathbf{k}).
\end{align}
\noindent
The renormalized energy in Eq.  \eqref{eqn:magnon-energy} may be interpreted as a quasiparticle  (magnon) energy and Eq.  \eqref{eqn:magnon-lifetime}  defines its lifetime (see below). Differentiating $\epsilon_\mathbf{k}$ with respect to $\epsilon^\ast_\mathbf{k}$ and recalling that the spin wave  group velocity is   $\mathbf{v}_g=\nabla_\mathbf{k}\epsilon_\mathbf{k}$, we obtain
\begin{equation}
\label{eqn:magnon-ratio}
\frac{d \epsilon_\mathbf{k} }{d \epsilon^\ast_\mathbf{k}  } = Z_\mathbf{k}^{-1} = \frac{\mathbf{v}_g}{\mathbf{v}_g^\ast}\nonumber,
\end{equation}
\noindent
which says that the group velocity of the spin wave is renormalized downward.

Finally, there is a general relationship, via the \emph{fluctuation-dissipation theorem}, that connects the pair correlation function of the magnon operators to the SDF (cf.  \ref{sec:appA}) through
\begin{equation}
\label{eqn:correlfun_sdf}
\langle b^\dagger_\mathbf{k}(t)b_\mathbf{k}(0) \rangle = \frac{1}{2\pi}\int_{-\infty}^\infty e^{-i\omega t}A(\mathbf{k},\omega) n_\eta(\omega)d\omega,
\end{equation}
\noindent
$n_\eta (\omega)= [\exp(\beta\omega)-\eta]^{-1}$ give the Bose $\eta=1$ and the Fermi $\eta=-1$ distributions, respectively.

\subsection{Averaging over impurity distribution}
\label{sec:av-field}
Our knowledge of the actual distribution of impurities in the lattice  may be poor, therefore an averaging of all possible impurity distributions is considered.  An average  Green function can be obtained by averaging over the distribution of impurities. The contribution of the term $\Sigma(\mathbf{k},\omega)$ in Eq. \eqref{eqn:self-energy-2} to the Green function \eqref{eqn:eom-green-sol1} is still a function of $c_{\mathbf{k}}$, which is the Fourier transform of the actual distribution of impurities in the lattice. There are $N!/n!(N-n)!$ distributions for $n$ impurities on $N$ lattice sites. In accordance with our first assumption in Sec. \ref{sec:intro}, in case of a disordered state of the lattice, all these distributions are equally probable. The average of a thermodynamic potential over fluctuations of the impurities, is equivalent to the expectation value in the space of the product of  $n\,c_{\mathbf{q}}$ variables, or their cumulant average; see Eq.  \eqref{eqn:cqn}.

Using  Eq.  \eqref{eqn:cqn}, we can calculate the average of $\Sigma(\mathbf{k},\omega)$. The physical content of the Green function \eqref{eqn:eom-green-sol1}  is related to its singularities, usually poles, i.e. the roots, real or complex, of the denominator in the expression \eqref{eqn:eom-green-sol1}. Therefore, the averaging over the impurity distribution is carried out in the denominator of the Green function to obtain physically reasonable equation for the position of the poles rather than to average the whole Green function, which may affect or modify its form.

Since the condition expressed by the $\delta$-functions in Eq. \eqref{eqn:cqn} is exactly satisfied for each term of the sum in Eq. \eqref{eqn:self-energy-2} then from these two equations  together with Eq. \eqref{eqn:pnc} we obtain
\begin{align}\begin{split}
\label{eqn:self-energy-av1}
\langle \Sigma(\mathbf{q},\omega)\rangle_{c}  &=   \sum_{n=1}^\infty \frac{p_{n+1}(c)}{N^n} \sum_{\mathbf{k}_1\dots\mathbf{k}_n}\frac{ \gamma_\mathbf{q}\gamma_{\mathbf{k}_1}\dots\gamma_{\mathbf{k}_n}}{(\epsilon_{\mathbf{k}_1}-\omega)\dots(\epsilon_{\mathbf{k}_n}-\omega)}\,.
\end{split} \end{align}
\noindent

Let us now define a base propagator  as
\begin{equation}
\label{eqn:base-green}
\mathfrak{g}^0(\omega) \equiv \frac{1}{N} \sum_{\mathbf{p}}\frac{\gamma_\mathbf{p}}{\epsilon_{\mathbf{p}}-\omega} =  \frac{2}{N} \sum_{\mathbf{p}}\frac{(1-c)^{-1}\epsilon_\mathbf{p}-B}{\epsilon_{\mathbf{p}}-\omega}
\end{equation}
\noindent
Then we write Eq. \eqref{eqn:self-energy-av1} in a compact form
\begin{equation}
\label{eqn:self-energy-av2}
\langle \Sigma(\mathbf{q},\omega)\rangle_{c}  = \frac{c(1-c) \gamma_\mathbf{q} \mathfrak{g}^0(\omega)}{[1-(1-c)\mathfrak{g}^0(\omega)][1+c\mathfrak{g}^0(\omega)]}.
\end{equation}
\noindent
The average  SDF is obtained by  taking the cumulant average of Eq.  \eqref{eqn:sdf-1} we have
\begin{equation}
\label{eqn:sdf-2}
\bar{A}(\mathbf{k},\omega) = \frac{2 \bar{\Gamma}( \mathbf{k},\omega) }{\big[\omega-\epsilon_\mathbf{k}+\bar{\Sigma}^\prime(\mathbf{k},\omega) \big]^2+\big[\bar{\Gamma}(\mathbf{k},\omega)\big]^2}.
\end{equation}
\noindent
Here and in sequel, we put $\bar{\Sigma}^\prime(\mathbf{k},\omega) \equiv \langle \Sigma^\prime(\mathbf{k},\omega)\rangle_{c}$, $\bar{\Gamma}(\mathbf{k},\omega) \equiv \langle \Gamma(\mathbf{k},\omega)\rangle_{c}$ and $\bar{A}(\mathbf{k},\omega) \equiv \langle A(\mathbf{k},\omega)\rangle_{c} $. For example, the Dyson equation \eqref{eqn:dyson-eq} after configurational averaging and  rearrangement reads
\begin{equation}
\label{eqn:av-dyson-eq}
\bar{G}(\mathbf{k},\omega) =  \big[G_0^{-1}(\mathbf{k},\omega)  - \bar\Sigma(\mathbf{k},\omega)\big]^{-1}.
\end{equation}
\noindent
 Diagrammatically,  configurational averaging corresponds to linking the crosses (see foregoing diagrams) by a bullet in all possible ways and associating a factor $p_m(c)$ with $m$ linked crosses \cite{jones1971impurity,edwards1958new}. For example, one  denotes the impurity-averaged double scattering event diagram by
  \begin{equation}
\begin{tikzpicture}[line width=1.0 pt, scale=0.65]
	\node at (-3.25,0.5) {$\boldsymbol{\Bigg[}$};
	\draw (-3.0,0) -- (-2,0);
	\draw (-2,0) -- (-1,0);
	\draw[fill=black] (-2,0) circle (.05cm);
	\draw[vector] (-2,0)--(-2,1.25);
	\node at (-2,1.25){$\boldsymbol{\times}$};
	\draw[fill=black] (-1,0) circle (.05cm);
	\draw[vector] (-1,0)--(-1,1.25);
	\node at (-1,1.25) {$\boldsymbol{\times}$};
	\draw  (-1,0)--(0,0);
	\node at (0.20,0.5){$\boldsymbol{\Bigg]}$};
	\draw[line width=0.5mm] (-3.25,1.6)--(0.20,1.6);
	\node at (0.85,0.25) {\Large=};
	\draw (1.5,0) -- (2.5,0);
	\node at (2.0, 0.30) {$\mathbf{k}$};
	\node at (2.4, 0.85) {$\mathbf{q}$};
	\draw[fill=black] (2.5,0) circle (.05cm);
	\draw (2.5,0) -- (3.5,0);
	\node at (3.0, -0.35) {$\mathbf{k}-\mathbf{q}$};
	\draw[vector] (2.5,0)--(3.0,1.5);
	\draw[fill=black] (3.0,1.5) circle (.125cm);
	\draw[vector] (3.0,1.5)--(3.5,0);
	\draw[fill=black] (3.5,0) circle (.05cm);
	\draw (3.5,0) -- (4.5,0);
	\node at (4.0, 0.30) {$\mathbf{k}$};
    \end{tikzpicture}
      \nonumber
 \end{equation}
 Note that after averaging the momentum is conserved, cf. $\S$8.6.2 in \cite{Coleman_2015}.
\subsection{Spectral density function}
\label{sec:cases}
In order to evaluate Eq. \eqref{eqn:sdf-2},  when the limit $\delta \to 0$ is taken, we consider the case that both $\bar{\Sigma}^\prime(\mathbf{k},\omega)$ and $\bar{\Gamma}(\mathbf{k},\omega) $ are nonzero:
\begin{align}
\label{eqn:case31}
\lim_{\delta\to 0} \bar{\Sigma}^\prime(\mathbf{k},\omega+i \delta) &=\bar{\Sigma}^\prime(\mathbf{k},\omega) \ne 0,\\
\label{eqn:case32}
\lim_{\delta\to 0}  \bar{\Gamma}(\mathbf{k},\omega+ i \delta) &= \bar{\Gamma}(\mathbf{k},\omega)  \ne 0,
\end{align}
\noindent
We note that if $\bar{\Gamma}(\mathbf{k},\omega+ i \delta)=0$, then

\begin{equation}
\label{eqn:sdf-case2}
\bar{A}(\mathbf{k},\omega) = 2\pi \delta(\omega - \omega^\ast_\mathbf{k}),
\end{equation}
\noindent
where $\omega^\ast_\mathbf{k}$ is the solution of the equation
\begin{equation}
\label{eqn:green-case2}
\omega - \epsilon_\mathbf{k} + \bar{\Sigma}^\prime(\mathbf{k},\omega) = 0,
\end{equation}
\noindent
for $\omega$, which  also leads to a dispersion relation $\omega=\omega^\ast_\mathbf{k}$ for magnons.  Regarding Eq.  \eqref{eqn:case32}, two situations are of interest, viz.
\begin{align}
\label{eqn:case33}
\bar{\Gamma}(\mathbf{k},\omega)  &\ll \omega^\ast_\mathbf{k},\\
\label{eqn:case34}
\bar{\Gamma}(\mathbf{k},\omega)  &\approx \omega^\ast_\mathbf{k}.
\end{align}
In the range of frequencies for which Eq. \eqref{eqn:case33} is satisfied, we can refer to \emph{quasi-stationary states}  or QSS in the sense that the system oscillates with the frequency $\omega^\ast_\mathbf{k}$ and a damping factor $\exp(-\lambda t)$, where $\lambda\equiv \lambda_\mathbf{k}$ is a positive parameter, the decay rate,  to be determined. Indeed, since the poles of  Eq. \eqref{eqn:sdf-2} are shifted from the real axis into the upper $\omega$ complex plane, we shall look for the roots of the equation ($\delta \to 0$)
\begin{equation}
\label{eqn:sdf-2_poles}
 \big[\omega-\epsilon_\mathbf{k}+\bar{\Sigma}^\prime(\mathbf{k},\omega) \big]^2+\big[\bar{\Gamma}(\mathbf{k},\omega)\big]^2 =0,
\end{equation}
\noindent
or
\begin{equation}
\label{eqn:sdf-2_poles-1}
 \omega-\epsilon_\mathbf{k} + \bar{\Sigma}^\prime(\mathbf{k},\omega) \pm i \, \bar{\Gamma}(\mathbf{k},\omega)=0.
\end{equation}
\noindent
If condition \eqref{eqn:case33} is satisfied, i.e. $\bar{\Gamma}(\mathbf{k},\omega) /\omega^\ast_\mathbf{k} \ll 1$, magnons are considered as  bona fide quasiparticles, and we can look for a solution in the form
\begin{equation}
\label{eqn:sdf-2_gensols}
 \omega = \omega^\ast_\mathbf{k} \pm i \lambda_\mathbf{k},
\end{equation}
\noindent
where $\lambda_\mathbf{k}$ is assumed to be a small parameter. Substituting  Eq. \eqref{eqn:sdf-2_gensols} into Eq. \eqref{eqn:sdf-2_poles-1}
\begin{equation}
\label{eqn:sdf-2_sols}
 \omega^\ast_\mathbf{k} \pm i \lambda_\mathbf{k}-\epsilon_\mathbf{k}+\bar{\Sigma}^\prime(\mathbf{k},\omega^\ast_\mathbf{k} \pm i \lambda_\mathbf{k}) \pm i \bar{\Gamma}(\mathbf{k},\omega^\ast_\mathbf{k}\pm i \lambda_\mathbf{k}) =0.
\end{equation}
\noindent
Considering  the argument  $+i\lambda_\mathbf{k}$ and expanding the self-energy components in powers of $\lambda_\mathbf{k}$
\begin{align}
\label{eqn:self-energy_real-expand}
\bar{\Sigma}^\prime(\mathbf{k},\omega^\ast_\mathbf{k} + i \lambda_\mathbf{k}) &= \bar{\Sigma}^\prime(\mathbf{k},\omega^\ast_\mathbf{k}) +i\lambda_\mathbf{k}\partial_\omega\bar{\Sigma}^\prime(\mathbf{k},\omega)\Big|_{\omega=\omega^\ast_\mathbf{k}} + O(\lambda_\mathbf{k}^2),\\
\label{eqn:self-energy_imag-expand}
\bar{\Gamma}(\mathbf{k},\omega^\ast_\mathbf{k}+i\lambda_\mathbf{k})  &= \bar{\Gamma}(\mathbf{k},\omega^\ast_\mathbf{k})+ O(\lambda_\mathbf{k}^2),
\end{align}
where by Eq. \eqref{eqn:case33}, $\bar{\Gamma}(\mathbf{k},\omega^\ast_\mathbf{k}) \sim  O(\lambda_\mathbf{k})$. Inserting  now Eq. \eqref{eqn:self-energy_real-expand}-\eqref{eqn:self-energy_imag-expand} in  Eq. \eqref{eqn:sdf-2_sols} with $+i\lambda_\mathbf{k}$, we obtain two relations [cf. Eqs. \eqref{eqn:magnon-energy}-\eqref{eqn:magnon-lifetime}]:
\begin{equation}
\label{eqn:sdf-2_sol1}
 \omega^\ast_\mathbf{k} -\epsilon_\mathbf{k}+\bar{\Sigma}^\prime(\mathbf{k},\omega^\ast_\mathbf{k})  = 0,
\end{equation}
\noindent
and
\begin{equation}
\label{eqn:sdf-2_sol2}
 \tau_\mathbf{k}^{-1} \equiv \lambda_\mathbf{k} = \frac{\bar{\Gamma}(\mathbf{k},\omega^\ast_\mathbf{k})}{1+\partial_\omega \bar{\Sigma}^\prime(\mathbf{k},\omega)\big|_{\omega=\omega^\ast_\mathbf{k}}}.
\end{equation}
\noindent
Then from  Eq.  \eqref{eqn:sdf-2}
\begin{equation}
\label{eqn:sdf-case3}
\bar{A}(\mathbf{k},\omega) = \frac{2 \bar{\Gamma}(\mathbf{k},\omega^\ast_\mathbf{k})}{\big(\omega-\omega^\ast_\mathbf{k})^2+\big[\bar{\Gamma}(\mathbf{k},\omega^\ast_\mathbf{k})\big]^2}.
\end{equation}
\noindent
Putting now this  relation in Eq. \eqref{eqn:correlfun_sdf}, for bosons, we write
\begin{equation}
\label{eqn:correl00_sdf}
\langle b^\dagger_\mathbf{k}(0)b_\mathbf{k}(0) \rangle = \frac{1}{\pi}\int_{-\infty}^\infty  \frac{d\omega}{e^{\beta\omega}-1}\cdot \frac{\bar{\Gamma}(\mathbf{k},\omega^\ast_\mathbf{k})}{\big(\omega-\omega^\ast_\mathbf{k})^2+\big[\bar{\Gamma}(\mathbf{k},\omega^\ast_\mathbf{k})\big]^2}.
\end{equation}
\noindent
Note that if relation \eqref{eqn:case33} holds, we can insert the identity
\begin{equation}
\label{eqn:dirac_fun_prop}
\delta(\omega-\omega^\ast_\mathbf{k}) = \frac{1}{\pi}\lim_{\bar{\Gamma}\to 0} \frac{\bar{\Gamma}(\mathbf{k},\omega^\ast_\mathbf{k})}{\big(\omega-\omega^\ast_\mathbf{k})^2+\big[\bar{\Gamma}(\mathbf{k},\omega^\ast_\mathbf{k})\big]^2},
\end{equation}
\noindent
in Eq. \eqref{eqn:correl00_sdf} to obtain
\begin{equation}
\label{eqn:correl00_sdf-lim}
\langle b^\dagger_\mathbf{k}(0)b_\mathbf{k}(0) \rangle =  \frac{1}{e^{\beta\omega^\ast_\mathbf{k}}-1}.
\end{equation}
\noindent
If  the condition \eqref{eqn:case33} does not hold, we cannot refer to QSS.  This puts a natural limit on considering magnon as a quasiparticle. Note that $\tau_\mathbf{k}$, given by  Eq. \eqref{eqn:sdf-2_sol2}, defines the lifetime of the quasiparticle with momentum $\mathbf{k}$.  It can be compared with Eq. \eqref{eqn:magnon-lifetime}, in which $\lambda_\mathbf{k}$ serves as the impurity averaged $\Gamma_\mathbf{k}^\ast$. This quantity  can be included in the expression for the Green function
\begin{equation}
\label{eqn:gen-green-omega}
\bar{G}(\mathbf{k},\omega+i\delta) = \frac{\Lambda_\mathbf{k}}{ \omega-\omega_\mathbf{k}^\ast + i \lambda_\mathbf{k}}, 
\end{equation}
\noindent
cf.  Eq. \eqref{eqn:gen-green}  with $\epsilon^\ast_\mathbf{k} \Leftrightarrow \omega_\mathbf{k}^\ast$, $\omega_\mathbf{k}^\ast  =\epsilon_\mathbf{k} + \Sigma^\prime(\mathbf{k},\omega_\mathbf{k}^\ast)$, $\lambda_\mathbf{k} = \Lambda_\mathbf{k} \bar{\Gamma}(\mathbf{k},\omega_\mathbf{k}^\ast)$, and $\Lambda_\mathbf{k}  \equiv  \big(1+\partial_\omega \bar{\Sigma}^\prime(\mathbf{k},\omega)\big|_{\omega=\omega^\ast_\mathbf{k}}\nonumber\big)^{-1}$.

 The inverse Fourier transform of  Eq. \eqref{eqn:gen-green-omega}, for sufficiently large $t$,  reads
\begin{equation}
\label{eqn:gen-green-long-t}
\bar{G}(\mathbf{k},t) \sim i a_{-1} e^{-i \omega_\mathbf{k}^\ast t - \lambda_\mathbf{k} t},
\end{equation}
\noindent
where $a_{-1}$ is the residue of $\bar{G}(\mathbf{k},\omega)$ at the pole; cf.  \S  7.3 in Ref.  \cite{AGD_1963}. Hence,   $\bar{G}(\mathbf{k},t)$ is the propagator of a magnon with energy $\omega_\mathbf{k}^\ast$, decaying in time with the rate $\lambda_\mathbf{k}$. This decaying Green function corresponds to a finite width of the spectral density function, viz.
\begin{align}
\label{eqn:spectral-function-width}
\bar{A}(\mathbf{k},\omega) &=-2 \mathrm{Im}\bar{G}(\mathbf{k},\omega+i\delta) =  -2 \mathrm{Im} \int_{-\infty}^\infty \bar{G}(\mathbf{k},t) e^{i\omega t}dt,\nonumber\\
\bar{A}(\mathbf{k},\omega)  &= \frac{2 a_{-1}\lambda_\mathbf{k}}{(\omega-\omega_\mathbf{k}^\ast)^2+\lambda_\mathbf{k}^2},
\end{align}
\noindent
where the width in the energy domain is $\lambda_\mathbf{k}$.

\subsection{Self-energy function}
\label{sec:selfenergy-calc}
We intend  to calculate $\bar{\Sigma}^\prime(\mathbf{k},\omega)$ and $\bar{\Gamma}(\mathbf{k},\omega)$. We start by separating the real and imaginary part of the self-energy and write  Eq. \eqref{eqn:self-energy-av1} or  Eq. \eqref{eqn:self-energy-av2} as
\begin{equation}
\label{eqn:self-energy-split1}
\bar{\Sigma}(\mathbf{k},\omega)  = c(1-c) \gamma_\mathbf{k} \big[f_1(\omega)-f_2(\omega)\big]
\end{equation}
\noindent
where
\begin{equation}
\label{eqn:self-energy-fi}
f_i(\omega) = \frac{\mathfrak{g}_i^0(\omega)}{1-\mathfrak{g}_i^0(\omega)},\quad i=1,2
\end{equation}
\noindent
$\mathfrak{g}_i^0(\omega)=s_i\mathfrak{g}^0(\omega)$, $s_1=1-c$,  $s_2=-c$, and recall that $\gamma_{\mathbf{k}} \equiv 2[B+(1-c)\varepsilon_\mathbf{k}]$, $\varepsilon_{\mathbf{k}}=S(J_0-J_{\mathbf{k}})$.

Thus the calculation of  $\bar{\Sigma}^\prime(\mathbf{k},\omega)$ and $\bar{\Gamma}(\mathbf{k},\omega)$  boils down to that of determining $\mathrm{Re} f_i(\omega)$ and $\mathrm{Im} f_i(\omega)$. After simple algebra, we write
\begin{align}
\label{eqn:Re-fi}
\mathrm{Re} f_i(\omega)  &= \frac{(1-\Re \mathfrak{g}^0_i)\Re \mathfrak{g}^0_i-(\Im \mathfrak{g}^0_i)^2}{(1-\Re \mathfrak{g}^0_i)^2 + (\Im \mathfrak{g}^0_i)^2},\quad i=1,2,\\
\label{eqn:Im-fi}
\mathrm{Im} f_i(\omega)  &= \frac{\Im \mathfrak{g}^0_i}{(1-\Re \mathfrak{g}^0_i)^2 + (\Im \mathfrak{g}^0_i)^2},\qquad i=1,2,
\end{align}
where $\Re \mathfrak{g}^0_i = \mathrm{Re}\mathfrak{g}^0_i$ and $\Im \mathfrak{g}^0_i = \mathrm{Im}\mathfrak{g}^0_i$. From Eq. \eqref{eqn:base-green} and by shifting $\omega \to \omega + i\delta$, we write
\begin{align}
\label{eqn:Re-base-green}
\Re \mathfrak{g}^0_j(\omega+i \delta) &= \frac{2s_j}{N(1-c)} \sum_{\mathbf{k}}\frac{[(1-c)^{-1}\epsilon_\mathbf{k}-B](\epsilon_{\mathbf{k}}-\omega)}{(\epsilon_{\mathbf{k}}-\omega)^2+\delta^2},\\
\Im \mathfrak{g}^0_j(\omega + i \delta) &= \frac{2s_j}{N(1-c)} \sum_{\mathbf{k}}\frac{[(1-c)^{-1}\epsilon_\mathbf{k}-B]\delta}{(\epsilon_{\mathbf{k}}-\omega)^2+\delta^2}.
\label{eqn:Im-base-green}
\end{align}
\noindent
In the limit $\delta \to 0$
\begin{align}
\label{eqn:Re-base-green0}
\lim_{\delta\to 0}\Re \mathfrak{g}^0_j(\omega+i\delta) &= \frac{2s_j}{N} \sum_{\mathbf{k}}\mathcal{P}\frac{[(1-c)^{-1}\epsilon_\mathbf{k}-B]}{\epsilon_{\mathbf{k}}-\omega}=\Re \mathfrak{g}^0_j(\omega),\\
\lim_{\delta\to 0}\Im \mathfrak{g}^0_j(\omega+i\delta) &= \frac{2\pi s_j}{N} \sum_{\mathbf{k}}[(1-c)^{-1}\epsilon_\mathbf{k}-B]\delta(\epsilon_{\mathbf{k}}-\omega)=\Im \mathfrak{g}^0_j(\omega),
\label{eqn:Im-base-green0}
\end{align}
\noindent
where $\mathcal{P}$ denotes the Cauchy principal value of an integral when it appears. Hence, we write the real and imaginary part  of the self-energy as
\begin{align}
\label{eqn:case-gen-re}
\bar{\Sigma}^\prime(\mathbf{k},\omega) &= c (1-c) \gamma_\mathbf{k}\big[\mathrm{Re} f_1(\omega)-\mathrm{Re} f_2(\omega)\big],\\
\label{eqn:case-gen-im}
\bar{\Gamma}(\mathbf{k},\omega) &= c (1-c) \gamma_\mathbf{k}\big[\mathrm{Im} f_1(\omega)-\mathrm{Im} f_2(\omega)\big].
\end{align}
\noindent
We next calculate $\Re \mathfrak{g}^0_i(\omega)$ and $\Im \mathfrak{g}^0_i(\omega)$ from Eqs.  \eqref{eqn:Re-base-green0} and \eqref{eqn:Im-base-green0} by replacing the summations over discrete values of $\mathbf{k}$ in the first Brillouin zone (BZ) with continuous integrations, viz.
\begin{align}
\label{eqn:Re-base-green0-int}
\Re \mathfrak{g}^0_1(\omega) &= 2\varv \fint \frac{d\mathbf{k}}{(2\pi)^3} \frac{\epsilon_\mathbf{k}-(1-c)B}{\epsilon_\mathbf{k}-\omega},\\
\Im \mathfrak{g}^0_1(\omega) &= 2\pi\varv \int  \frac{d\mathbf{k}}{(2\pi)^3} [\epsilon_\mathbf{k}-(1-c)B] \, \delta(\epsilon_\mathbf{k}-\omega),
\label{eqn:Im-base-green0-int}
\end{align}
\noindent
and
\begin{equation}
\label{eqn:G10toG20}
\mathfrak{g}^0_2(\omega) =  -\frac{c}{1-c}\mathfrak{g}^0_1(\omega),
\end{equation}
\noindent
where  $\varv=\mathcal{V}/N$ is the volume of the primitive cell and the symbol $\fint$ denotes the principal value integral in the Cauchy sense.\footnote{Recall that in a $d$-dimensional hypercube in momentum space with reciprocal  volume $L^{-d}$, lattice spacing $2\pi/L$,  $\sum_q \to L^{d}\int_{0<q_i<2\pi/L} \frac{d^dq}{(2\pi)^d}$; see  e.g.   $\S$2.5 in \cite{Coleman_2015}.}
\section{Results}
\label{sec:result}
In this section, we employ the formalism developed in  the foregoing sections in the simple cubic lattice medium.  Through lattice Green functions, we compute the spectral density function, the dispersion relation, and the magnon lifetime in terms of the mean concentration of impurity. From the spectral density function, we calculate  the magnon number density from which the internal energy is calculated. Furthermore, the thermal field (Matsubara) method formulated for the system in section \ref{sec:matsubara} is used to compute the thermodynamic potential in the lattice. 
\subsection{Lattice Green functions}
\label{sec:scl}
 In order to obtain concrete results, we apply the foregoing formalism to the case of the simple cubic (sc) lattice.  The unperturbed dispersion relation, but scaled with an impurity concentration dependent factor,  for the sc lattice  is expressed  as 
\begin{align}
\label{eqn:sc-dispersion}
\varepsilon_\mathbf{k}^\mathrm{sc}   &=  \bar\varepsilon_0(c)\Big[1-\frac{1}{3}\sum_{i=1}^3\cos(k_i a)\Big].
\end{align}
\noindent
Here, $\bar\varepsilon_0(c)=(1-c)^2\sigma$, $\sigma=2SZJ$, with $Z=6$ nearest neighbors, $k_i$'s are the projection of the $\mathbf{k}$-vector on the principal axes of the lattice, $a$ is the lattice constant,  $c$ is the mean concentration of nonmagnetic impurities in the lattice, and $J$ is the mean exchange integral; cf. relations below Eq. \eqref{eqn:ft-h}. In the presence of an external magnetic field $H$, we set $\epsilon_\mathbf{k} =\varepsilon_\mathbf{k}^\mathrm{sc} +2(1-c)B$,   $B=\mu_BH/2$.  The dispersion relation \eqref{eqn:sc-dispersion} is also called the mean lattice approximation and is manifestly translation invariant \cite{jones1971impurity}, and $\varepsilon_\mathbf{k}^\mathrm{sc}/(1-c)^2 \equiv \omega_\mathbf{k}^0$  represents the magnon energy band of the pure crystal. Putting now $\mathbf{k}a=\mathbf{q}$, $\varv=a^3$  the volume of the unit cell,  and considering the symmetry of  the first BZ,  we write Eqs. \eqref{eqn:Re-base-green0-int}-\eqref{eqn:Im-base-green0-int} as
\begin{align}
\label{eqn:Re-base-green0-int-1}
\frac{1}{2}\Re \mathfrak{g}^0_1(\omega) &=  \fint\frac{d^3q}{(2\pi)^3}\frac{\varepsilon^0_\mathbf{q}+(1-c)B}{\varepsilon^0_\mathbf{q}+2(1-c)B -\omega},\\
\label{eqn:Im-base-green0-int-1}
\frac{1}{2}\Im \mathfrak{g}^0_1(\omega) &=  \pi \int \frac{d^3q}{(2\pi)^3}\big[\varepsilon^0_\mathbf{q} + (1-c)B\big]\delta\big(\varepsilon^0_\mathbf{q}+2(1-c)B-\omega\big),
\end{align}
\noindent
\begin{equation}
\label{eqn:sc-dispersion-1a}
\text{where} \quad \varepsilon^0_\mathbf{q} \equiv \varepsilon^\mathrm{sc}_\mathbf{q}= \bar\varepsilon_0\Big[ 1-\frac{1}{3}\sum_{i=1}^3\cos(q_i) \Big]=  \frac{2 \bar\varepsilon_0}{3}\sum_{i=1}^3\sin^2\Big(\frac{q_i}{2}\Big).
\end{equation}
\noindent
Supposing  that the wave vector (momentum) is isotropic, then Eq.  \eqref{eqn:sc-dispersion-1a} gives us $\varepsilon^0(q) = 2\bar\varepsilon_0\sin^2(q/2)$.  Thus in the long wavelength (infrared) limit where $k$ is small, $\varepsilon^0_k \approx (\bar\varepsilon_0/2) a^2k^2$, as expected. Furthermore, the relation between the energy of a free particle and its momentum ($E= \hslash^2k^2/2m$), gives the magnon effective mass: $m^\ast=(\bar\varepsilon_0a^2)^{-1}$, where we put $\hslash=1$. Scaling the variables by $\bar\varepsilon_0(c)$:  $\mu_\mathbf{q}  = \varepsilon^0_\mathbf{q}/\bar\varepsilon_0$, $ z = \omega/\bar\varepsilon_0$, and $\text{\ss} = B/\bar\varepsilon_0$,  the integrals in Eqs. \eqref{eqn:Re-base-green0-int-1}-\eqref{eqn:Im-base-green0-int-1} can be expressed as
\begin{align}
\label{eqn:Re-base-green0-int-q}
\frac{1}{2}\Re \mathfrak{g}^0_1(z) &=  \iiint\limits_\mathfrak{q}\frac{d^3q}{(2\pi)^3}\Bigg(\frac{\mu_\mathbf{q}+(1-c)\text{\ss}}{\mu_\mathbf{q}+2(1-c)\text{\ss}-z}\Bigg),\\
\label{eqn:Im-base-green0-int-q}
\frac{1}{2}\Im \mathfrak{g}^0_1(z) &=  \pi \iiint\limits_\mathfrak{q} \frac{d^3q}{(2\pi)^3}\big[\mu_\mathbf{q}+(1-c)\text{\ss}\big]\delta\big(\mu_\mathbf{q}+2(1-c)\text{\ss}-z\big),
\end{align}
\noindent
where $\mathfrak{q}$ denotes the domain of integration ($ 0\le q_i \le \pi; i=1,2,3$) and  $\mu_\mathbf{q}=(2/3)\sum_i \sin^2(q_i/2)$ is the lattice structure factor \cite{nolting2009quantum}. We change the domain of integration from the 3D  $\mathfrak{q}$-domain (reciprocal space) to  1D energy-domain  $\mu$ by introducing a density function  $g(\mu)$ with the normalization condition:
\begin{align}
\label{eqn:dos-norm}
\int_0^{z_m} g(\mu)d{\mu} &=  1,
\end{align}
\noindent
where $z_m$ is the maximum spin-wave energy. Converting  now the momentum integral to an energy integral through $g(\mu)$ \cite{Ashcroft_Mermin_1976,Landau_Lifshitz_1980},  we write Eqs. \eqref{eqn:Re-base-green0-int-q} and \eqref{eqn:Im-base-green0-int-q} as
\begin{align}
\label{eqn:Re-base-green0-int-e}
\frac{1}{2}\Re \mathfrak{g}^0_1(z) &= 1+\big[z-(1-c)\text{\ss}\big]\int_0^{z_m} \frac{g(\mu)}{\mu+2(1-c)\text{\ss}-z} d\mu,\\
\frac{1}{2}\Im \mathfrak{g}^0_1(z)  & = \left\{\begin{array}{ll}
\pi \big[z-(1-c)\text{\ss}\big]  g\big(z-2(1-c\text{\ss}\big), \quad 0<z \le z_m\\
0. \qquad \qquad \qquad \qquad\quad z  > z_m
\end{array}\right.
\label{eqn:Im-base-green0-int-e}
\end{align}
\noindent
In a  lattice,  $g(\mu)$ is the density of states  and is proportional to the imaginary part of the lattice Green function or $G_\mathrm{I}(\mu)$ \cite{Wolfram_Callaway_1963,Jelitto_1969}. For a pure simple cubic lattice we may write $g(\mu) \equiv g_0(\mu)$ 
\begin{align}
\label{eqn:dos-scc-cal}
g_0(\mu) &= \frac{2}{\pi}\int\limits_{0}^\infty \cos[(3-2\mu)\lambda] J_0^3 (\lambda) d\lambda,
\end{align}
\noindent
where  $J_0(\bullet)$ is the Bessel function of the first kind of zeroth order; see  \cite{Wolfram_Callaway_1963} and \ref{sec:appD}. 

The Hilbert transform of $g(\mu)$, $\mathrm{\hat{H}}[g(\mu)]  \equiv h(z)$,  defined as
\begin{align}
\label{eqn:Re-base-g0-integral}
h(z) &= \fint_0^{\infty} \frac{g(\mu)}{\mu-z} d\mu,
\end{align}
\noindent
can be evaluated for the pure simple cubic lattice as described in \ref{sec:appD}:
\begin{align}
\label{eqn:hdos-scc-cal}
h_0(z) &= 2 \int\limits_{0}^\infty \sin[(3-2z)\lambda] J_0^3 (\lambda) d\lambda.
\end{align}
\noindent
In general,  $h(z)$  is related to the real part of the lattice Green function, namely   $h \sim G_\mathrm{R}$. The real and imaginary parts of the lattice  $G$ can also be expressed in terms of integrals of elliptic integrals \cite{Morita_Horiguchi_1971}, or the Heun functions  \cite{joyce1973simple}, which are listed in \ref{sec:appE} for the simple cubic lattice.  

\begin{figure*}[htbp]
\centering
\subfigure[]{\label{fig:dos-zw}
\includegraphics[width=0.45\textwidth]{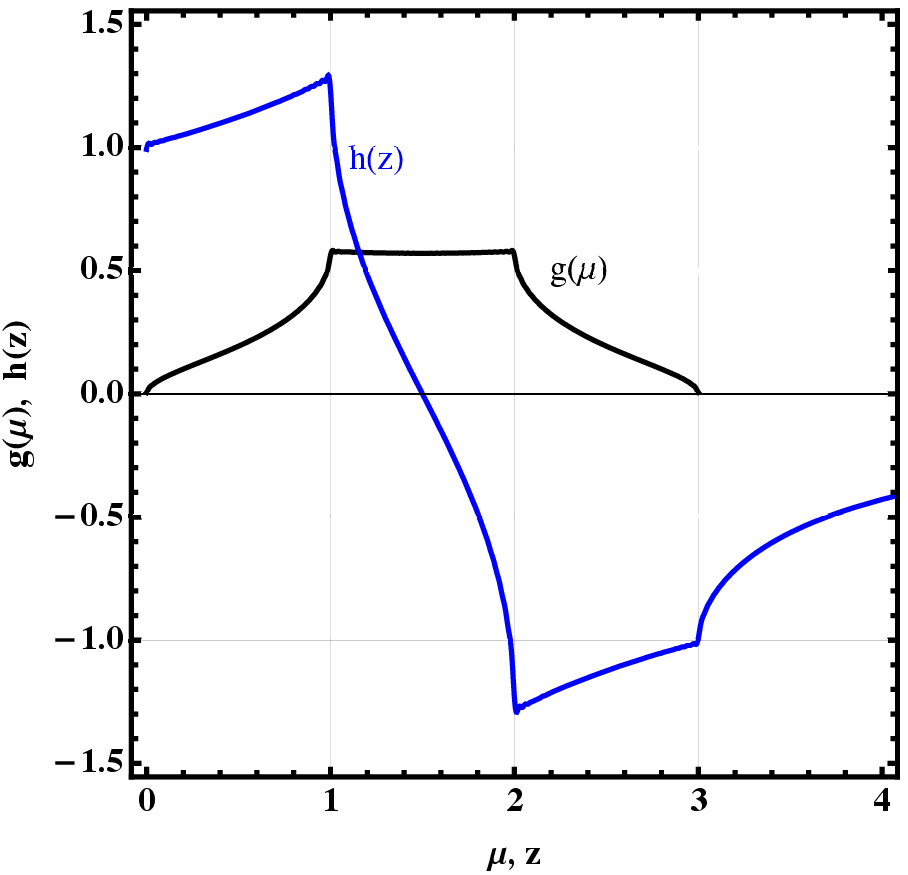}
}
\subfigure[]{\label{fig:psi-zw2i}
\includegraphics[width=0.45\textwidth]{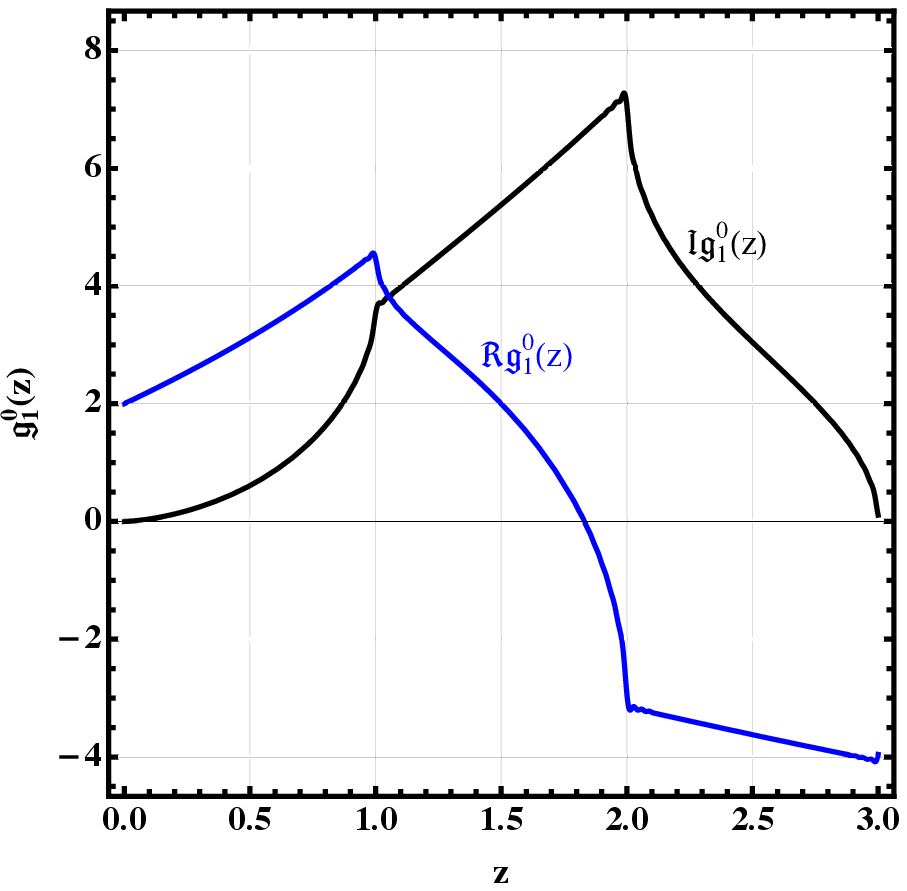}
}
\caption{(a) The pure cubic lattice density of states $g_0(\mu)$ and its Hilbert transform $h_0(z)$, as calculated by Eqs. \eqref{eqn:dos-scc-cal} and \eqref{eqn:hdos-scc-cal}, respectively. (b) The real  part  and the imaginary part  of the Green function  $\Re \mathfrak{g}^0_1(z)$  and  $\Im \mathfrak{g}^0_1(z)$ given by Eqs. \eqref{eqn:Re-base-green0-int-e0} and \eqref{eqn:Im-base-green0-int-e0}, respectively,   as a function of the scaled frequency parameter $z$,  with $z_m=3$. Recall that the dimensionless frequency $z$ contains $c$ as  $z=\omega/[\sigma(1-c)^2]$.}
\label{fig:dos-gfun}
\end{figure*}

Thus the problem of calculating $\Re \mathfrak{g}^0_1(\omega)$ and  $\Im \mathfrak{g}^0_1(\omega)$, and thereafter $\bar{\Sigma}^{\prime}(\mathbf{k},\omega)$ and $\bar{\Gamma}(\mathbf{k},\omega)$ through Eqs.  \eqref{eqn:case-gen-re} and \eqref{eqn:case-gen-im}, reduces to evaluating the density of states $g_0(\mu)$ with energy $\mu$ and its Hilbert transform $h_0(z)$ through Eqs.  \eqref{eqn:Re-base-green0-int-e} - \eqref{eqn:Im-base-green0-int-e}, where for $B=0$ we have
\begin{align}
\label{eqn:Re-base-green0-int-e0}
\frac{1}{2}\Re \mathfrak{g}^0_1(z) &= 1+ z h_0(z),\\
\frac{1}{2}\Im \mathfrak{g}^0_1(z) &= \left\{\begin{array}{ll}
\pi z  g_0\big(z\big), \quad 0<z  \le z_m\\
0. \qquad\qquad\quad z  >  z_m
\end{array}\right.
\label{eqn:Im-base-green0-int-e0}
\end{align}
\noindent
Figure  \ref{fig:dos-zw} shows plots of $g_0(\mu)$ and $h_0(z)$ versus $\mu$ and $z$, respectively. The real and imaginary parts of the Green function  given by Eqs. \eqref{eqn:Re-base-green0-int-e0} and \eqref{eqn:Im-base-green0-int-e0}  with $z_m=3$ are plotted in Fig. \ref{fig:psi-zw2i}.  The numerical integrations, which emanate from Eqs. \eqref{eqn:dos-scc-cal}-\eqref{eqn:hdos-scc-cal}, are carried out  in interval $\lambda =[0,100]$ in a  \texttt{Mathematica} package  \cite{Mathematica}.

We should mention that $g_0(\mu)$ is related to the Koster-Slater representation of the cubic lattice Green function, namely  $g_0(\mu)=\mathrm{Im}G_{11}(\mu)/\pi$, where $G_{11}(\mu)$ is the principal element of the Green function matrix $G_{ij}(\mu)$  as detailed in \cite{Wolfram_Callaway_1963}; and $h_0(z)=\mathrm{Re}G_{11}(z)$.

\subsection{Evaluation of spectral density function}
\label{sec:ev-sdf}

\begin{figure*}[htbp]
\centering
\subfigure[]{\label{fig:psi-zw1}
\includegraphics[width=0.45\textwidth]{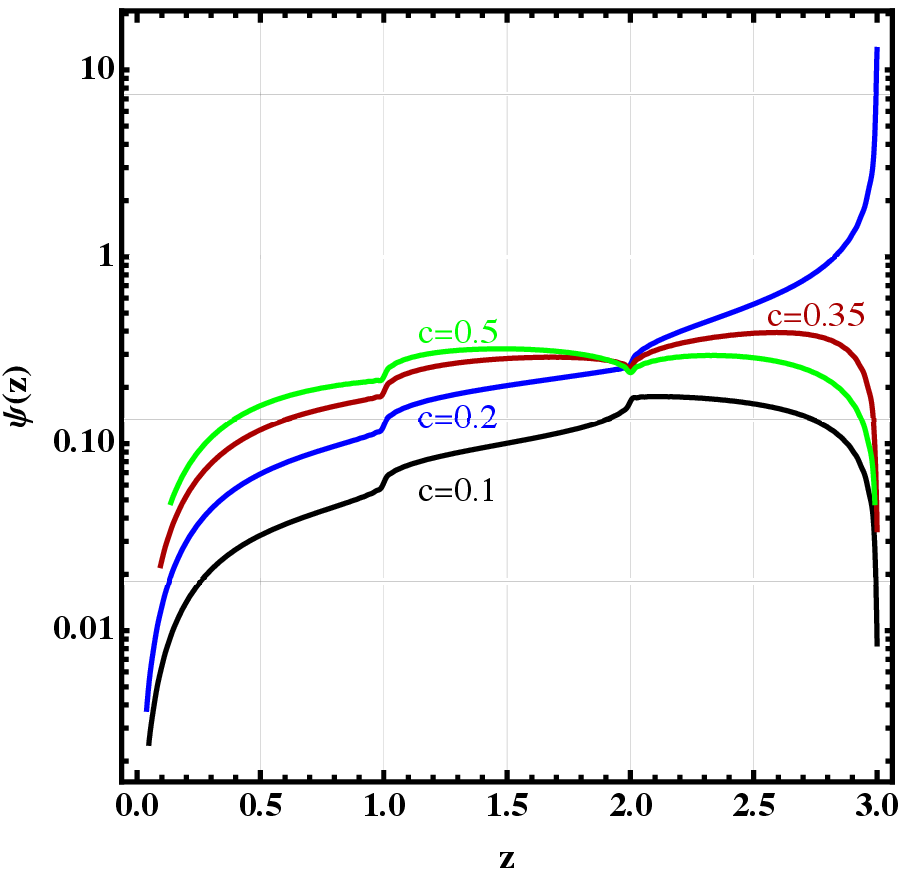}
}
\subfigure[]{\label{fig:zphi}
\includegraphics[width=0.425\textwidth]{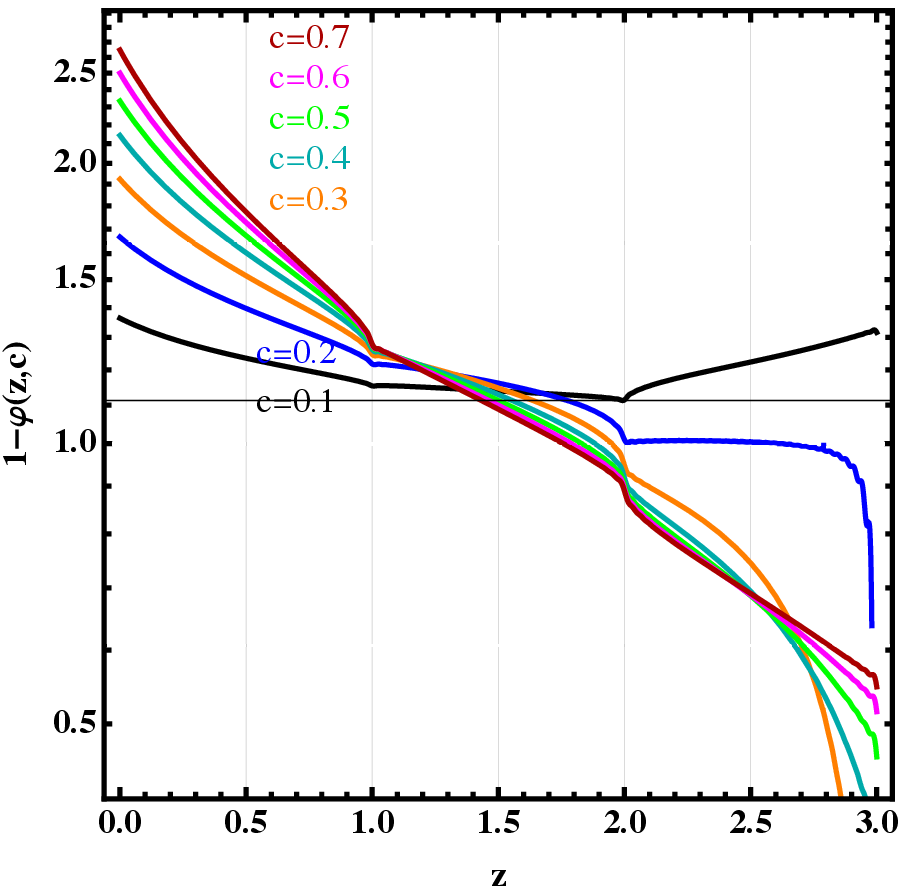}
}
\caption{(a) The imaginary part of the self-energy function  $\psi(z,c)$, Eq. \eqref{eqn:case-gen-psi},  as a function of the scaled frequency parameter $z=\omega/[\sigma(1-c)^2]$ for several values of the mean nonmagnetic ion concentration $c$. (b) $[1-\varphi(z,c)]$ versus $z$.}
\label{fig:psi-phi2}
\end{figure*}

We next calculate the spectral density function.  From Eqs. \eqref{eqn:case-gen-re}-\eqref{eqn:case-gen-im}, we first write
\begin{align}
\label{eqn:case-gen-re1}
\bar{\Sigma}^{\prime}(\mathbf{q},z) &= \bar\varepsilon_0\big[\mu_\mathbf{q}+(1-c)\text{\ss} \big]\varphi(z,c),\\
\label{eqn:case-gen-im1}
\bar{\Gamma}(\mathbf{q},z) &=  \bar\varepsilon_0\big[\mu_\mathbf{q}+(1-c)\text{\ss} \big]\psi(z,c).
\end{align}
\noindent
where we  rescaled and defined:
\begin{align}
\label{eqn:case-gen-phi}
\varphi(z,c) &= 2c\big[\mathrm{Re} f_1(z)-\mathrm{Re} f_2(z)\big],\\
\label{eqn:case-gen-psi}
\psi(z,c) &= 2c\big[\mathrm{Im} f_1(z)-\mathrm{Im} f_2(z)\big].
\end{align}
\noindent
Recall that  $\mathrm{Re} f_i(z)$ and  $\mathrm{Im} f_i(z)$ are per Eqs. \eqref{eqn:Re-fi}-\eqref{eqn:Im-fi}, expressed in terms of  $\Re \mathfrak{g}^0_i(\omega)$ and  $\Im \mathfrak{g}^0_i(\omega)$, which are related to the real and imaginary parts of the principal element of the Green function matrix $G_{ij}(\mu)$ as shown in the foregoing subsection. Figure \ref{fig:psi-zw1} shows semilog plots of $\psi(z,c)$ for several values of nonmagnetic ion concentration in the range of $c=0.1$ to $c=0.5$ and Fig. \ref{fig:zphi} shows the corresponding plots of $[1-\varphi(z,c)]$ including those for $c=0.6$ to $c=0.7$.  

We  express the spectral density function from Eq.  \eqref{eqn:sdf-2}, using Eqs. \eqref{eqn:case-gen-re1}-\eqref{eqn:case-gen-psi}, as 
\begin{equation}
\label{eqn:sdf-4}
 \tilde A(\mathbf{q},z,c) = \frac{2 E_\mathbf{q}\psi}{\big[z-(1-c)\text{\ss}-E_\mathbf{q}(1-\varphi) \big]^2+\big[E_\mathbf{q}\psi\big]^2},
\end{equation}
\noindent
where we put  $\tilde A=\sigma(1-c)^2\bar A$, $E_\mathbf{q} \equiv  [\mu_\mathbf{q}+(1-c)\text{\ss}]$, $\psi=\psi(z,c)$, and $\varphi=\varphi(z,c)$.   Now replacing  $\mu_\mathbf{q} \to \mu$ and setting $B=0$,  Eq. \eqref{eqn:sdf-4}  is reduced to
\begin{equation}
\label{eqn:sdf-b0}
 \tilde A(\mu,z,c) = \frac{2 \mu \psi}{\big[z-\mu(1-\varphi) \big]^2+\big[\mu\psi\big]^2}, \qquad B=0.
\end{equation}
\noindent

\begin{figure*}[htbp]
\centering
\subfigure[]{\label{fig:bessel-sdf1}
\includegraphics[width=0.45\textwidth]{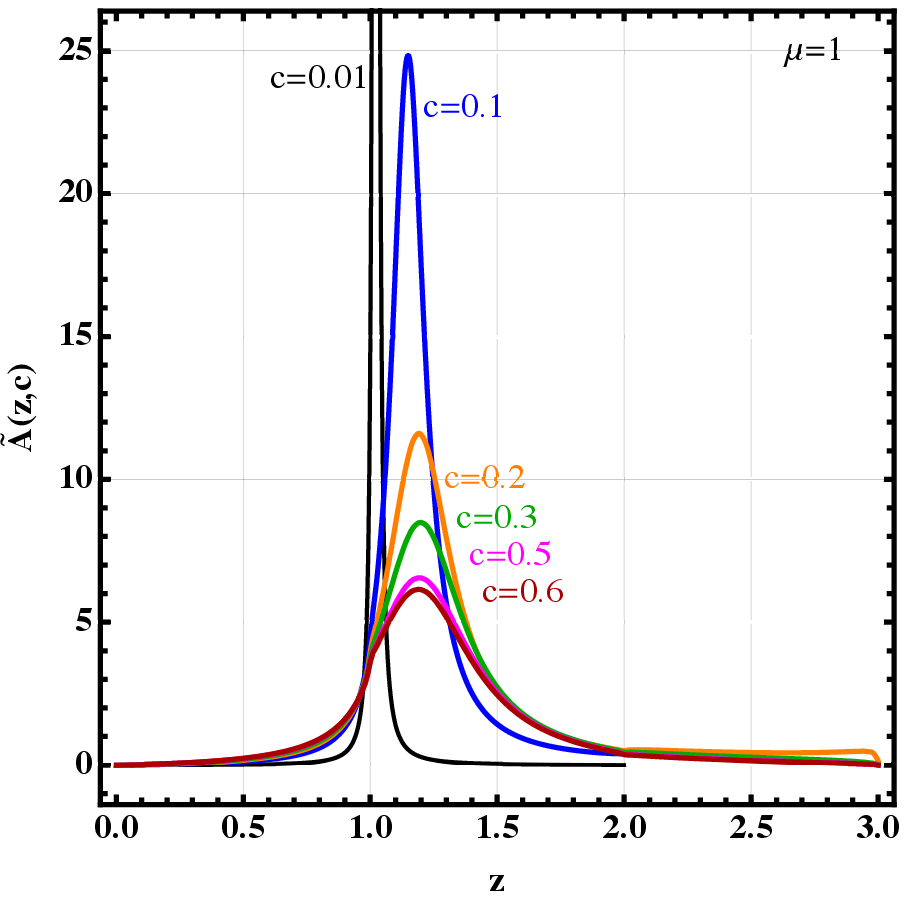}
}
\subfigure[]{\label{fig:bessel-sdf2}
\includegraphics[width=0.45\textwidth]{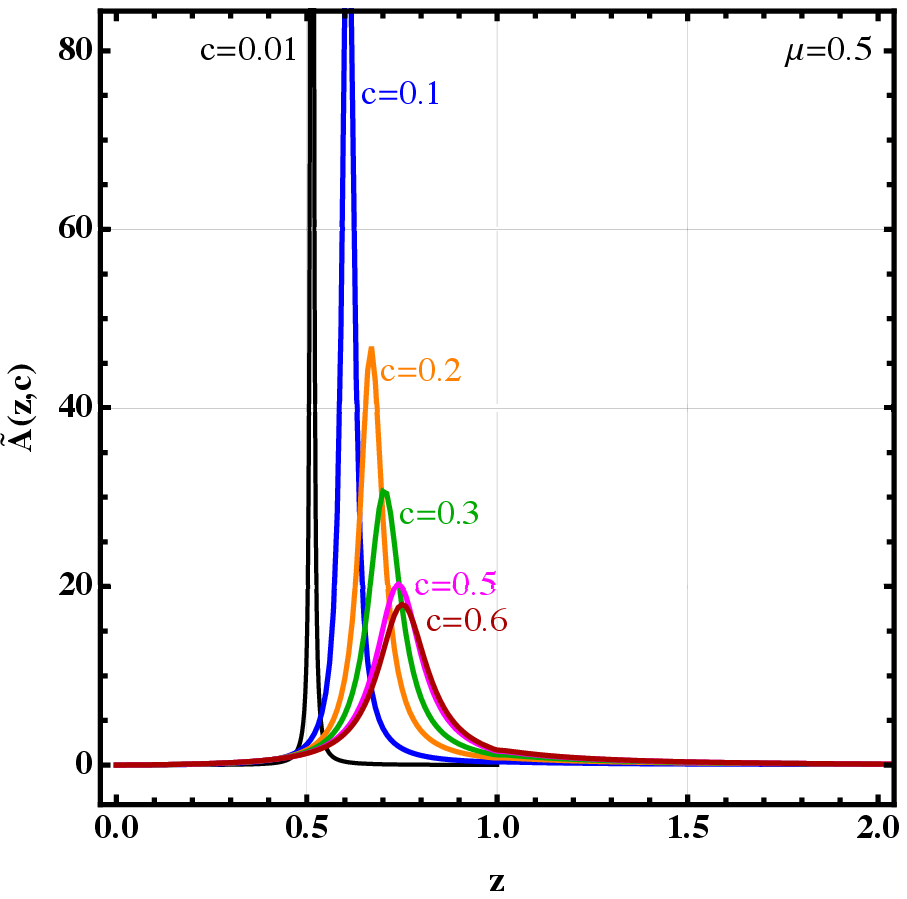}
}
\caption{The spectral density function  $\tilde A$  as a function of the scaled frequency parameter $z=\omega/[\sigma(1-c)^2]$ at $B=0$ for several values of the mean nonmagnetic ion concentration $c$  and energy parameter values (a) $\mu=1$ and  (b)  $\mu=0.5$.}
\label{fig:bessel-sdfi}
\end{figure*}

\begin{figure*}[htbp]
\centering
\includegraphics[width=0.75\textwidth]{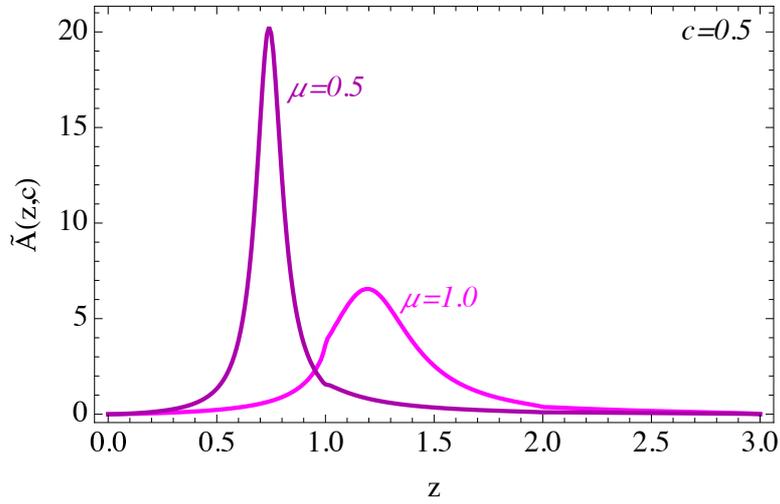}
\caption{Blow-up of two plots in figure \ref{fig:bessel-sdfi}  for $\mu=0.5$ and $\mu=1.0$  at $c=0.5$.}
\label{fig:bessel-sdfc-mu}
\end{figure*}

The dependence of $\tilde{A}$ on $z$ for several values of $c$ at zero external field and at two energy levels, $\mu=1$ and $\mu=0.5$, are calculated in the interval of $\Delta z=0.01$ by the \texttt{Mathematica} package \cite{Mathematica} and are displayed in Figs.  \ref{fig:bessel-sdf1} and \ref{fig:bessel-sdf2}, respectively. It is seen that as $c$ increases, the peaks of $\tilde{A}$ reduce height, shift to the right and  get broader.  Furthermore, at a smaller $\mu$, here $\mu=1/2$, the peaks occur at smaller $z$; cf. Fig. \ref{fig:bessel-sdfc-mu}.  As $c \to 0$, $\tilde{A}$ goes to a delta function, viz.   $\tilde A \to \delta(z-\mu)$. Note that, however, the shifts in the peaks against the rescaled  frequency, $\omega \propto z(1-c)^2$,  would tend to lower frequencies as $c$ increases. In Table  \ref{tab:sdf-vals} the values of $\tilde A(c,\mu,z)$, for $\mu=\{0.5,1\}$, in the vicinity of its maximum values, i.e. at positions $z_\mathrm{max}$, are tabulated with argument $c$  by using the \texttt{Mathematica} package algorithm \texttt{ArgMax[f,x]} \cite{Mathematica}.  

\begin{table}[!htb]
 \begin{center}
 \caption{Spectral density function $\tilde A(c,\mu,z)$  at its maximum values $z=z_\mathrm{max}$. }
\begin{tabular}[l]{|l|cc|cc|}
\hline
 $c$    & $z$ & $\tilde A(c,\mu,z)$ & $z$ & $\tilde A(c,\mu,z)$ \\
 \cline{2-5}
& \multicolumn{2}{c|} {$\mu=1.0$} & \multicolumn{2}{c|} {$\mu=0.5$}\\
\hline
 0.1 & 1.15705 & 24.5972 & 0.600269 & 86.8312 \\
0.2 & 1.2212 & 11.206 & 0.654712 & 40.6449 \\
0.3 & 1.2517 & 7.9349 & 0.685979 & 27.4599 \\
0.4 & 1.26873 & 6.55639 & 0.705497 & 21.6849 \\
 0.5 & 1.2794 & 5.82155 & 0.718604 & 18.4922 \\
 0.6 & 1.28663 & 5.36851 & 0.727929 & 16.5444 \\
 0.7 & 1.29184 & 5.05996 & 0.734866 & 15.2301 \\\hline
\end{tabular}
 \label{tab:sdf-vals}
 \end{center}
\end{table}

\subsection{Dispersion relation \& lifetime}
\label{sec:dispers}
\begin{figure*}[htbp]
\centering
\subfigure[]{\label{fig:tau-zw1}
\includegraphics[width=0.45\textwidth]{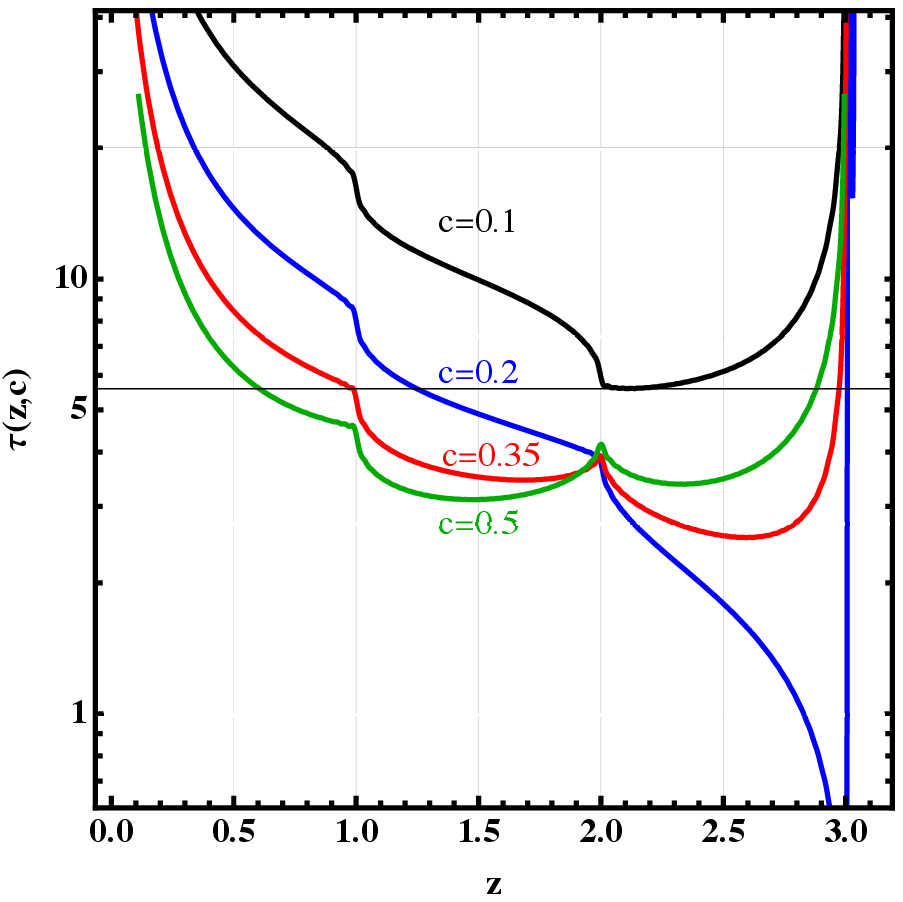}
}
\subfigure[]{\label{fig:tau-zw2}
\includegraphics[width=0.45\textwidth]{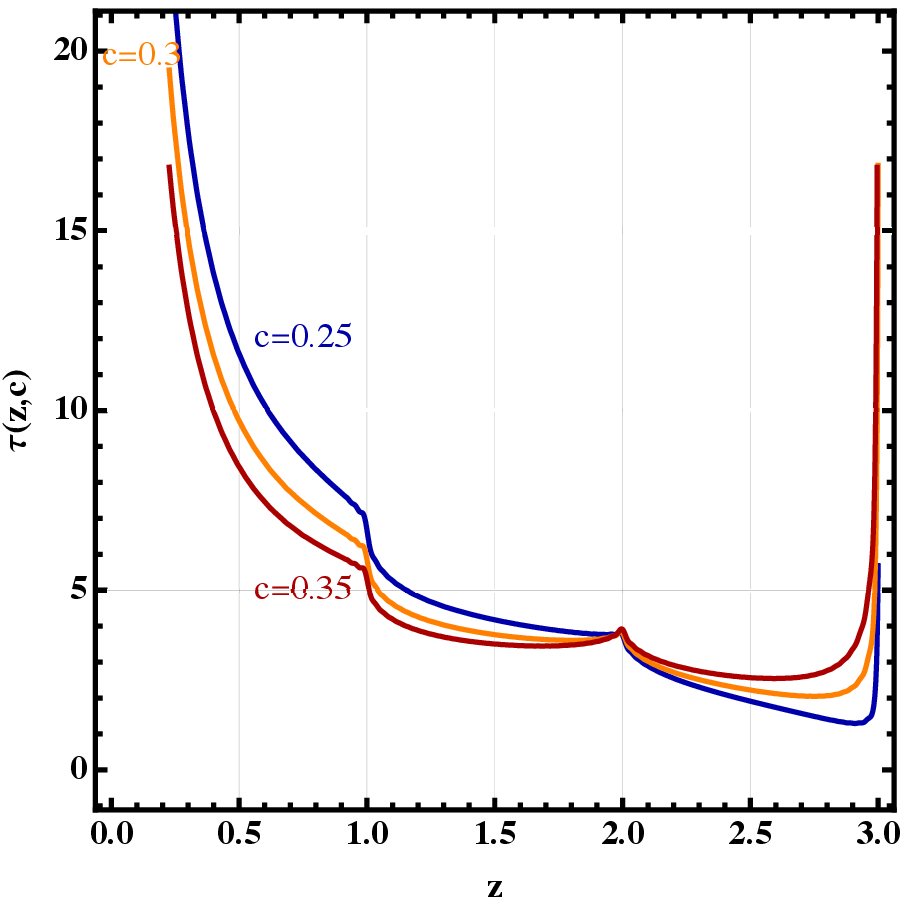}
}
\caption{(a) Semilog plots of magnon life-time $\tau$  ($B=0$ and $\mu=1/2$) as a function of the scaled frequency parameter $z$ for several values of the mean nonmagnetic ion concentration $c$. (b) A close-up of   $\tau$  near $c=0.3$.}
\label{fig:tau-zw}
\end{figure*}

In order to characterize the magnon energy and its lifetime as a function of impurity concentration, we  rewrite the spectral density function  \eqref{eqn:sdf-4} in a simple Lorentzian form
\begin{equation}
\label{eqn:sdf-5}
 \tilde{A}(\mu,z,c) = \frac{\lambda}{\big(z-\tilde E  \big)^2+(\lambda/2)^2},
\end{equation}
\noindent
where $\lambda=\lambda(z,c) =  2E\psi(z,c)$ is the magnon decay rate, $\tilde E \equiv  (1-c)\text{\ss}+E\big(1-\varphi(z,c)\big)$ is the renormalized energy, with $E=  \mu+(1-c)\text{\ss}$. The magnon lifetime  is defined as  the reciprocal of the decay rate $\tau \equiv \lambda^{-1}$. For $B=0$, $\lambda =  2\mu\psi$ and $\tilde E= \mu(1-\varphi)$. Note that, we have already plotted the decay rate (i.e. $\lambda/2\mu$ with $\mu=1/2$) in Fig. \ref{fig:psi-zw1};  see Figs. \ref{fig:tau-zw1}-\ref{fig:tau-zw2} for plots of  $\tau$.

\begin{figure*}[htbp]
\centering
\subfigure[]{\label{fig:eta-zw}
\includegraphics[width=0.425\textwidth]{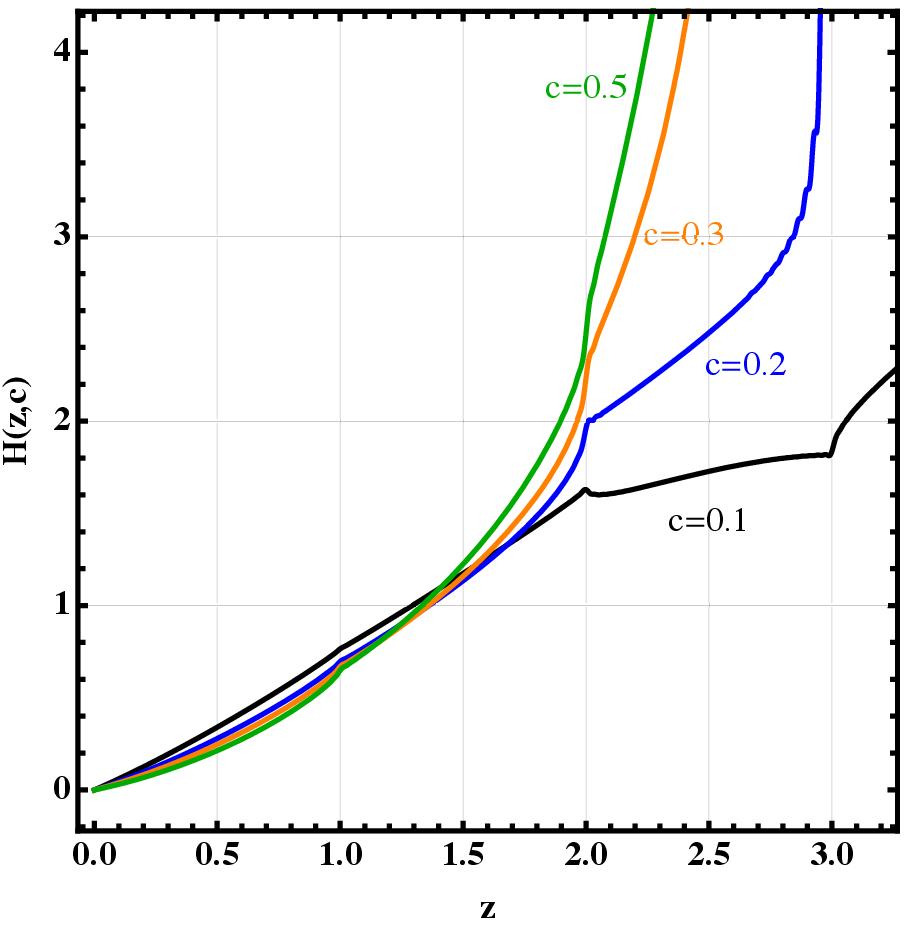}
}
\subfigure[]{\label{fig:eta-zw2}
\includegraphics[width=0.425\textwidth]{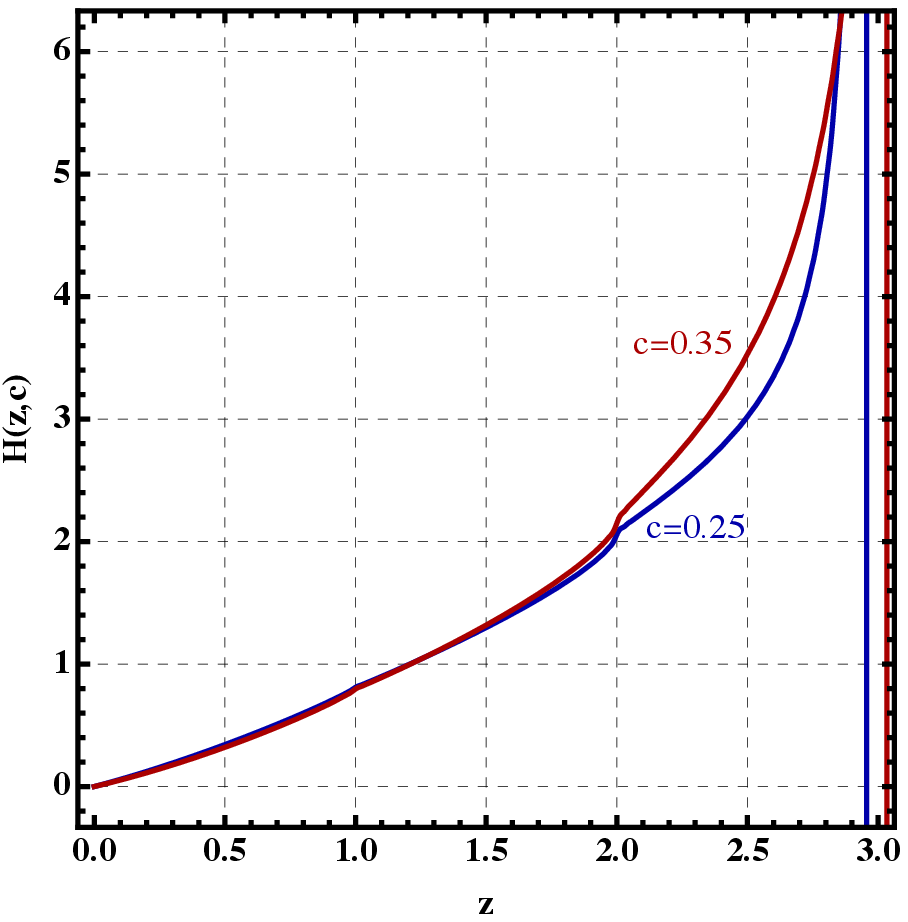}
}
\subfigure[]{\label{fig:eta-zw3}
\includegraphics[width=0.425\textwidth]{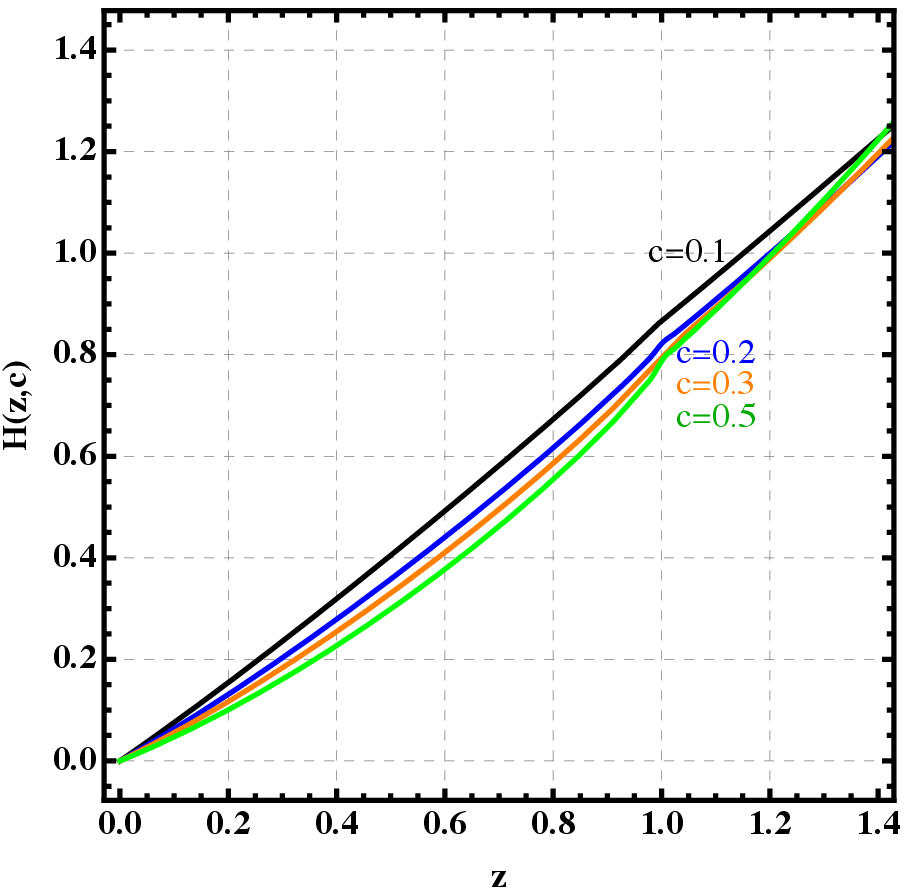}
}
\subfigure[]{\label{fig:muzcz}
\includegraphics[width=0.50\textwidth]{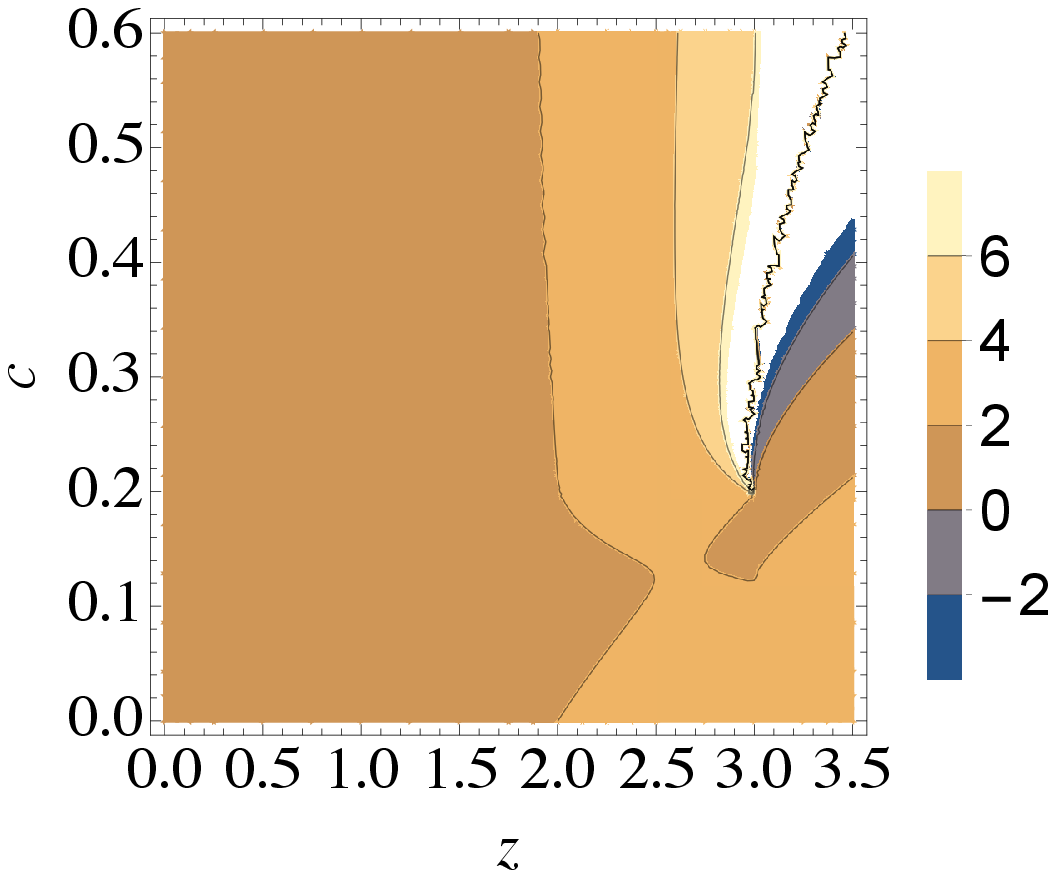}
}
\caption{(a)-(b) Plots of Eq. \eqref{eqn:pole-eq} for several values of the nonmagnetic impurity concentration $c$. (c) Blow-up of the plots in (a) in the region  $0 \le z \le 1.4$. (d) Contour plot of $\mathrm{H}(z,c)$ in the $zc$-plane.}
\label{fig:muvz}
\end{figure*}

 What's more, the equation $z=\mu[1-\varphi(z,c)]$ yields the poles $z \equiv z^\ast$  of the original Green function. Indeed, the excited energies of the spin waves as functions of the impurity concentration $c$ can be determined by solving the pole equation. We have used a Newton method to solve this pole equation for several levels of $\mu$ as a function of $c$.  The results of such calculations are summarized  in Table  \ref{tab:qp-energy}. For the sake of illustration, we have also plotted the ratio ($\mu \to \mathrm{H}(z,c)$)
\begin{equation}
\label{eqn:pole-eq}
\mathrm{H}(z,c)  = \frac{z}{1-\varphi(z,c)},
\end{equation}
\noindent
as a function $z$ in Figs.  \ref{fig:eta-zw}-\ref{fig:eta-zw2} for several values of nonmagnetic impurity concentration.  The crossings of a horizontal line drawn from the ordinate, corresponding to a constant $\mu$-value, with the plots in  Fig. \ref{fig:eta-zw3}  give the excitation energies  of the magnon.  As can be seen, anomalies appear in $\mathrm{H}(z,c)$ for $z\ge 1$,  and singularities occur for $c \gtrsim 0.35$ around $z=3$; cf. the contour plot in Fig. \ref{fig:muzcz}. Data in Table  \ref{tab:qp-energy} show that the excitation energies increase continuously by increasing $c$ for  $\mu\le1$. The anomalous or singular behavior of $\mathrm{H}(z,c)$  is a consequence of the van Hove singularities in the density of states $g(z)$ and its Hilbert transform $h(z)$, which appear in Eqs. \eqref{eqn:Re-base-green0-int-e0}-\eqref{eqn:Im-base-green0-int-e0}.

 We may also write Eq. \eqref {eqn:pole-eq} in terms of the wave vector $q$ for the sc lattice by replacements $\mathrm{H} \to \mu_q=2\sin^2(q/2)$, $z \to (1-c)^{-2}\sigma^{-1}\,\omega$ and $\omega \to \omega^\ast_q$ as
\begin{equation}
\label{eqn:dispersion-1}
\omega^\ast_q =2(1-c)^2\sigma\big[1-\varphi(\omega^\ast_q,c)\big]\sin^2(q/2),
\end{equation}
\noindent
being consistent with Eq. \eqref{eqn:sdf-2_sol1} for $B=0$. Eq.  \eqref{eqn:dispersion-1} serves as a renormalized dispersion relation for spin waves in a  ferromagnetic system where the spins occupy random positions on a simple cubic lattice  with nonmagnetic impurity concentration $c$.  Furthermore, in this setting, we can rewrite  Eq. \eqref{eqn:case-gen-im1} as
\begin{align}
\label{eqn:case-gen-im2}
\bar{\Gamma}^\ast_q &= 2(1-c)^2\sigma\,\psi(\omega^\ast_q,c) \sin^2(q/2).
\end{align}
\noindent
Magnon decay rate is $\lambda_q \sim \bar{\Gamma}^\ast_q$ and lifetime  $\tau_q =\lambda_q^{-1}$; cf. Eq. \eqref{eqn:sdf-2_sol2}. As can be  seen there is no energy gap in the dispersion relation at $q=0$ in the absence of external field. Both $\omega^\ast_q$ and $\bar{\Gamma}^\ast_q$, Eqs. \eqref{eqn:dispersion-1}-\eqref{eqn:case-gen-im2}, go to zero as $q \to 0$. This is the so-called gapless Goldstone mode, $\omega^\ast_q \sim (1-c)^2\sigma q^2$, which is a consequence of the broken rotation symmetry of the ferromagnetic ground state at nonzero temperature \cite{Auerbach_1994}. Hence in the absence of external magnetic field and anisotropy in the Heisenberg model, magnons act as Goldstone bosons in the infrared ($q \to 0$) or  at long wavelengths.

Let's check the wave-vector dependence of the decay rate $\lambda_q$ and also that of $\bar{\Sigma}^{\prime}(\mathbf{q},z\to \omega^\ast_q)$ from the present formalism. In the long-wave limit and zero external field ($B=0$), we write
\begin{align}
\label{eqn:case-gen-re1-inf}
\bar{\Sigma}^{\prime}(\mathbf{q},\omega^\ast_q) & \approx \frac{1}{2}(1-c)^2\sigma q^2 \varphi(\omega^\ast_q,c),\\
\label{eqn:case-gen-im1-inf}
\bar{\Gamma}(\mathbf{q},\omega^\ast_q) & \approx   \frac{1}{2}(1-c)^2\sigma q^2 \psi(\omega^\ast_q,c).
\end{align}
Because in the considered limit $\omega^\ast_q \to q^2$, calculations show that $\psi(\omega^\ast_q,c) \to c q^3$ and $\varphi(\omega^\ast_q,c) \to c (C_0+ q^2)$, with $C_0$ being a constant; see \ref{sec:appF}. Hence
\begin{align}
\label{eqn:case-gen-re-inf}
\bar{\Sigma}^{\prime}(\mathbf{q},\omega^\ast_q) & \sim c(1-c)^2\sigma q^2 +O(q^4),\\
\label{eqn:case-gen-im-inf}
\bar{\Gamma}(\mathbf{q},\omega^\ast_q) & \sim  c (1-c)^2\sigma q^5+O(q^7).
\end{align}
These formulae are similar to the results obtained in \cite{murray1966low,kaneyoshi1969contribution,edwards1971green} using different methods; cf. \cite{mano1982discrepancy,chakraborty2015long}.

\begin{table}[!htb]
  \begin{center}
    \caption{Magnon energy spectrum $z^\ast$ determined from  $z^\ast =\mu[1-\varphi(z^\ast,c)]$. }
    \begin{tabular}[l]{l|cccc}
      \hline
  $c$    & $\mu=0.25$ & $\mu=0.5$ & $\mu=0.75$ & $\mu=1.0$ \\
  \hline
 0.1 & 0.311653 & 0.600244 & 0.881053 & 1.15699 \\
 0.2 & 0.350984 & 0.65453 & 0.942079 & 1.22035 \\
 0.3 & 0.376776 & 0.685293 & 0.972854 & 1.24937 \\
 0.4 & 0.39437 & 0.70406 & 0.990206 & 1.26477 \\
 0.5 & 0.406863 & 0.716319 & 1.00094 & 1.27392 \\
 0.6 & 0.416065 & 0.72479 & 1.00806 & 1.27981 \\
 0.7 & 0.423059 & 0.730916 & 1.01305 & 1.28385 \\\hline
  \end{tabular}
  \label{tab:qp-energy}
  \end{center}
\end{table}

\subsection{Density of states \& thermodynamics}
\label{sec:dos-thermo}

We start by computing the number of magnon modes excited at  finite temperature. From Eqs. \eqref{eqn:sdf-case3} and \eqref{eqn:correl00_sdf} for bosons or from the fluctuation-dissipation theorem (\ref{sec:appA}), the mean occupation number in the $\mathbf{q}$-state ($\mathbf{q} \equiv \mathbf{k}a$) at temperature $\beta^{-1}$ is
\begin{equation}
\label{eqn:correl00-simp}
\langle n_\mathbf{q} \rangle \equiv \langle b^\dagger_\mathbf{q}b_\mathbf{q} \rangle = \frac{1}{2\pi}\int_{-\infty}^\infty  \frac{\bar A(\mathbf{q},\omega)}{e^{\beta\omega}-1}d\omega.
\end{equation}
\noindent
Integrating $\langle n_\mathbf{q} \rangle$ over all $\mathbf{q}$  in the first BZ, we obtain the mean occupation number at a lattice site
\begin{equation}
\label{eqn:correl00-mean}
\bar{n} = \int \frac{ d\mathbf{q}}{(2\pi)^3}\langle n_\mathbf{q} \rangle.
\end{equation}
Here, the integration limits are: $ -\pi \le q_i \le \pi; i=1,2,3$. Changing the variables from the reciprocal space $\mathbf{q}$ to the energy $\mu$-domain through the normal mode DOS, $g(\mu)$, we write $\bar{n}$ as
\begin{equation}  
\label{eqn:correl00-mean-energy}
\bar{n} = \int_0^{\mu_m} d\mu g(\mu) \langle n(\mu) \rangle.
\end{equation}
From  Eqs. \eqref{eqn:correl00-simp}--\eqref{eqn:correl00-mean-energy}, we express $\bar{n}$ in terms of $\tilde{A}(\mu,z,c)$ as
\begin{equation}
\label{eqn:correl00-mean-energy-spec}
\bar{n} (c,\vartheta) = \frac{1}{2\pi}\int_0^{\mu_m}d\mu\,g(\mu) \int_{-\infty}^\infty dz \frac{\tilde{A}(\mu,z,c)}{e^{z/\vartheta(c)}-1}.
\end{equation}
where $\vartheta(c)=T/[(1-c)^2\sigma]$. Substituting for $\tilde{A}(\mu,z,c)$ from Eq. \eqref{eqn:sdf-b0} for $B=0$, we write
\begin{equation}
\label{eqn:correl00-mean-energy-spec0}
\bar{n} (c,\vartheta) = \frac{1}{2\pi}\int_0^{\mu_m}d\mu\, g(\mu) \int_{-\infty}^\infty dz \frac{2\mu\psi(z,c)(e^{z/\vartheta}-1)^{-1}}{[z-\mu(1-\varphi(z,c))]^2+[\mu\psi(z,c)]^2}.
\end{equation}

In the limit $c \to 0$, i.e.  a negligible concentration of impurities, both $\psi$ and  $\varphi$  go to zero,  see Eqs.  \eqref{eqn:case-gen-phi}-\eqref{eqn:case-gen-psi}. So using the identity $\lim_{\epsilon \rightarrow 0} \frac{\epsilon}{x^2+\epsilon^2}=\pi\delta(x)$,  Eq.  \eqref{eqn:sdf-b0}  in the negligible $c$ limit becomes $\tilde{A}(\mu,\omega,c=0)=2\pi\delta(z-\mu)$, and  Eq. \eqref{eqn:correl00-mean-energy-spec0} yields the number of magnon modes excited at temperature $\vartheta(0)=T/\sigma$:
\begin{equation}
\label{eqn:correl00-lim}
\bar{n}\big\vert_{c=0} = \int_0^{\mu_m}d\mu\,\frac{g_0(\mu) }{e^{\mu/\vartheta(0)}-1}.
\end{equation}
The function  $g_0(\mu)$ is calculated (\ref{sec:appD}) by asymptotic expansion ($0 <\mu \le 1$), resulting in
\begin{equation}
\label{eqn:dos-ae}
g_0(\mu) =  \frac{2}{\pi^2}\mu^{1/2} +  \frac{1}{\pi^2}\mu^{3/2} +O(\mu^{5/2}).
\end{equation}
Substituting now for  $g_0(\mu)$  from Eq.  \eqref{eqn:dos-ae}  into Eq. \eqref{eqn:correl00-lim} and evaluating the integral by extending its upper limit to infinity, since we are interested in the region $\vartheta \ll \mu_m$; so we obtain
\begin{equation}
\label{eqn:correl00-lim-final}
\bar{n}(\vartheta)\big\vert_{c=0} = \frac{2}{\pi^2}\Gamma\Big(\frac{3}{2}\Big)\zeta\Big(\frac{3}{2};1\Big)\vartheta^{3/2}+\frac{1}{\pi^2}\Gamma\Big(\frac{5}{2}\Big)\zeta\Big(\frac{5}{2};1\Big)\vartheta^{5/2} + O(\vartheta^{7/2}).
\end{equation}
Here, $\vartheta \equiv T/\sigma$ for $c=0$, $\Gamma(x)$ is the Euler gamma function and $\zeta(s;a)$ is the generalized Riemann zeta function \cite{Whittaker_Watson_1927}. The DOS is further discussed in the subsequent section.

 The internal energy of the magnon gas in thermal equilibrium at temperature $\vartheta$ is given by
 \begin{equation}
\label{eqn:Umag-gen}
\mathcal{U}(c,\vartheta) = \int_0^{\mu_m}  d\mu \, \mu \, \bar n(c,\vartheta)
\end{equation}
Using Eq. \eqref{eqn:correl00-lim-final} and integrating with $\mu_m=\infty$, in the region $\vartheta \ll \mu_m$
\begin{equation}
\label{eqn:Umag}
\mathcal{U}(\vartheta)\vert_{c=0}   =  \frac{2}{\pi^2}\Gamma\Big(\frac{5}{2}\Big)\zeta\Big(\frac{5}{2};1\Big)\vartheta^{5/2}+\frac{1}{\pi^2}\Gamma\Big(\frac{7}{2}\Big)\zeta\Big(\frac{7}{2};1\Big)\vartheta^{7/2} + O(\vartheta^{9/2}).
\end{equation}
The heat capacity $\mathcal{C} =\partial \mathcal{U} /\partial \vartheta$ at $c=0$ is
\begin{equation}
\label{eqn:Cmag}
\mathcal{C}(\vartheta)\vert_{c=0}   = \frac{5}{\pi^2}\Gamma\Big(\frac{5}{2}\Big)\zeta\Big(\frac{5}{2};1\Big)\vartheta^{3/2}+\frac{7}{2\pi^2}\Gamma\Big(\frac{7}{2}\Big)\zeta\Big(\frac{7}{2};1\Big)\vartheta^{5/2} + O(\vartheta^{7/2}).
\end{equation}
We note:  $\Gamma(3/2)=\sqrt{\pi}/2$, $\Gamma(5/2)=3\sqrt{\pi}/4$,  $\Gamma(7/2)=15\sqrt{\pi}/8$ and  $\zeta(3/2)=2.612375348$, $\zeta(5/2)=1.341487257$, $\zeta(7/2)=1.1267338673$, with $\zeta(s) \equiv \zeta(s;1)$.

The magnetization in the region $\vartheta \ll \mu_m$ can be computed directly from Eq.  \eqref{eqn:correl00-lim-final}. To do that, consider a ferromagnetic state at zero temperature, where all spins are up. Then introduce a number of magnon $\bar n$ into the the system by raising its temperature. The total magnetization is 
\begin{equation}
\label{eqn:totmag}
M = M_0-\bar n(\vartheta) = NS-\bar n(\vartheta),
\end{equation}
where $M_0=NS$ represents the saturation magnetization of $N$ spins of length $S$, and the magnons are present due to thermal fluctuations at temperature $\vartheta$. Using Eq.  \eqref{eqn:correl00-lim-final}, the magnetization for $c=0$, in units of $M_0$,  is
\begin{equation}
\label{eqn:magneti}
\mathcal{M}(\vartheta)\big\vert_{c=0} =1- 2\mathfrak{B}_{3/2}\vartheta^{3/2}-\mathfrak{B}_{5/2}\vartheta^{5/2} -\mathfrak{B}_{7/2}\vartheta^{7/2}- O\big(\vartheta^{9/2}\big),
\end{equation}
with $\mathcal{M} \equiv M/M_0$ and  $\mathfrak{B}_{n/2} \equiv (1/\pi^2)\Gamma\big(n/2)\zeta\big(n/2\big)$, which is equivalent to the result in \cite{Dyson_1956b} at low temperaures. In order to compute the foregoing thermodynamic quantities for nonzero $c$, one may evaluate the integrals in Eq. \eqref{eqn:correl00-mean-energy-spec0}  numerically for a given $g(\mu)$ and plot the results; however, this is of little theoretical interest.

 In order to see the effect of external magnetic field  on thermodynamic quantities, we can appeal to Eq. \eqref{eqn:helm-tot}, i.e.  the total Helmholtz free energy per ion, which we write as
\begin{equation}\begin{split}
\frac{F}{N} & = -\big[(1-c)SB+\frac{1}{2}(1-c)^2S^2J\big] + \frac{\varv}{\beta}\int \frac{d^3k}{(2\pi)^3} \log\Big[1-e^{-\beta \big[2(1-c)B+(1-c)^2\varepsilon_\mathbf{k}\big]}\Big] \\
 & - 4c(1-c)^{-1} \varv^2 \int \frac{d^3k}{(2\pi)^3}  \int \frac{d^3k_1}{(2\pi)^3}\,\frac{[B+(1-c)\varepsilon_{\mathbf{k}}] [B+(1-c)\varepsilon_{\mathbf{k}_1}] }{\varepsilon_\mathbf{k}-\varepsilon_\mathbf{k_1}}n_\mathbf{k},
 \label{eqn:helm-tot-1st}
\end{split}\end{equation}
where  only the second-order term ($n=2$) contribution to the interacting free energy $\langle  F_{\mathrm{I}}^{(2)} \rangle_c$, from Eq.  \eqref{eqn:helmI-mf2-1},  was considered, $\varv=V/N$, $n_\mathbf{k}$ is given by Eq. \eqref{eqn:amn},  and we used $\gamma_{\mathbf{k}}=2[B+(1-c)\varepsilon_{\mathbf{k}}]$ with $\varepsilon_{\mathbf{k}}=\sigma[1-\cos(ka)]$. The second term on the right-hand side of  Eq. \eqref{eqn:helm-tot-1st} is the Bloch free energy of ideal Bose gas of magnons at temperature $\beta^{-1}$, viz.
\begin{equation}
\label{eqn:bloch0}
F_\mathrm{bloch} \equiv   \frac{\varv}{\beta}\int \frac{d^3k}{(2\pi)^3} \log\Big[1-e^{-\beta \big[2(1-c)B+(1-c)^2\varepsilon_\mathbf{k}\big]}\Big].\nonumber
\end{equation}
 For $\varepsilon_{\mathbf{k}}\approx(\sigma/2)(ka)^2$ upon  integration, we get
\begin{equation}
\label{eqn:bloch}
F_\mathrm{bloch} \approx  -\sqrt{2}\pi^{-2}(\vartheta(c))^{3/2}\beta^{-1}\frac{\sqrt\pi}{4}\mathrm{Li}_{\frac{5}{2}}(e^{-\beta L(c)}),
\end{equation}
where  $L(c)=2(1-c)B$  and $\mathrm{Li}_s(z)$ is the polylogarithm \cite{NIST_Math_Handbook}, defined as
\begin{equation}
\label{eqn:bloch1}
\mathrm{Li}_{s}(z) = \sum_{n=1}^\infty \frac{z^n}{n^s}.
\end{equation}
If we expand the difference between  $\varepsilon_{\mathbf{k}}$ and its quadratic approximation in powers of $(ka)$ as in \cite{Dyson_1956b} in the Bloch energy integrand, and then integrate term by term, we obtain Dyson's well-known formula for low-temperature expansion,  viz.
\begin{equation}
\label{eqn:bloch-dyson}
F_\mathrm{bloch} =  -\sqrt{2}\pi^{-2}\vartheta^{3/2} \beta^{-1} \frac{\sqrt\pi}{4}\Big[\mathrm{Li}_{\frac{5}{2}}(e^{-\beta L})+\lambda_1 \vartheta \mathrm{Li}_{\frac{7}{2}}(e^{-\beta L})+\lambda_2 \vartheta^2\mathrm{Li}_{\frac{9}{2}}(e^{-\beta L})+O(\vartheta^3)\Big],
\end{equation}
with $\lambda_1=3/4$ and $\lambda_2=33/32$ for the sc lattice; cf. \cite{Vaks1968thermodynamics}. Recall that in Eq. \eqref{eqn:bloch-dyson} both $L$ and $\vartheta$ are $c$-dependent as designated above and the special case of polylog is $\mathrm{Li}_s(1)=\zeta(s)$.

The third term on the right hand-side of  Eq. \eqref{eqn:helm-tot-1st} which accounts for the magnon nonmagnetic ion interaction  to second order may be written in a simplified form as 
\begin{equation}
\label{eqn:mag-nonmag-inter}
F_I^{(2)} =  -2c(1-c)^{-1} \varv \int \frac{d^3k}{(2\pi)^3}  \big[B+(1-c)\varepsilon_{\mathbf{k}}\big]  \mathcal{G} (\mathbf{k}) n_\mathbf{k},
\end{equation}
where $n_\mathbf{k}$ is given by Eq.  \eqref{eqn:amn} and we introduced an interaction Green function defined as
\begin{equation}
\label{eqn:inter-green}
\mathcal{G} (\mathbf{k}) =  2\varv \int \frac{d^3k_1}{(2\pi)^3} \Bigg[\frac{B+(1-c)\varepsilon_{\mathbf{k}_1} } { \varepsilon_\mathbf{k}-\varepsilon_{\mathbf{k}_1}}\Bigg].
\end{equation}
In the absence of external field ($B=0$), we write
\begin{align}
\label{eqn:magnonmag-inter-c}
F_I^{(2)}\Big|_{B=0}  &=  - 2c \varv \int \frac{d^3k}{(2\pi)^3}  \varepsilon_{\mathbf{k}}  \mathcal{G} (\mathbf{k}) n(\mathbf{k}),\\
\label{eqn:inter-green-c}
\mathcal{G} (\mathbf{k})\Big|_{B=0}  &= 2(1-c) \varv \int \frac{d^3k_1}{(2\pi)^3} \Bigg(\frac{\varepsilon_{\mathbf{k}_1} } { \varepsilon_\mathbf{k}-\varepsilon_{\mathbf{k}_1}}\Bigg),\\
n(\mathbf{k})\Big|_{B=0} &=\frac{1}{\exp[\beta(1-c)^2\varepsilon_{\mathbf{k}}]-1},
\end{align}
with $\varepsilon_{\mathbf{k}}=2\sigma\sin^2(ka/2)$ for the simple cubic lattice with a lattice constant $a$. The magnon-magnon interaction term is not included in the expression for the total free energy because at low temperatures  its contribution is negligible; see e.g.  \cite{Dyson_1956a,Dyson_1956b,Vaks1968thermodynamics,Vaks1968spinwaves}.

\section{Discussion}
\label{sec:discuss}
 The  characteristics of unperturbed density of states $g_0(\mu)$ for pure Heisenberg ferromagnet in the simple cubic lattice shown in Fig. \ref{fig:dos-zw} is well known; see e.g.  \cite{Hone_et_al_1966,Mattis_ii_1985}. The plot of  $g_0(\mu)$ given by Eq. \eqref{eqn:dos-scc-cal}  in the domain  $0 \le \mu \le 3$, Fig.  \ref{fig:dos-zw}, displays the  van Hove singularities (for $c=0$) at $\mu=\{1,2\}$, and at its minima $\mu=\{0,3\}$, where $dg(\mu)/d\mu=\infty$. These singularities  can be understood from Eq. \eqref{eqn:dos-gen} of \ref{sec:appD}; see  e.g. \cite{Ashcroft_Mermin_1976}. Perhaps  less well known  is  the  behavior of the Hilbert transform of $g_0(\mu)$, i.e.  $h_0(z)$, as calculated by Eq. \eqref{eqn:hdos-scc-cal},  also shown in  Fig. \ref{fig:dos-zw},  which exhibits the van Hove of singularities at $z={0,1,2,3}$. This plot  shows that  $h(z)$ tends slowly to zero as  $z$ increases beyond the value 3.
 
 As we alluded in Sec. \ref{sec:dispers}, the real and imaginary parts of  the base impurity-averaged  Green function  given by Eqs. \eqref{eqn:Re-base-green0-int-e0} and \eqref{eqn:Im-base-green0-int-e0} for the case of the sc lattice exhibit the van Hove singularities at  certain frequencies,  Fig. \ref{fig:psi-zw2i}. Evidently, the singularities show up in all functions or quantities that are related to $g_0(\mu)$ and  $h_0(z)$, which also include the mean nonmagnetic ion concentration $c$;  see Figs. \ref{fig:tau-zw}--\ref{fig:muvz}.

In general the DOS for our system can be expressed in terms of the spectral density function: 
\begin{equation}
\label{eqn:dos-sdf-1}
g(\omega) = \frac{1}{2\pi}\sum_{\mathbf{q}} \bar A(\mathbf{q},\omega,c) = \frac{1}{2\pi}\int \frac{d^3q}{(2\pi)^3} \bar A(\mathbf{q},\omega,c)
\end{equation}
\noindent
where $\bar A(\mathbf{q},z,c) $ is given by Eq. \eqref{eqn:sdf-4} . By setting $B=0$, we can write Eq. \eqref{eqn:dos-sdf-1} as
\begin{equation}
\label{eqn:dos-sdf-2}
g(z) = \frac{1}{\pi\sigma(1-c)^2}\int \frac{d^3q}{(2\pi)^3}  \frac{ \mu_\mathbf{q}\psi(z,c)}{\big[z-\mu_\mathbf{q}\big(1-\varphi(z,c)\big) \big]^2+\big[\mu_\mathbf{q}\psi(z,c)\big]^2}.
\end{equation}
\noindent
Since both  $\psi(z,c)$, and $\varphi(z,c)$ tend to zero as $c\to 0$ and $z \to \omega/\sigma$, in the limit of zero impurity concentration, we have   
\begin{equation}
\label{eqn:dos-sdf-0}
g(z)\Big|_{c \to 0} = g_0(\omega) =  \frac{1}{\sigma}\int \frac{d^3q}{(2\pi)^3}   \delta \big(\omega/\sigma -\mu_\mathbf{q}\big),
\end{equation}
which is a standard result.  Evaluating now the integral in Eq. \eqref{eqn:dos-sdf-0} by putting $\sigma=2JSZ$ and taking the shape of the Brillouin zone as a sphere, in the long-wavelength limit where $\mu_\mathbf{q} \sim q^2/2$, we obtain the well-known formula \cite{izyumov1966spin,edwards1971green}
\begin{equation}
\label{eqn:dos-sdf-00}
 g_0(\omega) =  \frac{1}{4\pi^2 (JSZ)^{3/2}} \omega^{1/2}.
\end{equation}

The physical quantities of interest evaluated  in the foregoing section for magnons  were expressed in terms $g_0(\omega)$ and its Hilbert transform $h_0(\omega)$ or the primary diagonal element of the perfect cubic lattice Green function matrix (imaginary and real parts). In more detail, the cubic lattice Green functions  emanate from the point symmetry of the crystal,  as described in  \cite{Wolfram_Callaway_1963} though for a single impurity ferromagnet. The lattice Dyson equation  in a matrix form  is $\mathbf{G}=\mathbf{G}^0+\mathbf{G}^0 \mathbf{V} \mathbf{G}$, where $\mathbf{G}$ is the Green function perturbed by impurity, $\mathbf{G}^0$ the Green function of a perfect lattice, and $\mathbf{V}$ is  the perturbation matrix due to impurity,   cf. Eq. \eqref{eqn:dyson-eq}. Solving this  equation,  by using the full cubic symmetry of the lattice, one can show that the  Green function at sites $n m$, $G_{nm}(\omega)$, breaks up into a sum of the contributions corresponding to irreducible representations of the point group of the lattice \cite{Wolfram_Callaway_1963,izyumov1967incoherent,izyumov1973magnetically}. For the sc crystal one may write
\begin{equation}
\label{eqn:green-sc-split}
 G_{nm} =  G_{nm}^0+G_{nm}^{(s)} + G_{nm}^{(p)} + G_{nm}^{(d)},
\end{equation}
where $G_{nm}^0$ is the Green function of a perfect crystal and superscripts $s, p, d$ denote the irreducible representations of the cubic point group $O_h$, viz.  $\Gamma_1, \Gamma_{15}, \Gamma_{12}$; see e.g.  \cite{izyumov1973magnetically,kittel1987quantum,dresselhaus2007group}. For the case of s-wave symmetry in a more refined notation \cite{izyumov1967incoherent,izyumov1973magnetically}:
\begin{eqnarray}
\label{eqn:green-swave}
 G_{nm}^{(s)} (\omega) &=& \frac{\omega}{1-\omega G_{11}^0(\omega)}G_{n1}^0(\omega)G_{m1}^0(\omega), \qquad n \ne 1,  \quad m \ne 1,\\
 G_{nm}^{0} (\omega) &=& \frac{1}{N}\sum_\mathbf{k} \frac{\exp[-i \mathbf{k}\cdot(\mathbf{r}_n-\mathbf{r}_m)]}{\omega-\varepsilon_\mathbf{k}-i0^+},
\end{eqnarray}
where $\varepsilon_\mathbf{k}$ is the spin-wave energy in the perfect lattice. The expressions for $G_{nm}^{(p)}$ and  $G_{nm}^{(d)}$ are given in \cite{izyumov1966theory,izyumov1973magnetically}, where they are expressed in terms of $G_{nm}^{0}$ over the nearest-neighbor sites.

In the same manner, the density of states of a ferromagnet containing an impurity is the sum of the contributions corresponding to the irreducible representations of s-, p-, and d-waves \cite{izyumov1966spin,izyumov1973magnetically}, viz.
\begin{equation}
\label{eqn:dos-sc-split}
 g(\omega) = g_0(\omega)+\frac{1}{N\pi} \Im \Big[ \frac{d\ln D_s(\omega)}{d\omega}+3 \frac{d\ln D_p(\omega)}{d\omega}+2 \frac{d\ln D_d(\omega)}{d\omega}\Big],
\end{equation}
where the respective DOS components for the nonmagnetic impurity are
\begin{align}
\label{eqn:dos-sc-s}
D_s(\omega) &= 1-\omega G_{11}^0(\omega),\\
D_p(\omega) &= 1 +\sigma \Big(G_{11}^0(\omega)- G_{23}^0(\omega)\Big),\\
D_d(\omega) &= 1 +\sigma \Big( G_{11}^0(\omega)+ G_{23}^0(\omega)-2G_{24}^0(\omega)\Big),
\end{align}
with $\Im[\bullet] \equiv \mathrm{Im}[\bullet]$ and $\sigma \equiv 2SJZ$ as before. As can be seen the necessary Green functions for calculation of  the DOS in the simple cubic lattice  are  $G_{11}^0, G_{23}^0$ and $G_{24}^0$,   which can explicitly be expressed in terms of the Bessel functions \cite{Wolfram_Callaway_1963}:
\begin{subequations}
\begin{eqnarray}
\label{eqn:green-s}
G_{11}^0(\omega) &=&  i \int_0^\infty dt e^{-i\omega t} J_0^3(t),\\
\label{eqn:green-p}
 G_{23}^0(\omega) &=& -i \int_0^\infty dt e^{-i\omega t}J_2(t) J_0^2(t),\\
\label{eqn:green-d}
G_{34}^0(\omega) &=& -i \int_0^\infty dt e^{-i\omega t}J_1^2(t) J_0(t),
\end{eqnarray}
\label{eqn:green-spd}
\end{subequations}
We should note that in the presence of many impurities in the crystal, $1/N$ in Eq. \eqref{eqn:dos-sc-split} shall be replaced by the concentration of impurities  $c$ as in \cite{izyumov1966spin,izyumov1973magnetically}. Our formalism naturally accommodates these extra terms in the DOS, but  that formulation requires much extended numerical computations than those presented in the foregoing section.

It is worth here to discuss the results of the DOS computation made by other authors using other analytical approaches.  For example, Edwards and Jones \cite{edwards1971green} using the Born series approach to leading order for the sc lattice containing nonmagnetic impurities near  the bottom of the energy band  computed
\begin{equation}
\label{eqn:dos-ej-1}
 g(\omega) = \Big[1-c\frac{1+I/2}{1-I/2} \Big]^{-3/2} g_0(\omega),
\end{equation}
where $I=0.42$ is an integration constant. Taylor expanding the square bracket in Eq. \eqref{eqn:dos-ej-1}, 
\begin{equation}
\label{eqn:dos-izme-1}
 g(\omega) = \Big[1+\frac{3}{2}c\frac{1+I/2}{1-I/2} \Big] g_0(\omega) + O(c^2),
\end{equation}
 which is the result obtained earlier by  Izyumov and Medvedev \cite{izyumov1966peculiarities,izyumov1966spin} using a perturbative scheme near the bottom of the energy band. In view of the aforementioned  contributions  of s- p- and d-states to the  Izyumov-Medvedev formula \eqref{eqn:dos-izme-1}, these are: $g^{(s)}(\omega)=-(3/2)cg_0(\omega)$, $g^{(p)}(\omega)=3c(1-I/2)^{-1} g_0(\omega)$, and $g^{(d)}(\omega)\approx 0$; cf. Eq. \eqref{eqn:dos-sc-split}.
 
 Edwards and Jones \cite{edwards1971green}  include also an interference scattering term in their impurity averaged Green function, represented by a series of diagrams,   which may have some effect on the density of states, but it would not affect the positions of the poles of $\bar{G}(\mathbf{k},\omega)$, i.e. the dispersion relation for the system. Indeed, our computations (not presented here) using the Edwards-Jones low frequency, low impurity concentration formula for DOS indicate that in the range $0<\omega/2JS<0.5$ and $0<c<0.2$, the contribution of the interference diagrams is negligible.
 
Regarding the dispersion relation for the three modes (s-, p-, d-waves) of the sc lattice, corresponding to the poles of the impurity averaged  Green function, Jones \cite{jones1971impurity} has derived a relation to linear order in $c$ in the form
\begin{equation}
\label{eqn:disper-sc-split}
 \omega(\mathbf{k}) = (1-c)\omega_\mathbf{k}^0 + c  \Big[ \Delta_s(\mathbf{k},\omega)+ \Delta_p(\mathbf{k},\omega)+ \Delta_d(\mathbf{k},\omega)\Big],
\end{equation}
 where $\omega_\mathbf{k}^0$ is the perfect lattice dispersion law  and $ \Delta_\alpha(\mathbf{k},\omega), \alpha = (s,p,d)$ are given in \cite{jones1971impurity}. By Taylor expanding the right-hand side of Eq. \eqref{eqn:disper-sc-split} to lowest order in $k$, Jones found
 \begin{equation}
\label{eqn:disper-sc-split-taylor}
 \omega(\mathbf{k}) = 2SJ\Bigg[1-c\frac{1+I/2}{1-I/2}\Bigg] (ak)^2- i \frac{SJ}{2\pi} \Bigg[1+\frac{4}{3(1-I/2)^2}\Bigg] (ak)^5
\end{equation}
Comparing our calculations by recalling Eq. \eqref{eqn:sdf-2_poles-1} and rewriting it in a new setting
\begin{equation}
\label{eqn:disp_poles-1}
 \omega(\mathbf{k}) = (1-c)^2\omega_\mathbf{k}^0 - \bar{\Sigma}^\prime(\mathbf{k},\omega^\ast_k) - i \, \bar{\Gamma}(\mathbf{k},\omega^\ast_k),
\end{equation}
\noindent
and then inserting  Eqs. \eqref{eqn:case-gen-re-inf} and \eqref{eqn:case-gen-im-inf}, in the long-wavelength limit, we obtain
\begin{equation}
\label{eqn:disp_poles-lwl}
 \omega(\mathbf{k}) \sim  (1-c)^3 SJ (ak)^2- ic(1-c)^2SJ(ak)^5.
\end{equation}
 
\begin{figure*}[htbp]
\centering
\subfigure[]{\label{fig:qp-am}
\includegraphics[width=0.45\textwidth]{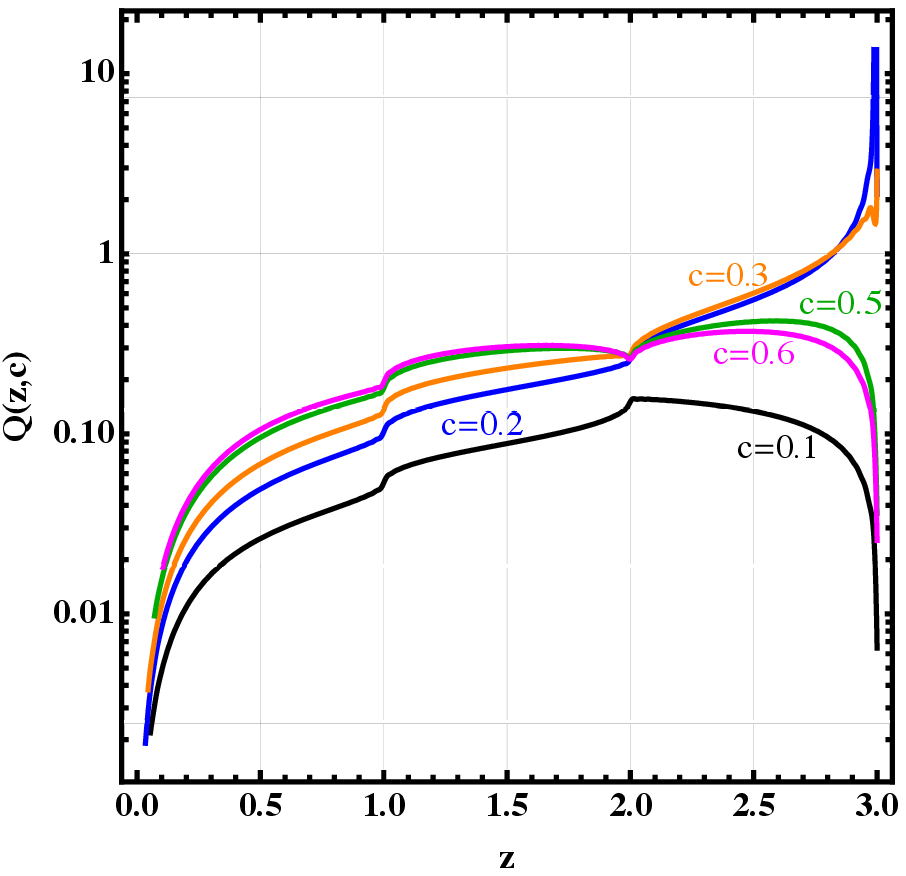}
}
\subfigure[]{\label{fig:qpcz}
\includegraphics[width=0.45\textwidth]{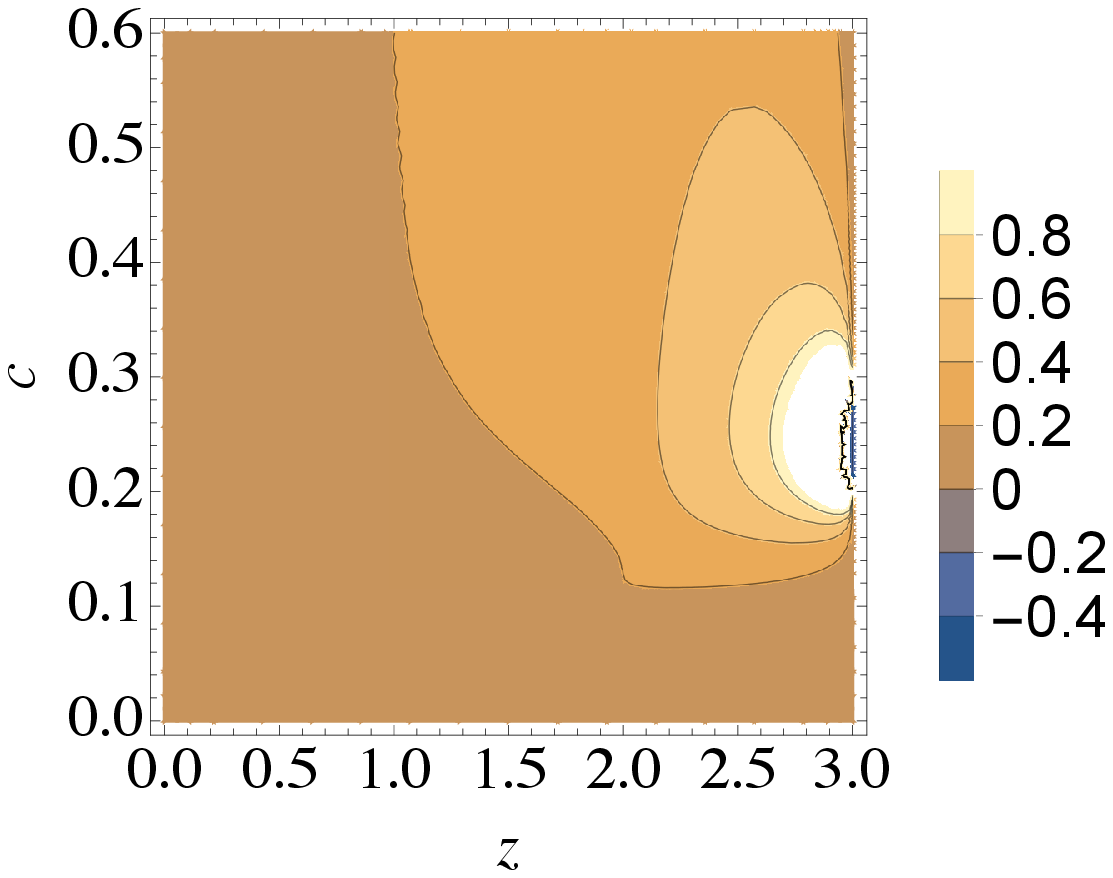}
}
\caption{The ratio $Q(z,c)=\psi(z,c)/\big[1-\varphi(z,c)\big]$  as  a function of $z$ for several values of the nonmagnetic ion concentration $c$. If $Q(z,c) \ll 1$, true  quasiparticles can be identified.}
\label{fig:quasiparts}
\end{figure*}

\begin{figure*}[htbp]
\centering
\includegraphics[width=0.5\textwidth]{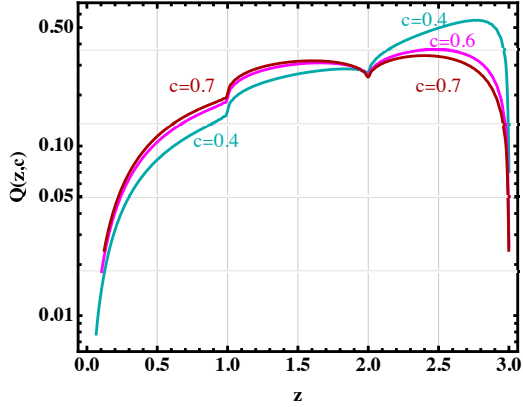}
\caption{Blow-up of the  plot for $c=0.6$ in figure \ref{fig:qp-am}  compared with those of  $c=0.4$  and $c=0.7$.}
\label{fig:qp-am2}
\end{figure*}

Now  to check the existence of magnon as a quasiparticle, we recall  Eq.  \eqref{eqn:case33} as the condition for this attribute; which states that the ratio $\bar\Gamma_q/\omega_q$ should be less than unity. From Eqs. \eqref{eqn:dispersion-1} and \eqref{eqn:case-gen-im2}, we can write this ratio as
\begin{equation}
\label{eqn:qp-rat}
Q(z,c) \equiv \frac{\bar\Gamma_q}{\omega_q}  =  \frac{\psi(z,c)}{1-\varphi(z,c)}
\end{equation}
\noindent
One may compute $Q(z,c)$ as a function $z$ and $c$ to locate  the region that this ratio is much less than unity, which means that the energy level width is much less than the energy \cite{mattuck1964lifetime,mattuck1992guide}. The merit of using this ratio is its independence from the energy parameter $\mu$, hence it can serve as a criterium for the existence of  quasiparticles. Here, we plot the function $Q(z,c)$ as a function of $z$ for several values of $c$   in Figs. \ref{fig:quasiparts}-\ref{fig:qp-am2} in order to be compatible with the foregoing  plots. It is seen that $Q(z,c)$ is sufficiently small for $z \le 2$, however in the region $ 2.5< z < 3$,  it gets large ($\gtrapprox 1$), and about $z \approx 3$, it becomes singular. The numerator and denominator of Eq. \eqref{eqn:qp-rat}  as a function of $z$ were displayed  in Figs. \ref{fig:psi-zw1} and \ref{fig:zphi}, respectively.

The calculations discussed in the foregoing section for spin waves were confined to the simple cubic lattice. It is, however, straightforward task to extend these evaluations to body-centered  cubic and face-centered cubic lattice with nearest neighbor  interaction. Also, the computations presented in Section \ref{sec:dos-thermo} correspond to low temperature limit of ferromagnet by assuming that the contribution of the magnon-magnon interaction in the Heisenberg hamiltonian is negligible at the outset.  The Green function formalism described in Section \ref{sec:spinwave}, however, is not \textit{per se} restricted to  the low-temperature limit. Hence, this surmise can be avoided at the expense of some additional calculations. Indeed, the effect of the contribution of the magnon-magnon interaction  through the Heisenberg hamiltonian of  pure ferromagnet has been a subject of extensive studies using various techniques over the years; see e.g. \cite{Callen_1963,Vaks1968thermodynamics,Oguchi_1960,bloch1962magnon,kaganov1987interacting,zhitomirsky2013colloquium}. Nevertheless, as argued in \cite{kittel1987quantum} magnon-magnon interactions do not have much effect on the temperature dependence of the saturation magnetization, except near the Curie temperature. Similarly, our calculations can be extended to case of low-dimensional lattices such as  square and triangular lattices through their corresponding density of states or lattice Green functions, cf. Eqs. \eqref{eqn:green-lattice-cube-lmn}-\eqref{eqn:green-lattice-cube-bessel} of \ref{sec:appE}. For the pure Heisenberg ferromagnet this case has been studied over the years \cite{mubayi1969phase,oguchi1971phase,takahashi1986quantum,takahashi1987few,takahashi1987classical,takahashi1987two,takahashi1990dynamics}.

 A point worth commenting is on the behavior or dwelling of magnons as Goldstone bosons in lower spatial dimensions, e.g. in a square or triangular lattice.
To this end, let's check the attribute of the relative magnetization $M$ from Eq. \eqref{eqn:totmag}. The deviation from the ground state magnetization at finite temperature in $d$-dimension (cf. Eqs.  \eqref{eqn:amn} and \eqref{eqn:correl00-mean}) can be expressed as 

\begin{equation}
\label{eqn:del-mag}
\Delta M = -\int \frac{ d^dq}{(2\pi)^d} \frac{1}{\exp[\beta\epsilon_\mathbf{q}]-1} \simeq  -K_d \int \frac{ dq q^{d-1}}{(2\pi)^d} \frac{1}{\exp[\beta \tilde S q^2]-1},
\end{equation}
where in the absence of the external field and small wave vector, $\epsilon_\mathbf{q} \simeq \tilde S q^2$ with $\tilde S=(1-c)^2\sigma/2$; and that the integration was taken over a a unit hypersphere with $K_d=2\pi^{d/2}/\Gamma(d/2)$ as its area. As can be seen, there is an infrared singularity in the integrand \eqref{eqn:del-mag}, $\exp[\beta \tilde S q^2]-1 \to \beta \tilde S q^2$, for $d \le 2$ rendering the spin wave concept anomalous. That is,  proliferation of  magnons  or Goldstone modes destroys the long range order of magnetization at any finite temperature, and the rotation symmetry is restored. This makes $d=2$ the lower critical dimension for the isotropic Heisenberg model at zero external field with or without nonmagnetic impurities.  We should also note that at $c=1$, the integrand of \eqref{eqn:del-mag} becomes infinite, i.e. the system is in complete disorder.

The breakdown of the spin wave concept, or the Goldstone mode associated with spontaneous symmetry breaking, is of course consistent with the Mermin-Wagner theorem \cite{mermin1966absence,nolting2009quantum}, which excludes spontaneous broken symmetry in $d=1, 2$ at nonzero temperature.\footnote{The theorem states that in one and two dimensions, the isotropic Heisenberg model with finite-range at nonzero temperature does not have any spontaneous magnetization.} Evidently, this is only true for  an ideal magnetic system with an isotropic and short-range interaction. The presence of anisotropy in the system overrides the Mermin-Wagner theorem and thereby opens a spin wave gap. Moreover, real two-dimensional magnetic systems, like ferromagnetic monolayers, small magnetic anisotropies or dipolar interactions are able to stabilize a long-range magnetic order; see e.g. \cite{zakeri2014elementary,burch2018magnetism,ortmanns2021magnon}.  Magnetic excitations in  the dilute two-dimensional ferromagnet K$_2$Cu$_{1-x}$Zn$_{x}$F$_4$ have been observed   with spin-1/2 for the Cu ions \cite{wagner1978spin}.   More specifically, Wagner and Krey \cite{wagner1978spin} studied this system by inelastic neutron scattering for $x$ = 0, 0.08 and 0.22, at $T$= 2 K. They obtained the dispersion of the frequency and of the linewidth  for wavevectors $\mathbf{q}$ in the direction of the strongest dispersion, q=(1,1,0), within the plane containing the strong exchange interaction. In the dilute samples, $x$ = 0.08 and 0.22, they observed the effect of the nonmagnetic impurities on both the frequency and the linewidth \cite{wagner1979spin}.

An interesting recent study \cite{evers2015spin} investigated the effect of disorder (nonmagnetic impurities) on magnonic transport in low-dimensional magnetic materials in the frame of a classical spin (Heisenberg-type) model. Evers et al. \cite{evers2015spin} studied the out-of-equilibrium, long-time spin-wave dynamics by integrating the Landau-Lifshitz equations of motion numerically. The authors examined the influence of randomly distributed impurities on the propagation of  spin waves in low-dimensional disordered magnets and found evidence for Anderson localization of spin waves in a one-dimensional spin chain. Furthermore, in a two-dimensional disordered (square) lattice, their computed spin-wave scattering intensity shows the presence of weak localization;  for a review and a textbook description of localizations see. e.g. \cite{wolfle2010self,girvin2019modern}.
\section{Summary and conclusions}
\label{sec:conclude}
Nonmagnetic inert ions randomly distributed in a ferromagnetic crystal lattice interrupt short-range spin-spin interactions between magnetic ions. In this paper, a Heisenberg hamiltonian was selected for this system and the Dyson-Maleev transformation was employed to express the spin operators in terms of the Bose operators of the second quantization. Then using methods of quantum statistical field theory, the partition function and the free energy were derived. The Matsubara thermal perturbation method has been adopted to that part of the hamiltonian which describes (in the second quantization picture)  the interaction between magnons and the stationary  field of nonmagnetic ions. The random distribution of nonmagnetic atoms was included in the  hamiltonian by a random variable $c_j$, which takes values of 0 or 1 depending on whether the lattice site $j$ is occupied by a magnetic or by a nonmagnetic impurity, respectively. Upon  averaging over all possible distributions of impurities, the free energy of the system was expressed as a function of the mean impurity concentration $c$. Applications of the presented theory to other problems that include stochastic variables in the hamiltonian should be possible. 

In the second part of the paper, we have set up the double-time single particle Green function at temperature $T$ in momentum (wave-vector) space in terms of magnon operators and derived the equation of motion for the Green function through the Heisenberg equation of motion. Assuming time invariance,  the coupled equation of motion for the Green function was Fourier transformed  from the time domain to the frequency domain, which included a momentum dependent field variable describing the distribution of nonmagnetic ions.  The equation of motion was split into two equations in terms of a kernel function associated with the Green function, and then the equations were solved. From that, the self-energy function and subsequently the spectral density function for the system were calculated. Averaging over impurity concentration was performed  and  all the quantities of interest were expressed in terms of the nonmagnetic ion (impurity) mean concentration $c$.  We have presented the main results of our calculations by applying the formalism to the case of the simple cubic lattice.  The density of states, the spectral density function and the lifetime of the magnons were expressed as a function of energy (frequency) for several values of the mean impurity concentrations. We have calculated the magnon energy spectrum as a function of $c$, which shows that for low lying states, the excitation energy increases continuously with $c$ in the studied range $0.1 \le c \le 0.7$. The spectral density function was used to compute some thermodynamical quantities through the magnon occupation number. Moreover, the impurity-averaged general formula for the Helmholtz free energy, derived by the Matsubara perturbation scheme, was used to compute to second-order approximation in regard to the magnetic-impurity interaction. Finally, we have discussed the results  our calculations with those in the literature. 

We have obtained closed form expressions for the configurationally averaged physical quantities of interest  in a unified fashion as functions of the mean concentration  of nonmagnetic impurities $c$ to any order of $c$  applicable below the critical percolation concentration $c_p$. The quantities of interest comprise the thermodynamic potential (free energy), the spin-wave self-energy and the spectral density function from which other quantities can be derived. The range of applicability of our method is: $0 \le c <c_p$, where for the simple cubic lattice $c_p\approx 0.7$.

In the foregoing section, we alluded to some possible generalization of the formalism to account for other effects or to other conditions. An additional extension may include, generalizing the putative Heisenberg hamiltonian to the case where the impurity has a spin $S^\prime$ which may differ from that of the host $S$ and its coupling constant  to the host $J^\prime$ which may also differ from that between nearest neighbors $J$ in the pure host system. Yet another extension would be to modify the considered Heisenberg hamiltonian so that is applicable to disordered antiferromagnets for which the spectral density function may be calculated and compared with the neutron scattering data made on compounds such as (Mn, Co)F$_2$ and K(Mn, Co)F$_3$. 


\section*{Acknowledgments}
Original calculations were done by Z.W. with further analyses and completion  by A.R.M..
\appendix
\renewcommand{\thefigure}{\Alph{section}.\arabic{figure}}
\setcounter{figure}{0}    
\section{Evaluation of integrals $\Xi_1$ and $\Xi_2$ of connected graphs}
\label{sec:appA0}

We expand Eq. (\ref{eqn:agd-sav}) in the form
\begin{equation}
\langle \mathcal{S}(\beta) \rangle_\beta  = \sum_{n=1}^\infty S_n = \exp\Big(\sum_{j=1}^\infty \Xi_j\Big),
\label{eqn:agd-sav-app0}
\end{equation}
\noindent
where $S_n$ are related to  $\Xi_n$ through
\begin{equation}
S_n = \sum_{k_1,k_2,\dots}\frac{n!}{k_1!k_2!\dots}\Xi_1^{k_1}\Xi_2^{k_2}\dots
\label{eqn:agd-sav-app1}
\end{equation}
\noindent
with condition $\sum_{j=1}jk_j=n$. Recalling  Eq. (\ref{eqn:matsu-sav}), we write
\begin{equation}
S_n = \frac{1}{N^{n/2}}\int_0^\beta d\tau_1 \int_0^{\beta} \dots \int_0^{\beta}\!\!\!\!d\tau_n
 \langle \mathrm{T} \big[\widetilde{\mathcal{H}}_{\mathrm{I}}(\tau_1) \dots \widetilde{\mathcal{H}}_{\mathrm{I}}(\tau_n)\big]\rangle_\beta.
\label{eqn:matsu-Sn}
\end{equation}

Let us first show that $S_1 \equiv \Xi_1=0$. From (\ref{eqn:matsu-Sn})
\begin{equation}
S_1 = \frac{1}{N^{1/2}}\int_0^\beta d\tau_1
 \langle \mathrm{T} \big[\widetilde{\mathcal{H}}_{\mathrm{I}}(\tau_1)\big]\rangle_\beta.
\label{eqn:matsu-S1}
\end{equation}
Using  Eqs.  \eqref{eqn:matsu-hI} and \eqref{eqn:matsu-htilde}, we write
\begin{equation}
 \langle \mathrm{T} \big[\widetilde{\mathcal{H}}_{\mathrm{I}}(\tau_1)\big]\rangle_\beta = -\frac{1}{N^{1/2}}
  \langle \sum_{\mathbf{k},\mathbf{q}} \gamma_{\mathbf{k}} \mathrm{T}[b_{\mathbf{k}+\mathbf{q}}^\dagger (\tau_1) b_{\mathbf{k}}(\tau_1)]c_{\mathbf{q}}\rangle_\beta.
\label{eqn:T-order2}
\end{equation}
where as in the Heisenberg picture for  operators, cf. Eq. (\ref{eqn:matsu-htilde}), we express the time-dependence of the Pauli operators as
\begin{equation}
b_{\mathbf{k}}(\tau) = e^{\tau \mathcal{H}_0} b_{\mathbf{k}} e^{-\tau \mathcal{H}_0},
\label{eqn:operator-time}
\end{equation}
and the corresponding relation for $b_{\mathbf{k}}^\dagger (\tau)$. Differentiating $b_{\mathbf{k}}(\tau)$, $b_{\mathbf{k}}^\dagger (\tau)$ with respect to $\tau$, making use of Eq. (\ref{eqn:matsu-h0}), and the commutation relations for the Pauli operators, we find
\begin{align}
\label{eqn:operator-traj}
b_\mathbf{k}(\tau) &= b_\mathbf{k}e^{-(\alpha+\varepsilon_\mathbf{k})\tau},\quad  b_\mathbf{k}^\dagger(\tau) = b_\mathbf{k}^\dagger e^{(\alpha+\varepsilon_\mathbf{k})\tau}.
\end{align}
with the initial conditions $b_\mathbf{k}(0) = b_\mathbf{k}$ and $b_\mathbf{k}^\dagger(0) = b_\mathbf{k}^\dagger$, and for brevity, we put $\epsilon_\mathbf{k} = \alpha+\varepsilon_\mathbf{k}$.

We should next recall that the T-product of two operators, e.g. $A$ and $B$, can be related to their normal product via the relation

\begin{equation}
\contraction{}{A}{}{B}
AB = \mathrm{T}[AB]-\mathrm{N}[AB],
\label{eqn:pair}
\end{equation}
where the over-bracket denotes the contraction between the two operators $A$ and $B$ and $\mathrm{N}[\bullet]$ is the normal product which puts all the annihilation operators on the right, see e.g. Ref. \cite{AGD_1963}. Furthermore, the normal product's average over the ground state vanishes \cite{AGD_1963}, thus resulting
\begin{equation}
\contraction{ \langle}{A}{}{B}
 \langle AB \rangle \equiv \contraction{}{A}{}{B}
AB = \langle \mathrm{T}[AB]\rangle.
\label{eqn:pair-av}
\end{equation}
In our case, however, this may not be directly applicable, since the ground state here consists of free magnon gas in the average field of nonmagnetic impurity. Therefore, we employ the Matsubara decomposition \cite{Matsubara_1955} to write
\begin{subequations}
\begin{align}
\label{eqn:mats-decomp-1}
b_\mathbf{k}(\tau) &= b_\mathbf{k+}(\tau) + b_\mathbf{k-}(\tau),\\
\label{eqn:mats-decomp-2}
b_\mathbf{k}^\dagger(\tau) &= b_\mathbf{k+}^\dagger(\tau) + b_\mathbf{k-}^\dagger(\tau),
\end{align}
\end{subequations}
where
\begin{subequations}
\begin{align}
\label{eqn:mats-var1}
b_\mathbf{k-}(\tau) &= f_\mathbf{k}b_\mathbf{k}e^{-(\alpha+\varepsilon_\mathbf{k})\tau},\quad b_\mathbf{k+}(\tau) = (1-f_\mathbf{k})b_\mathbf{k}e^{-(\alpha+\varepsilon_\mathbf{k})\tau},\\
\label{eqn:mats-var2}
b_\mathbf{k-}^\dagger(\tau) &= g_\mathbf{k}b_\mathbf{k}^\dagger e^{(\alpha+\varepsilon_\mathbf{k})^\dagger\tau},\quad b_\mathbf{k+}^\dagger(\tau) = (1-g_\mathbf{k})b_\mathbf{k}^\dagger e^{(\alpha+\varepsilon_\mathbf{k})\tau}.
\end{align}
\end{subequations}
Here, $f_\mathbf{k}$ and $g_\mathbf{k}$ are functions of $\mathbf{k}$ to be determined in sequel. If we postulate that the normal product will always rearrange the operators with $+$sign to the left and those with the $-$sign to the right, after simple algebra, we obtain
\begin{subequations}
\begin{align}
\label{eqn:n-prod-1}
\mathrm{N}[b_\mathbf{k_1}(\tau_1) b_\mathbf{k_2}^\dagger(\tau_2)] &= b_\mathbf{k_1}(\tau_1) b_\mathbf{k_2}^\dagger(\tau_2)-
[b_\mathbf{k_1-}(\tau_1),b_\mathbf{k_2+}^\dagger(\tau_2) ],\\
\label{eqn:n-prod-2}
\mathrm{N}[b_\mathbf{k_1}^\dagger(\tau_1) b_\mathbf{k_2}(\tau_2)] &= b_\mathbf{k_1}^\dagger(\tau_1) b_\mathbf{k_2}(\tau_2)-
[b_\mathbf{k_1-}^\dagger(\tau_1),b_\mathbf{k_2+}(\tau_2)].
\end{align}
\end{subequations}
The commutators on the right hand side, utilizing (\ref{eqn:mats-decomp-1})-(\ref{eqn:mats-var2}), can be written as
\begin{subequations}
\begin{align}
\label{eqn:comm-1}
[ b_\mathbf{k_1-}(\tau_1),b_\mathbf{k_2+}^\dagger(\tau_2) ] &= f_\mathbf{k_1}(1-g_\mathbf{k_1}) e^{-\epsilon_\mathbf{k_1}(\tau_1-\tau_2)}\delta_{\mathbf{k_1},\mathbf{k_2}},\\
\label{eqn:comm-2}
[ b_\mathbf{k_1-}^+(\tau_1),b_\mathbf{k_2+}(\tau_2) ] &= -g_\mathbf{k_1}(1-f_\mathbf{k_1}) e^{\epsilon_\mathbf{k_1}(\tau_1-\tau_2)}\delta_{\mathbf{k_1},\mathbf{k_2}}.
\end{align}
\end{subequations}
 By definition, the thermal average  of the normal product over the ground state (its vacuum expectation value) vanishes \cite{AGD_1963}; thus (\ref{eqn:n-prod-1})-(\ref{eqn:comm-2}) yield
\begin{subequations}
\begin{align}
\label{eqn:av-prod-1}
\langle b_\mathbf{k_1}(\tau_1)b_\mathbf{k_2}^\dagger(\tau_2)\rangle_\beta  &= f_\mathbf{k_1}(1-g_\mathbf{k_1}) e^{-\epsilon_\mathbf{k_1}(\tau_1-\tau_2)}\delta_{\mathbf{k_1},\mathbf{k_2}},\\
\label{eqn:av-prod-2}
\langle b_\mathbf{k_1}^\dagger(\tau_1) b_\mathbf{k_2}(\tau_2) \rangle_\beta &= -g_\mathbf{k_1}(1-f_\mathbf{k_1}) e^{\epsilon_\mathbf{k_1}(\tau_1-\tau_2)}\delta_{\mathbf{k_1},\mathbf{k_2}}.
\end{align}
\end{subequations}
Inserting now (\ref{eqn:operator-traj}) in the expectations (average) values, we obtain
\begin{subequations}
\begin{align}
\label{eqn:av-prodred-1}
\langle b_\mathbf{k}b_\mathbf{k}^\dagger\rangle_\beta  &= f_\mathbf{k}(1-g_\mathbf{k}),\\
\label{eqn:av-prodred-2}
\langle b_\mathbf{k}^\dagger b_\mathbf{k}\rangle_\beta &= -g_\mathbf{k}(1-f_\mathbf{k}).
\end{align}
\end{subequations}
Recalling that $\langle b_\mathbf{k} b_\mathbf{k}^\dagger\rangle_\beta=1+n_\mathbf{k}$ and $\langle b_\mathbf{k}^\dagger b_\mathbf{k}\rangle_\beta=n_\mathbf{k}$, where $n_\mathbf{k}$ is the average number of magnons in states with momentum $\mathbf{k}$, which depends on  temperature and chemical potential (Sec. 11.2 in Ref. \cite{AGD_1963}),  lets us to reduce (\ref{eqn:av-prodred-1})-(\ref{eqn:av-prodred-2}) to
\begin{subequations}
\begin{align}
\label{eqn:fgk-1}
f_\mathbf{k}(1-g_\mathbf{k})  &= 1+n_\mathbf{k},\\
\label{eqn:fgk-2}
-g_\mathbf{k}(1-f_\mathbf{k}) &= n_\mathbf{k},
\end{align}
\end{subequations}
where $n_\mathbf{k}=[\exp\beta\epsilon_\mathbf{k}-1]^{-1}$ for bosons. Using (\ref{eqn:pair}) and (\ref{eqn:av-prod-1})-(\ref{eqn:av-prod-2}) plus  (\ref{eqn:fgk-1})-(\ref{eqn:fgk-2}), after some algebra, we arrive at
\begin{subequations}
\begin{align}
\label{eqn:contract-1}
 \contraction{}{b_\mathbf{k_1}}{(\tau_1}{b_\mathbf{k_2}^\dagger}
b_\mathbf{k_1}(\tau_1) b_\mathbf{k_2}^\dagger(\tau_2)  &=  e^{-\epsilon_\mathbf{k_1}(\tau_1-\tau_2)}\delta_{\mathbf{k_1},\mathbf{k_2}}
 [(1+n_\mathbf{k_1})\theta(\tau_1-\tau_2)+n_\mathbf{k_1}\theta(\tau_2-\tau_1)],\\
\label{eqn:contract-2}
\contraction{}{b_\mathbf{k_1}^\dagger}{(\tau_1}{b_\mathbf{k_2}}
b_\mathbf{k_1}^\dagger(\tau_1) b_\mathbf{k_2}(\tau_2)  &=  e^{\epsilon_\mathbf{k_1}(\tau_1-\tau_2)}\delta_{\mathbf{k_1},\mathbf{k_2}}
[n_\mathbf{k_1}\theta(\tau_1-\tau_2)+(1+n_\mathbf{k_1})\theta(\tau_2-\tau_1)],
\end{align}
\end{subequations}
where $\theta(\tau)$ is the Heaviside step-function for continuous variables.

Let us get back to the evaluation of $\Xi_1=S_1$. Combining (\ref{eqn:matsu-S1}), (\ref{eqn:T-order2}) and (\ref{eqn:pair-av}), we write
\begin{equation}
\Xi_1 = -\frac{1}{N}\int_0^\beta d\tau_1 \sum_{\mathbf{k},\mathbf{q}} \gamma_{\mathbf{k}} \contraction{}{b}{_{\mathbf{k}+\mathbf{q}}^\dagger\tau_1}{b_\mathbf{k}()}
 b_{\mathbf{k}+\mathbf{q}}^\dagger(\tau_1) b_{\mathbf{k}}(\tau_1)\,c_{\mathbf{q}}
\label{eqn:matsu-Gamma1}
\end{equation}
Moreover, (\ref{eqn:contract-2}) gives
\begin{equation}
\label{eqn:autocontract-2}
\contraction{}{b}{_{\mathbf{k}+\mathbf{q}}^\dagger\tau_1}{b_\mathbf{k}()}
b_\mathbf{k+q}^\dagger(\tau_1) \, b_\mathbf{k}(\tau_1)
 = (n_\mathbf{k}+1)\delta_{\mathbf{k+q},\mathbf{k}}.
\end{equation}
Inserting now (\ref{eqn:autocontract-2}) into (\ref{eqn:matsu-Gamma1}) and integrating over $\tau_1$
\begin{align}
\Xi_1 &= -\frac{\beta}{N}\sum_{\mathbf{k},\mathbf{q}} \gamma_{\mathbf{k}}(n_\mathbf{k}+1)\delta_{\mathbf{k+q},\mathbf{k}}\,c_{\mathbf{q}}\nonumber\\
&= -\frac{\beta}{N}\sum_{\mathbf{k}} \gamma_{\mathbf{k}} (n_\mathbf{k}+1) c_{\mathbf{q=0}}
\label{eqn:Gamma1-inter}
\end{align}
Recalling $c_\mathbf{q} =  N^{-1/2}\sum_{j}  \tilde{c}_j e^{-i \mathbf{q} \cdot \mathbf{r}_j}$  with $\tilde c_j \equiv (c_j-c)$ from Sec. \ref{sec:model}, we have
\begin{equation}
\label{eqn:ft-c0}
 c_\mathbf{q=0} = \frac{1}{\sqrt{N}}\sum_{j}(c_j-c) = 0.
\end{equation}
Thus by virtue of relation (\ref{eqn:ft-c0}), (\ref{eqn:Gamma1-inter}) becomes $\Xi_1 = 0$. This means that no impulse has been transferred from the impurity to the magnon.

Next, we evaluate $S_2=2\Xi_2$. From (\ref{eqn:matsu-Sn}), we have
\begin{equation}\begin{split}
S_2 = \frac{1}{N}\sum_{\substack{\mathbf{k}_1\mathbf{k}_2 \\ \mathbf{q}_1\mathbf{q}_2}}
 \!\!\gamma_{\mathbf{k_1}}\gamma_{\mathbf{k_2}}c_{\mathbf{q_1}}c_{\mathbf{q_2}}
   \int_0^\beta \!\!\! d\tau_1 \!\!\int_0^{\beta}\!\!\!\!d\tau_2
 \langle \mathrm{T} \big[b_{\mathbf{k_1}+\mathbf{q_1}}^\dagger(\tau_1) b_{\mathbf{k_1}}(\tau_1) b_{\mathbf{k_2}+\mathbf{q_2}}^\dagger(\tau_2) b_{\mathbf{k_2}}(\tau_2)\big]\rangle_\beta.
\label{eqn:matsu-S2-1}
\end{split}\end{equation}
Employing (\ref{eqn:pair-av}) this becomes
\begin{equation}
S_2 = \frac{1}{N}\sum_{\substack{\mathbf{k}_1\mathbf{k}_2 \\ \mathbf{q}_1\mathbf{q}_2}}
 \!\!\gamma_{\mathbf{k_1}}\gamma_{\mathbf{k_2}}c_{\mathbf{q_1}}c_{\mathbf{q_2}}
  \int_0^\beta\!\!\!d\tau_1\!\!\int_0^{\beta}\!\!\!\!\!d\tau_2
  \,
\contraction{WWW\quad}{X}{Y}{Z\quad\quad\quad\;}
\contraction[2ex]{}{b}{_{\mathbf{k}+\mathbf{q}}^\dagger\tau_1}{b_\mathbf{k}(\qquad\qquad\qquad\qquad\qquad\qquad\qquad\qquad)}
b_{\mathbf{k_1}+\mathbf{q_1}}^\dagger(\tau_1)\, b_{\mathbf{k_1}}(\tau_1)\,b_{\mathbf{k_2}+\mathbf{q_2}}^\dagger(\tau_2)\,b_\mathbf{k_2}(\tau_2)
\label{eqn:matsu-S2-2}
\end{equation}
This expression or $\Xi_2$  can be represented as a closed loop (vacuum) graph (cf. Fig. \ref{fig:mag-defect-scat-1})
\[
\begin{tikzpicture}[line width=1.5 pt, scale=1]
 \node at (0,0) {$\Xi_2$};
 \node at (0.45,0) {\large=};
	\draw[vector] (0:2)--(1,0);
	\draw[fill=black] (1,0) circle (.075cm);
	\draw[fermion] (1,0) arc (180:0:1cm);
	\draw[fermion] (3,0) arc (0:-180:1cm);
	\draw[vector] (2,0) --(3,0);
	\draw[fill=black] (3,0) circle (.075cm);
	\node[red,very thick] at (2,0) {\LARGE$\boldsymbol{\times}$} ;
	\node at (1.5, 0.2) {\small$\mathbf{q}_1$};
	\node at (2.5, 0.2) {\small$\mathbf{q}_2$};
	\node at (2, 1.25) {\small$\mathbf{k}_1+\mathbf{q}_1$};
	\node at (2, -1.25) {\small$\mathbf{k}_2+\mathbf{q}_2$};
	\node at (0.80, 0) {\small$\tau_1$};
	\node at (3.25, 0) {\small$\tau_2$};
\end{tikzpicture}
\]

Making now use of (\ref{eqn:contract-1})-(\ref{eqn:contract-2})
\begin{equation}\begin{split}
\Xi_2 &= \frac{1}{2N}\sum_{\substack{\mathbf{k}_1\mathbf{k}_2 \\ \mathbf{q}_1\mathbf{q}_2}}
 \!\!\gamma_{\mathbf{k_1}}\gamma_{\mathbf{k_2}}c_{\mathbf{q_1}}c_{\mathbf{q_2}}\;\delta_{\mathbf{k_1+q_1},\mathbf{k_2}}
 \delta_{\mathbf{k_1-k_2},\mathbf{q_2}}
   \int_0^\beta \!\!\! d\tau_1 \!\!\int_0^{\beta}\!\!\!\!d\tau_2 \;e^{(\epsilon_\mathbf{k_1+q_1}-\epsilon_\mathbf{k_2})(\tau_1-\tau_2)}\\
 &\times \big[\theta(\tau_1-\tau_2)n_\mathbf{k_1+q_1}(n_\mathbf{k_1}+1)+\theta(\tau_2-\tau_1)(n_\mathbf{k_1+q_1}+1)n_\mathbf{k_1}\big].
\label{eqn:matsu-S2-3}
\end{split}\end{equation}
Evaluating the integrals over $\tau_1$, $\tau_2$ and simplifying, we obtain Eq. (\ref{eqn:Gamma-2}). These integrals simply reduce to
\begin{subequations}
\begin{align}
\label{eqn:integration1}
 \int_0^\beta \!\!\! d\tau_1 \!\!\int_0^{\beta}\!\!\!\!d\tau_2 \;e^{\Delta(\tau_1-\tau_2)} \theta(\tau_1-\tau_2) &= -\frac{1}{\Delta}
 \Big[\beta-\frac{1}{\Delta}(e^{\beta\Delta}-1)\Big],\\
 \int_0^\beta \!\!\! d\tau_1 \!\!\int_0^{\beta}\!\!\!\!d\tau_2 \;e^{\Delta(\tau_2-\tau_1)} \theta(\tau_2-\tau_1) &= \frac{1}{\Delta}
 \Big[\beta+\frac{1}{\Delta}(e^{-\beta\Delta}-1)\Big]
 \label{eqn:integration2}
\end{align}
\end{subequations}
where $\Delta=\epsilon_\mathbf{k_1+q_1}-\epsilon_\mathbf{k_2}$. In a similar manner,  one can evaluate the higher order terms in (\ref{eqn:agd-sav-app1})-(\ref{eqn:matsu-Sn}) and obtain  the result  given by Eq. (\ref{eqn:Gamma-n}) for $\Xi_n$.
\section{Cumulant average in random lattice}
\label{sec:cumav}
\setcounter{figure}{0}    
In our computations, we carried out the averaging  over a random distribution of impurities (nonmagnetic ions) by replacing the concentration $c_j$, at the lattice site $j$, with  $c_j-c$, where $c=\langle c_j \rangle=n/N$.  This simplified  the subsequent computations. If  $\tilde c_j \equiv c_j-c$, the average of the products of $\tilde c_j$ are
\begin{align}
\label{eqn:av-c}
\langle \tilde c_j\rangle &= 0,\\
\label{eqn:av-cc}
\langle \tilde c_{j_1} \tilde c_{j_2} \rangle &= \delta_{j_1j_2}P_2(c),\\
\label{eqn:av-ccc}
\langle \tilde c_{j_1} \tilde c_{j_2} \tilde c_{j_3} \rangle &= \delta_{j_1j_2j_3}P_3(c),\\
\label{eqn:av-cccc}
\langle \tilde c_{j_1} \tilde c_{j_2} \tilde c_{j_3}  \tilde c_{j_4} \rangle &= \delta_{j_1j_2j_3j_4}P_4(c) + (\delta_{j_1j_2} \delta_{j_3j_4} + \delta_{j_1j_3} \delta_{j_2j_4} + \delta_{j_1j_4} \delta_{j_2j_3})P_2^2(c),
\end{align}
where $\delta_{j_1j_2j_3}=\delta_{j_1j_2} \delta_{j_2j_3}$, etc. and $P_n$ is the cumulant average of $c_j$, namely the Matsubara-Yonezawa polynomials \cite{Matsubara_Yonezawa_1965,Yonezawa_1968},  but with $P_1(c)=0$, are generated through
\begin{align}
\label{eqn:matyon_formula}
P_n(c) &= \Big[\frac{\partial^n}{\partial x^n}\big[\ln(1-c +c e^x)-c x\big]\Big]_{x=0}.
\end{align}
\noindent
It should be noted that in averaging $\Pi_j \tilde c_j$ no value is obtained unless every index has been set equal to at least one other, through the Kronecker delta. For example, in Fourier transformed form,  with $ c_{\mathbf{q}}  = N^{-1/2}\sum_j \exp(-i\mathbf{q} \cdot r_j) \tilde c_j $, \eqref{eqn:av-cccc} reads
\begin{align}
\label{eqn:cq4-prod}
\langle c_{\mathbf{q}_1} c_{\mathbf{q}_2} c_{\mathbf{q}_3}  c_{\mathbf{q}_4} \rangle_{c}  &= \frac{P_4(c) + 3P_2^2(c)}{N}
\delta\Big(\sum_{i=1}^{4}\mathbf{q}_i\Big),\\
\label{eqn:P1c}
 P_1(c) &= 0,\\
\label{eqn:P2c}
P_2(c) &= c(1-c),\\
\label{eqn:P3c}
 P_3(c) &= c(1-c)(1-2c),\\
\label{eqn:P4c}
P_4(c) &= c(1-c)(1-6c+6c^2).
\end{align}
The polynomial $p_n$  in  Eq. \eqref{eqn:pnc} would give $p_4=P_4+ 3P_2^2$ in Eq. \eqref{eqn:cq4-prod}. The  two polynomials  $p_n$  and $P_n$ are the moments and the cumulants, respectively. The first four moments are
\begin{align}
\label{eqn:p1-c}
p_1(c) &=P_1(c) \equiv 0,\\
\label{eqn:P2-c}
p_2(c) &= P_1^2(c)+P_2(c),\\
\label{eqn:p3-c}
p_3(c) &= P_1^3(c)+3P_1(c)P_2(c)+P_3(c),\\
\label{eqn:p4-c}
p_4(c) &=  P_1^4(c) +6P_1(c)^2P_2(c)+3P_2^2(c)+4P_1(c)P_3(c)+P_4(c).
\end{align}
We have plotted  $p_n$  as a  function of $c$ for several values of $n$; Fig. \ref{fig:CumAv}.
\begin{figure*}[htbp]
\centering
\subfigure[]{\label{fig:CumAv-a}
\includegraphics[width=0.55\textwidth]{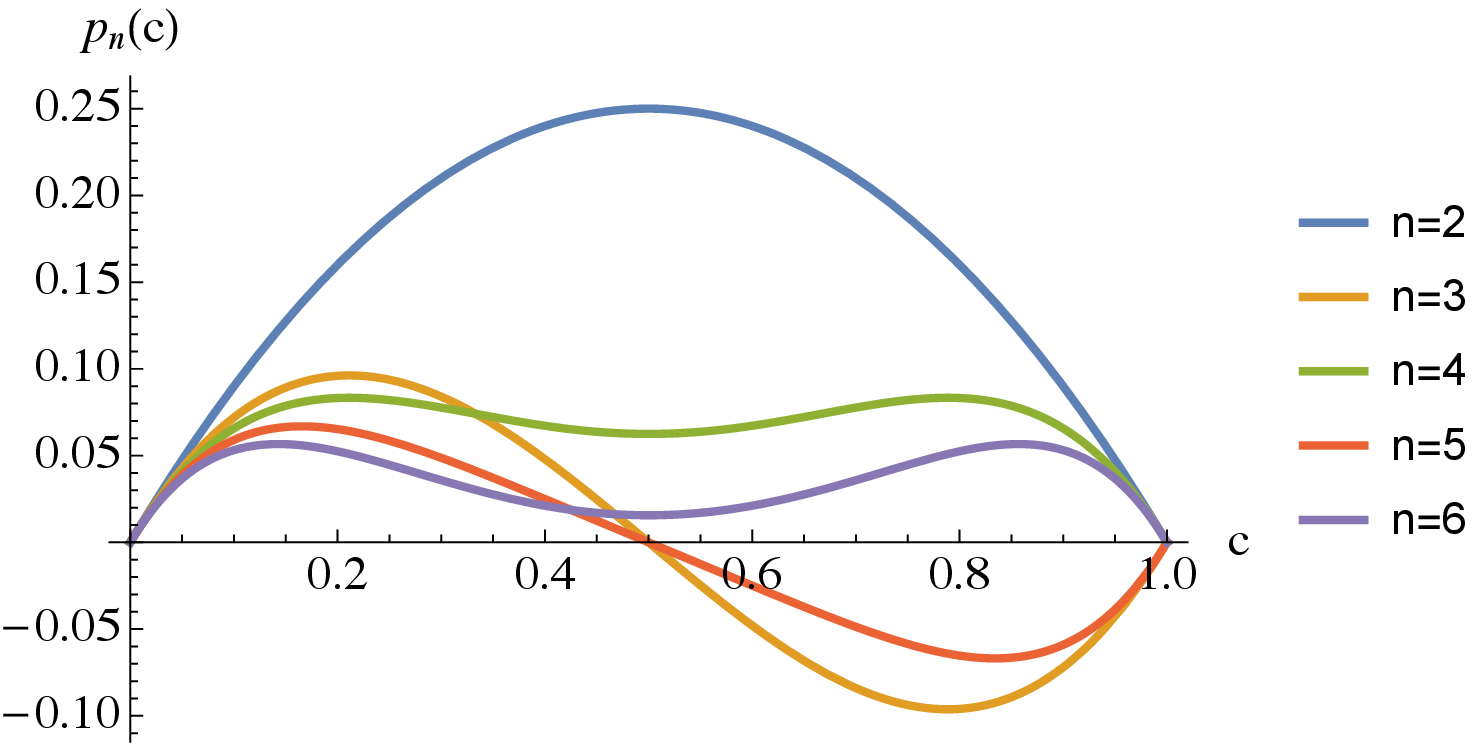}
}
\subfigure[]{\label{fig:CumAv-b}
\includegraphics[width=0.55\textwidth]{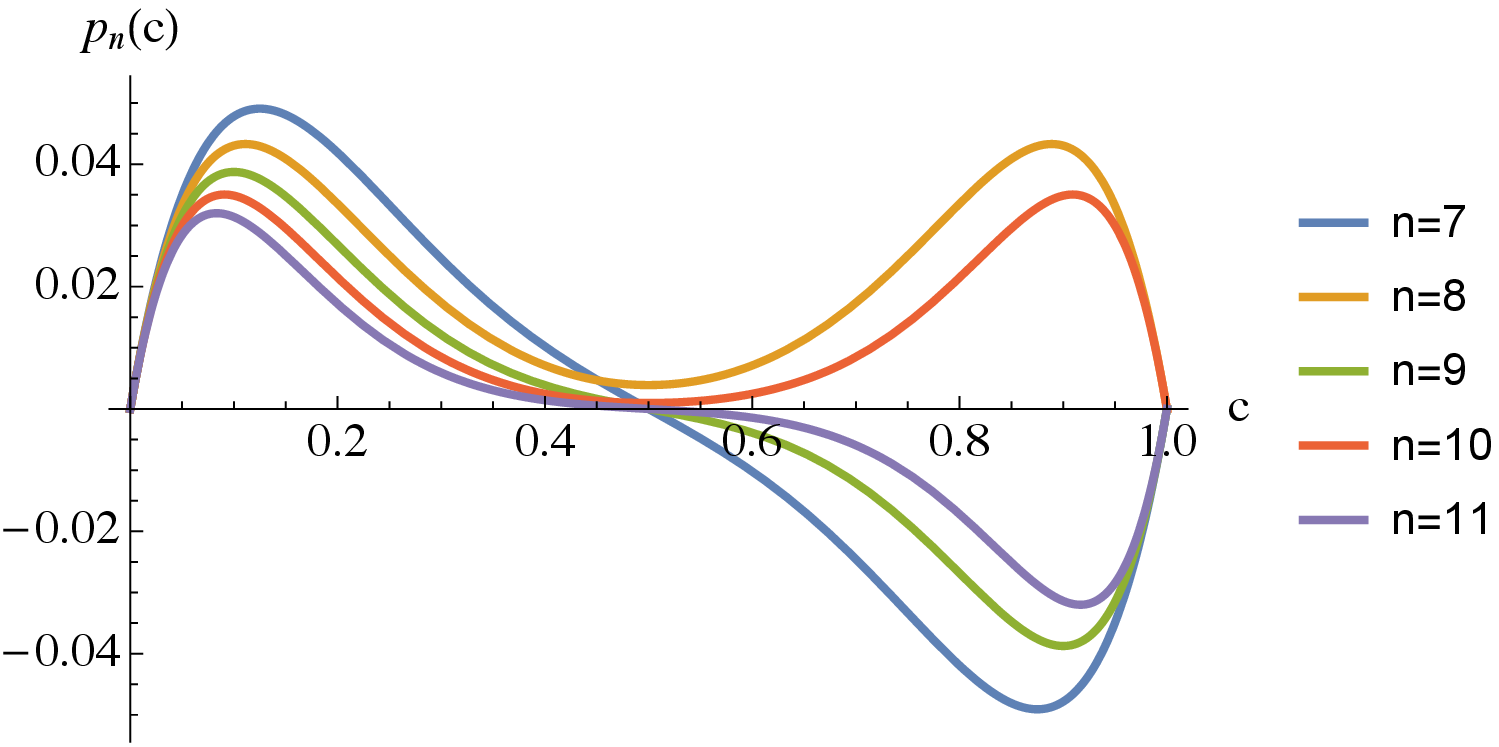}
}
\caption{Plots of the cumulative average function $p_n(c) =  [c(1-c)^n+ (-1)^n (1-c)  c^n]$.}
\label{fig:CumAv}
\end{figure*}
\section{Fluctuation-dissipation theorem}
\label{sec:appA}
There are several manifestations of the fluctuation and dissipation theorem. Here,  we outline  a variant applicable to bosonic and fermionic particles \cite{Bruus_Flensberg_2004} . The basic pair correlation functions or the thermal averages of the Pauli operators are
\begin{align}
\langle b^\dagger_\mathbf{k}(t)b_\mathbf{k}(0) \rangle &= -\frac{\eta}{2\pi i}\int_{-\infty}^\infty e^{-i\omega t} G^<(\mathbf{k},\omega) d\omega,\\
\langle b_\mathbf{k}(t) b^\dagger_\mathbf{k}(0) \rangle &= \frac{1}{2\pi i}\int_{-\infty}^\infty e^{-i\omega t} G^>(\mathbf{k},\omega) d\omega.
\label{eqn:pop_corr-t}
\end{align}
\noindent
where $G^\gtrless$are the lesser and greater Green functions. Furthermore,  we write the retarded Green function in momentum-time coordinates in the form
\begin{equation}
G^R(\mathbf{k},\mathbf{k^\prime};t,t^\prime) = \theta(t-t^\prime)\big[G^>(\mathbf{k},\mathbf{k^\prime};t,t^\prime)  -  G^<(\mathbf{k},\mathbf{k^\prime};t,t^\prime)\big],
\label{eqn:ret-green_kt}
\end{equation}
\noindent
where $\theta(\bullet)$ is the Heaviside step-function and $G^\gtrless$ are expressed in terms of  thermal averages of the Pauli operators
\begin{align}
\label{eqn:green_great_def}
G^>(\mathbf{k},\mathbf{k^\prime};t,t^\prime)  &= -i\langle b_\mathbf{k}(t) b^\dagger_\mathbf{k^\prime}(t^\prime)\rangle,\\
G^<(\mathbf{k},\mathbf{k^\prime};t,t^\prime) &= -i\eta\langle b^\dagger_\mathbf{k^\prime}(t^\prime)b_\mathbf{k}(t) \rangle,
\label{eqn:green_small_def}
\end{align}
\noindent
where $\eta=1$  specifies boson and $\eta=-1$  fermion. In the Heisenberg picture,  the Pauli operators are time dependent, viz.
\begin{equation}
b_\mathbf{k}(t) = e^{i\mathcal{H}t}b_\mathbf{k}e^{-i\mathcal{H}t}; \quad b_\mathbf{k}^\dagger(t) = b^{i\mathcal{H}t}b_\mathbf{k}^\dagger e^{-i\mathcal{H}t}.
\label{eqn:pop-evolut}
\end{equation}
\noindent
Now in the so-called Lehmann representation a set of eigenstates $\{|\mu\rangle\}$ is used as a basis  for the hamiltonian $\mathcal{H}$ with the property $\sum_\mu |\mu\rangle\langle \mu| =1$ and $\mathcal{H}|\mu\rangle=E_\mu|\mu\rangle$. A spectral decomposition of the greater Green function gives
\begin{align}
G^>(\mathbf{k};t,t^\prime)  &= -\frac{i}{\mathcal{Z}}\sum_\mu\langle \mu |e^{-\beta\mathcal{H}}b_\mathbf{k}(t) b^\dagger_\mathbf{k}(t^\prime)|\mu\rangle,\nonumber\\
 &= -\frac{i}{\mathcal{Z}}\sum_{\mu,\nu}e^{-\beta E_\mu}| \langle \nu | \,b^\dagger_\mathbf{k}\,|\mu\rangle|^2 e^{i(E_\mu-E_\nu)(t-t^\prime)},
\label{eqn:green_great_time}
\end{align}
\noindent
In the frequency domain, Fourier transformation yields
\begin{align}
G^>(\mathbf{k},\omega)  &= -\frac{2\pi i}{\mathcal{Z}}\sum_{\mu,\nu}e^{-\beta E_\mu}| \langle \nu |b^\dagger_\mathbf{k}|\mu\rangle|^2 \delta(E_\mu-E_\nu-\omega),
\label{eqn:green_great_freq}
\end{align}
\noindent
where $\mathcal{Z}=\exp(-\beta F)$, and $F$ is the Helmholtz free energy. Similarly, the "lesser" Green function
\begin{align}
 G^<(\mathbf{k},\omega)  &= -\eta \frac{2\pi i}{\mathcal{Z}}\sum_{\mu,\nu}e^{-\beta(E_\mu+\omega)}|\langle \nu | b_\mathbf{k}|\mu\rangle|^2 \ \delta(E_{\nu}-E_\mu-\omega).
\label{eqn:green_less_freq}
\end{align}
\noindent
The retarded Green function with $\omega \to \omega + i\delta$ becomes
\begin{equation}
G^R(\mathbf{k},\omega+ i\delta) = -i\int_0^\infty dt e^{i(\omega+i\delta)t}  \sum_{\mu,\nu}e^{-\beta (E_\mu-F)}\Big[ |\langle \nu| b_\mathbf{k}|\mu\rangle |^2 e^{i(E_\mu-E_\nu)t}
  +\eta | \langle \nu | b^\dagger_\mathbf{k}|\mu\rangle |^2 e^{-i(E_\mu-E_\nu)t}\Big].
\label{eqn:ret-green_komega}
\end{equation}
Carrying out the integration and simplifying, we find
\begin{equation}
G^R(\mathbf{k},\omega+i\delta) = e^{\beta F} \sum_{\mu,\nu}\frac{|\langle \nu| b_\mathbf{k}|\mu\rangle |^2}{\omega+ E_\mu-E_\nu+i\delta}\big(e^{-\beta E_\mu}-\eta e^{-\beta E_\nu}\big)
\label{eqn:ret-green_komega-simp}
\end{equation}
Taking the imaginary part of this quantity by using the Plemelj formula $(\omega+i\delta)^{-1}=\mathcal{P}(1/\omega)-i\pi\boldsymbol{\delta}(\omega)$ and recalling the definition of spectral density function \eqref{eqn:sdf-def}
\begin{align}
A(\mathbf{k},\omega) &= 2\pi e^{\beta F} \sum_{\mu,\nu}|\langle \nu| b_\mathbf{k}|\mu\rangle |^2 \big(e^{-\beta E_\mu}-\eta e^{-\beta E_\nu}\big) \delta(\omega+E_\mu-E_\nu)\nonumber\\
A(\mathbf{k},\omega)& = i(1 -\eta e^{-\beta\omega}) G^>(\mathbf{k};\omega) = -i(1-\eta e^{\beta\omega}) G^<(\mathbf{k};\omega).
\label{eqn:ret-green_komega-imag}
\end{align}
Alternatively, we may write
\begin{align}
\label{eqn:G<=A}
i G^>(\mathbf{k},\omega)  &= A(\mathbf{k},\omega)\big[1 + \eta n_\eta(\omega)\big],\\
i G^<(\mathbf{k},\omega) &= \eta A(\mathbf{k},\omega)n_\eta(\omega),
\label{eqn:G>=A}
\end{align}
\noindent
where $ n_\eta (\omega)= 1/[\exp(\beta\omega)-\eta]$ gives the Bose $\eta=1$ and the Fermi  $\eta=-1$ functions. These last two relations are statements of the fluctuation-dissipation theorem. Finally, from Eqs.  \eqref{eqn:G<=A}-\eqref{eqn:G>=A}, we have
\begin{equation}
G^>(\mathbf{k},\omega)  =  \eta e^{\beta\omega} G^<(\mathbf{k},\omega).
\label{eqn:great-less}
\end{equation}
\noindent
\section{Evaluations of density of states and its integral for a simple cubic crystal}
\label{sec:appD}
\setcounter{figure}{0}    
The  density of states (DOS) associated with magnons is expressed as
\begin{equation}
\label{eqn:dos-gen}
g(E) =   \int_\mathrm{BZ} \frac{d\mathbf{k}}{(2\pi)^3}   \delta\big(E-\omega(\mathbf{k})\big).
\end{equation}
\noindent
where the integral is over the Brillouin zone (BZ)  and $\omega(\mathbf{k})$ is the magnon excitation energy characterized by the wave vector $\mathbf{k}$. For the perfect simple cubic (sc) lattice with a lattice constant $a$,  this energy is given by \cite{Callaway_1963,takeno1963spin}
\begin{align}
\label{eqn:sc-spectra}
\omega_0(\mathbf{k})   &=  \epsilon_0 -\frac{ \epsilon_0}{3}\sum_{i=1}^3\cos(k_i a) =  \frac{2 \epsilon_0}{3}\sum_{i=1}^3\sin^2\Big(\frac{k_i a}{2}\Big)
\end{align}
\noindent
with $\epsilon_0=2JSZ$ and the constant $J$ is the nearest neighbor exchange integral in the Heisenberg model,  $S$ is the spin quantum number, $Z$ stands for the number of nearest neighbors, and $\mathbf{k}=(k_1,k_2,k_3)$; cf. Eq.  \eqref{eqn:sc-dispersion}.

\begin{figure}
\begin{center}
 \includegraphics[width=0.50\textwidth]{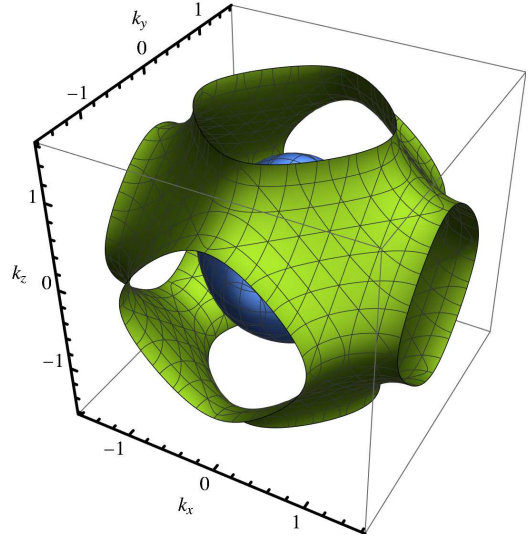}\\
\caption{ A contour plot of $\big|\nabla_\mathbf{k}\omega_0(\mathbf{k})\big|$  per equation \eqref{eqn:sc-spectra} in the range $-\pi/2 \le \mathbf{k} \le \pi/2$ for  $\mathbf{k}=(k_x,k_y,k_z)$.}
 \label{fig:va-sc}
 \end{center}
\end{figure}

A similar problem was studied  by Van Hove \cite{van1953occurrence} for the phonon density of states in a crystalline solid in $d$-dimensional hypercube  by
\begin{equation}
\label{eqn:dos-gen-hove}
g(E) =   \sum_{\alpha} \int_\mathrm{BZ} \frac{d^dk}{(2\pi)^d}   \delta\big(E-\omega_\alpha(\mathbf{k})\big)=  \sum_{\alpha}  \int \frac{d[S(E)]}{(2\pi)^d}  \frac{1}{\big|\nabla_\mathbf{k}\omega_\alpha(\mathbf{k})\big|}.
\end{equation}
\noindent
where the sum index $\alpha$ is over the branches of phonon frequencies. Here, $\alpha$  may be assigned to the magnon frequencies in a pure sc cubic lattice plus the three irreducible representations of the cubic point group $\mathrm{O}_h=(\Gamma_{1}, \Gamma_{15}, \Gamma_{12})$ or s, p, and d, representing the spin-wave  symmetry of the impurity (e.g. nonmagnetic ion) in the lattice. In the right-hand side of  Eq. \eqref{eqn:dos-gen-hove}, the integration is over the surface  $S(E)$ in $k$-space with energy $\omega_\alpha(\mathbf{k})=E$ and $\nabla_\mathbf{k}\omega_\alpha(\mathbf{k})$ is the group velocity of the spin-wave. Singularities in DOS can occur when $\nabla_\mathbf{k}\omega_\alpha(\mathbf{k})$ vanishes. In more detail,  Van Hove by utilizing the Morse inequalities showed that the homology groups of the Brillouin zone (topologically a $d$-torus) render such singular or saddle points in $\omega_\alpha(\mathbf{k})$ unavoidable. Moreover,  by analyzing the saddle points and applying the Morse Lemma, he showed  that the resulting singularities are non-integrable for space dimension $d = 2$, giving rise to a logarithmic divergence in the density of states, and are integrable for $d = 3$, causing divergences only in $dg/dE$. Figure  \ref{fig:va-sc} shows a contour plot of $\big|\nabla_\mathbf{k}\omega_0(\mathbf{k})\big|$ in the $(k_x,k_y,k_z)$-domain.  A generalization of the van Hove singularities to wave kinetics has recently been made in \cite{shi2016resonance}.

From Eq. \eqref{eqn:dos-gen}, we note that $g(E)$ vanishes exterior to the range of values of $\omega(\mathbf{k})$, i.e. per Eq. \eqref{eqn:sc-spectra}, $ 0 \le \omega_0(\mathbf{k})\le 2 \epsilon_0$. Moreover, $g(E)$  must be regular except at the  points of the van Hove singularities, at which $\omega_0(\mathbf{k})$ becomes extremal points on the surface of constant energy $S(E)$ . The singular points of the perfect sc lattice are
\begin{equation}
\label{eqn:vanhove-pts}
E =\big \{0,\pm\epsilon_0,\pm3\epsilon_0\big\}.
\end{equation}
\noindent

 It can been shown  that $g(E)$ corresponds to the imaginary part of the lattice Green function. For example, for the perfect sc lattice
\begin{equation}
\label{eqn:dos-sc-green}
g_0(E )=\frac{2}{\pi}G_\mathrm{I} (3-2E),
\end{equation}
\noindent
 where $G_\mathrm{I} (E)$ are given by Eqs. \eqref{eqn:green-sc-i1}, \eqref{eqn:green-sc-i2}, \eqref{eqn:green-sc-i3} in \ref{sec:appE} for the energy domains $E>3$, $0<E\le 3$, $0\le E\le 1$, respectively. Morita and Horiguchi \cite{morita1972analytic} using analytical function theory have shown that  lattice Green functions, i.e. $G_\mathrm{I} (E)$  in $d=1,2,3$, exhibit the same singularities as those derived by Van Hove for the density of states.

Let us now compute $g(E)$ for small values of its argument, i.e. $E <1$. We first,  after some elementary transformations using Eq. \eqref{eqn:sc-spectra}, write Eq. \eqref{eqn:dos-gen} as
\begin{equation}
\label{eqn:dos-scc}
g_0(\mu) =  \Big(\frac{2}{\pi}\Big)^3\iiint\limits dk_1dk_2dk_3 \delta\Big(\mu-\sum_{i=1}^3\sin^2 k_i\Big),
\end{equation}
\noindent
with the domain of integration:  $ 0\le k_i \le \pi/2; i=1,2,3$. Here, for convenience, we defined $\mu \equiv 3E/\epsilon_0$, $k_ia \to k_i$  and put $\epsilon_0a^3=1.5$. Next, we introduce an integral representation of $\delta$-function in the integrand
\begin{equation}
\label{eqn:dirac-scc}
\delta\Big(\mu-\sum_{i=1}^3\sin^2 k_i\Big) = \frac{1}{2\pi}\int\limits_{-\infty}^\infty e^{-i\mu\lambda} e^{i\lambda\sum_{j=1}^3 \sin^2k_j}\,d\lambda,
\end{equation}
\noindent
and express  \eqref{eqn:dos-scc} as
\begin{align}
\label{eqn:dos-scc-1}
g_0(\mu) &=  \Big(\frac{2}{\pi}\Big)^3 \Big(\frac{1}{\pi}\Big)\int\limits_{0}^\infty e^{-i\mu\lambda}\Big(\int_0^{\pi/2} e^{i\lambda\sin^2k} dk\Big)^3 \,d\lambda,\\
\text{or} \qquad g_0(\mu) &= \frac{1}{\pi}\int\limits_{0}^\infty e^{i(3/2-\mu)\lambda} J_0^3\big(\frac{\lambda}{2}\big) d\lambda,
	   \label{eqn:dos-scc-2}
\end{align}
\noindent
where  $J_0(\bullet)$ is the Bessel function of the first kind of zeroth order. Taking  the real part of $g_0(\mu)$
\begin{align}
\label{eqn:dos-scc-real}
g_0(\mu) &= \frac{1}{\pi}\int\limits_{0}^\infty \cos[(3/2-\mu)\lambda] J_0^3\big(\frac{\lambda}{2}\big) d\lambda,
\end{align}
\noindent
where $\mathrm{Re}[g_0(\mu)] \equiv g_0(\mu)$. Let us also evaluate  the Hilbert transform of  Eq. \eqref{eqn:dos-scc-real}, viz.
\begin{align}
\label{eqn:Re-base-g0-integral-b}
h_0(\omega) &= \int_0^{\infty} \frac{g_0(\mu)}{\mu-\omega} d\mu.
\end{align}
\noindent
We may express the denominator in the integrand as
\begin{align}
\label{eqn:sing-integ}
\lim_{\eta\to 0} \frac{1}{\mu -\omega -i\eta}  &= \lim_{\eta\to 0} i \int_0^{\infty} e^{-i(\mu-\omega-i\eta)\lambda} d\lambda.
\end{align}
\noindent
Substituting now Eqs. \eqref{eqn:dos-scc-2} and \eqref{eqn:sing-integ} into Eq. \eqref{eqn:Re-base-g0-integral-b} and then taking its real part, we obtain
\begin{align}
\label{eqn:f-integral-real}
h_0(\omega) &= \int_0^{\infty} \sin[(3/2 -\omega)\lambda] J_0^3\Big(\frac{\lambda}{2}\Big)d\lambda.
\end{align}
\noindent
This quantity is equivalent to the real part of the sc lattice Green function $G_\mathrm{R}$  discussed in \ref{sec:appE}, namely $h_0(\omega)=\eta_0G_\mathrm{R}(3-2\omega)$, with $\eta_0=G_\mathrm{R}^{-1}(3)$ a constant. Moreover, $h_0(\omega)=\pi \mathrm{Im}g_0(\omega)$.

We now compute $g_0(\mu)$ in the range $0 <\mu \le 1$ by first evaluating its Laplace transform
\begin{equation}
\label{eqn:dos-Lt}
\bar{g}_0(s) = \int_0^\infty g_0(\mu)e^{-\mu s}d\mu.
\end{equation}
Next, using \eqref{eqn:dos-scc-2} for $g_0(\mu)$, we obtain
\begin{equation}
\label{eqn:dos-Lt-ev1}
\bar{g}_0(s) = \frac{1}{2\pi}\int_{-\infty}^\infty \frac{e^{i \frac{3}{2}\lambda} J_0^3\big(\frac{\lambda}{2}\big)}{s+i\lambda} d\lambda.
\end{equation}
Thereafter, the residue theorem gives
\begin{equation}
\label{eqn:dos-Lt-ev2}
\bar{g}_0(s) = e^{-\frac{3}{2}s}\Big[I_0\big(\frac{s}{2}\big)\Big]^3,
\end{equation}
where $I_0(z)$ is the modified Bessel function of zeroth order with the property $J_0(iz)=I_0(z)$. We now carry out an asymptotic expansion of the right-hand side of Eq. \eqref{eqn:dos-Lt-ev2} for large $s$ to obtain
\begin{equation}
\label{eqn:dos-Lt-ae}
\bar{g}_0(s) = \frac{1}{\pi^{3/2}}\Big[(1-3e^{-2s})s^{-3/2}+\frac{3}{4}(1+e^{-2s})s^{-5/2}\Big] + O(s^{-7/2}).
\end{equation}
Finally, the inverse Laplace transform of Eq. \eqref{eqn:dos-Lt-ae} yields
\begin{equation}
\label{eqn:dos-iLt-ae}
g_0(\mu) = \frac{2}{\pi^{2}}\mu^{1/2} + \frac{1}{\pi^{2}}\mu^{3/2} + O\big(\mu^{5/2}\big).
\end{equation}

\section{Formulae of the lattice Green function for the simple cubic crystal}
\label{sec:appE}
\setcounter{figure}{0}    
\setcounter{table}{0}    

Let's start by writing Eq. \eqref{eqn:dos-gen} in the form
\begin{equation}
\label{eqn:dos-gen-sc}
g_0(E) =  \frac{3}{\epsilon_0} \iiint\limits_\mathrm{BZ} \frac{dk_1 dk_2 dk_3}{(2\pi)^3}   \delta(E+\cos k_1+\cos k_2+\cos k_3),
\end{equation}
\noindent
where we substituted for $\omega_\mathbf{k}$ from  \eqref{eqn:sc-spectra} and put $a=1$ for convenience. The domain of integration (first Brillouin zone) is a cube with sides $2\pi$, lying symmetrically to the origin of coordinates.  We also drop the factor $3/\epsilon_0$ for simplicity; cf. \cite{Jelitto_1969}.

Putting $l_i=\cos k_i$ and noting the symmetry, Eq. \eqref{eqn:dos-gen-sc} is transformed into
\begin{equation}
\label{eqn:dos-gen-sc-3l}
g_0(E) =  \frac{1}{\pi^3} \int_{-1}^{+1} dl_1 \int_{-1}^{+1} dl_2\int_{-1}^{+1} dl_3 \frac{ \delta(E+l_1+l_2+l_3)}{\sqrt{(1-l_1^2)(1-l_2^2)(1-l_3^2)}}.
\end{equation}
\noindent
Integrating over $l_3$ gives
\begin{equation}
\label{eqn:dos-gen-sc-2l}
g_0(E) =  \frac{1}{\pi^3} \iint\limits_\mathrm{L} dl_1dl_2\Big\{(1-l_1^2)(1-l_2^2)\big[1-(E+l_1+l_2)^2\big]\Big\}^{-1/2},
\end{equation}
\noindent
where the domain of integration L is  the section of a square with the sides 2 lying symmetrically to the origin, and a band defined by $|E+l_1+l_2 |\le 1$ \cite{Jelitto_1969}. Three conditions are possible: (i) for $|E| > 3$, the domain is empty, hence $g(E)=0$, (ii)  for $1\le |E| \le 3$, the section is  a triangle, and (iii) for $|E| < 1$, it is a band within the square. 

As can be checked, $g_0(E)$ is an even function, so one can restrict the computations in the range of $0 \le  E <3$. We set $\mathcal{I}(l_1,l_2)=\big\{(1-l_2^2)\big[1-(E+l_1+l_2)^2\big]\big\}^{-1/2}$ and write \eqref{eqn:dos-gen-sc-2l} as
\begin{equation}
\label{eqn:dos-gen-sc-ii}
g_0(E) =  \frac{1}{\pi^3} \int_{-1}^{2-E} \!\! \! \! \!  \frac{dl_1} {\sqrt{1-l_1^2}} \int_{-1}^{1-E-l_1} \!\! \! \! \!  dl_2 \, \mathcal{I}(l_1,l_2),
\end{equation}
\noindent
for $ 1 \le E < 3$,  and
\begin{equation}
\label{eqn:dos-gen-sc-iii}
g_0(E) =  \frac{1}{\pi^3}\Bigg[ \int_{-1}^{-E} \!\! \! \! \!  \frac{dl_1} {\sqrt{1-l_1^2}} \int_{-1-E-l_1}^{1} \!\! \! \! \!  dl_2 \, \mathcal{I}(l_1,l_2) + \! \int_{-E}^{1} \!\! \!   \frac{dl_1} {\sqrt{1-l_1^2}} \int_{-1}^{1-E-l_1} \!\! \! \! \!  dl_2 \, \mathcal{I}(l_1,l_2) \Bigg],
\end{equation}
for $ 0 \le E < 1$. Integrations over $l_2$ lead to \cite{Jelitto_1969}
\begin{equation}
\label{eqn:dos-gen-sc-iie}
g_0(E) =  \frac{1}{\pi^3} \int_{-1}^{2-E} \!\! \! \! \!  \frac{dl} {\sqrt{1-l^2}} K^\prime\Big( \frac{E+l}{2} \Big),
\end{equation}
\noindent
for $ 1 \le E < 3$,  and
\begin{equation}
\label{eqn:dos-gen-sc-iiie}
g_0(E) =  \frac{1}{\pi^3}\Bigg[ \int_{-1}^{-E} \!\! \! \! \!  \frac{dl} {\sqrt{1-l^2}} K^\prime\Big( \frac{-E-l}{2} \Big) + \! \int_{-E}^{1} \!\!   \frac{dl} {\sqrt{1-l^2}}\, K^\prime\Big( \frac{E+l}{2} \Big) \Bigg],
\end{equation}
for $ 0 \le E < 1$. Here,  $K^\prime$ is the first complete elliptic integral of the complementary module which is related to the  first complete elliptic integral $K$ via $K^\prime(k)=K(k^\prime)=K\big(\sqrt{1-k^2}\big)$; see e.g.  \cite{NIST_Math_Handbook}. These integrals need to be evaluated numerically.

Next, we  consider  the simple cubic lattice Green function as \cite{morita1971lattice}
\begin{equation}
\label{eqn:sc-greenfun}
G(t) =   \frac{1}{\pi^3}\iiint_0^\pi\frac{dx_1 dx_2 dx_3}{t-(\cos x_1+\cos x_2 + \cos x_3)}.
\end{equation}
\noindent
A closely related lattice Green function  is  defined as \cite{joyce1973simple}.
\begin{equation}
\label{eqn:sc-greenfun-j}
P(z) = \frac{3}{z} G\Bigg(\frac{3}{z}\Bigg) =  \frac{1}{\pi^3}\iiint_0^\pi\frac{dx_1 dx_2 dx_3}{1-\frac{1}{3}z(\cos x_1+\cos x_2 + \cos x_3)}.
\end{equation}
\noindent
The integrand in this equation represents a single-valued analytical function throughout the $z^2$-domain cut along the real axis from $+1$ to $\infty$ \cite{morita1971lattice,joyce1973simple}.
The Green function as defined by \eqref{eqn:sc-greenfun} is real for $t > 3$ and can be expressed  in terms of the complete integral of the first kind $K(\bullet)$, viz.
\begin{equation}
\label{eqn:green-ellipitic-int}
G(t) =   \frac{1}{\pi^2}\int_0^\pi k K(k) dx,
\end{equation}
\noindent
where
\begin{equation}
k =   2/(t-\cos x).
\label{eqn:k-arg}
\end{equation}
\noindent
Relation \eqref{eqn:green-ellipitic-int} holds on the entire complex $t$-plane.

Putting now $t=s-i \epsilon$, where $s$ is a real variable and $\epsilon$ is an infinitesimal positive number, $G(s-i\epsilon)$  is a complex function in $-3<s<3$. Denoting the real and imaginary parts as $G_\mathrm{R}(s)$ and $G_\mathrm{I}(s)$, respectively, we write
\begin{equation}
\label{eqn:lattice-green-re-im}
G(s-i\epsilon) =   G_\mathrm{R}(s)+iG_\mathrm{I}(s).
\end{equation}
\noindent
Note that for the sc lattice, $G_\mathrm{R}(s)$ is an odd function of $s$ and $G_\mathrm{I}(s)$ is an even function:
\begin{equation}
\label{eqn:lattice-green-even-odd}
 G_\mathrm{R}(-s) =   -G_\mathrm{R}(s), \qquad G_\mathrm{I}(-s)= G_\mathrm{I}(s).
\end{equation}
\noindent

The lattice Green function for the simple cubic lattice is expressed as a piecewise continuous function in three positive domains, (I) $s>3$,  (II) $1<s\le 3$, and (III) $0< s \le 1$. It has been shown in  \cite{Morita_Horiguchi_1971}  that  the real  and the imaginary  parts  of $G(s) $  for sc lattice are:

Domain I. ($s>3$)
\begin{align}
\label{eqn:green-sc-r1}
G_\mathrm{R}(s) &= \frac{2}{\pi^2}\int\limits_{-1}^{1} \frac{(1-l^2)^{-1/2}}{s+l} K\Big(\frac{2}{s+l}\Big)dl,\\
\label{eqn:green-sc-i1}
G_\mathrm{I}(s) &= 0.
\end{align}
\noindent
Domain II. ($1< s \le 3$)
\begin{align}
\label{eqn:green-sc-r2}
G_\mathrm{R}(s) &= \frac{1}{\pi^2} \Bigg[\int\limits_{-1}^{2-s}(1-l^2)^{-1/2} K\Big(\frac{s+l}{2}\Big)dl +
  2\negthickspace\int\limits_{2-s}^{1} \frac{(1-l^2)^{-1/2}}{s+l} K\Big(\frac{2}{s+l}\Big)dl\Bigg] ,\\
\label{eqn:green-sc-i2}
G_\mathrm{I}(s)  &= \frac{1}{\pi^2}\int\limits_{-1}^{2-s}(1-l^2)^{-1/2} K^\prime \Big(\frac{s+l}{2}\Big)dl .
\end{align}
\noindent
Domain III. ($ 0 < s \le 1$)
\begin{align}
\label{eqn:green-sc-r3}
G_\mathrm{R}(s)  &= \frac{1}{\pi^2}\Bigg[ \int\limits_{-s}^{1}(1-l^2)^{-1/2} K\Big(\frac{s+l}{2}\Big)dl -\int\limits_{-1}^{-s}(1-l^2)^{-1/2} K\Big(\frac{s+l}{2}\Big)dl\Bigg],\\
\label{eqn:green-sc-i3}
 G_\mathrm{I}(s) &= \frac{1}{\pi^2}\Bigg[\int\limits_{-1}^{s}(1-l^2)^{-1/2} K^\prime \Big(\frac{-s-l}{2}\Big)dl +\negthickspace \int\limits_{-s}^{1}(1-l^2)^{-1/2} K^\prime \Big(\frac{s+l}{2}\Big)dl \Bigg].
\end{align}
Comparing now Eqs. \eqref{eqn:green-sc-i2} and \eqref{eqn:green-sc-i3} with Eqs. \eqref{eqn:dos-gen-sc-iie} and \eqref{eqn:dos-gen-sc-iiie}, we see that $ G_\mathrm{I}(s)=\pi g(s)$.

For computations, one may also eliminate  singularities in the integrands that arise from $\sqrt{1-l^2}$ by substituting $l=-\cos x$,  resulting in:
Domain I. ($s>3$)
\begin{align}
\label{eqn:green-sc-r1s}
G_\mathrm{R}(s) &= \frac{1}{\pi^2}\int\limits_{0}^{\pi} k K(k)dx,\\
\label{eqn:green-sc-i1s}
G_\mathrm{I}(s) &= 0.
\end{align}
\noindent
where $k=2/(s-\cos x)$, which is identical to  relations \eqref{eqn:green-ellipitic-int} and \eqref{eqn:k-arg}.

Domain II. ($1< s \le 3$)
\begin{align}
\label{eqn:green-sc-r2s}
G_\mathrm{R}(s) &= \frac{1}{\pi^2} \Bigg[\int\limits_{0}^{\arccos(s-2)}K(k_1)dx +\int\limits_{\arccos(s-2)}^{\pi} k K(k)dx \Bigg] ,\\
\label{eqn:green-sc-i2s}
G_\mathrm{I}(s)  &= \frac{1}{\pi^2}\int\limits_{0}^{\arccos(s-2)}K(k_1^\prime)dx .
\end{align}
Domain III. ($ 0 < s \le 1$)
\begin{align}
\label{eqn:green-sc-r3s}
G_\mathrm{R}(s)  &= \frac{1}{\pi^2}\Bigg[\; \int\limits_{\arccos s}^{\pi}K(k_1)dx - \int\limits_{0}^{\arccos s} K(|k_1|)dx \Bigg],\\
\label{eqn:green-sc-i3s}
 G_\mathrm{I}(s) &= \frac{1}{\pi^2}\Bigg[\;\int\limits_{0}^{\arccos(-s)}\negthickspace K(-k_1^\prime)dx +\negthickspace\int\limits_{\arccos s}^{\pi}K(k_1^\prime)dx \Bigg].
\end{align}
Here, $k_1=(s-\cos x)/2=1/k$ and $k_1^\prime = \sqrt{1-k_1^2}$, cf. \cite{Morita_Horiguchi_1971} .

Joyce \cite{joyce1973simple} has shown that $G_\mathrm{R}(s)$ and $G_\mathrm{I}(s)$ can be expressed in terms of the Heun function, which may be more convenient  for computation than above numerical integrations. They are:
Domain I. ($s>3$)
\begin{align}
\label{eqn:green-sc-rj1}
G_\mathrm{R}(s) &= \frac{1}{3}P(1) \Big[F\Big(\frac{9}{8},\frac{7}{128}; \frac{1}{4},\frac{1}{4},\frac{1}{2},\frac{1}{2};\frac{9-s^2}{8}\Big)\Big]^2\nonumber\\
& -\frac{(9-s^2)}{16\pi^2P(1)} \Big[F\Big(\frac{9}{8},\frac{75}{128}; \frac{3}{4},\frac{3}{4},\frac{3}{2},\frac{1}{2};\frac{9-s^2}{8}\Big)\Big]^2\nonumber\\
& -\frac{(s^2-9)^{1/2}}{2\pi\sqrt{3}} \Big[F\Big(\frac{9}{8},\frac{7}{128}; \frac{1}{4},\frac{1}{4},\frac{1}{2},\frac{1}{2};\frac{9-s^2}{8}\Big)\Big]^2\nonumber\\
&\times F\Big(\frac{9}{8},\frac{75}{128}; \frac{3}{4},\frac{3}{4},\frac{3}{2},\frac{1}{2};\frac{9-s^2}{8}\Big),\\
\label{eqn:green-sc-ij1}
G_\mathrm{I}(s) &= 0.
\end{align}
\noindent
where $F(a,q;\alpha,\beta,\gamma,\delta;z)$ is the general Heun function, which is the solution of the Heun equation \cite{NIST_Math_Handbook}, and it is  also denoted by $H\ell \equiv F$. Note that here $q=-b$ in Joyce's notation \cite{joyce1973simple}. And $P(1)$ is evaluated through \eqref{eqn:sc-greenfun-j} with the result 
\begin{align}
\label{eqn:watson-1}
P(1) &=12\pi^{-2}(18+12\sqrt{2} -10\sqrt{3}-7\sqrt{6})(K_2)^2,\nonumber\\
& \approx 1.516\,386\,059\,151\,978,
\end{align}
\noindent
where $K_2$ denotes a complete integral of the first kind with a modulus $k=(2-\sqrt{3})(\sqrt{3}-\sqrt{2})$.

Domain II. ($1< s \le 3$)
\begin{align}
\label{eqn:green-sc-r2h}
G_\mathrm{R}(s) &= G_\mathrm{R}(1) \Big[F\Big(-8,-\frac{1}{16}; \frac{1}{4},\frac{1}{4},\frac{1}{2},\frac{1}{2};1-s^2\Big)\Big]^2\nonumber\\
&+\frac{3(1-s^2) }{16\pi^2 G_\mathrm{R}(1)}\Big[F\Big(-8,-\frac{29}{16}; \frac{3}{4},\frac{3}{4},\frac{3}{2},\frac{1}{2};1-s^2\Big)\Big]^2,\\
\label{eqn:green-sc-i2h}
G_\mathrm{I}(s)  &=  \sqrt{2}G_\mathrm{R}(1) \Big[F\Big(-8,-\frac{1}{16}; \frac{1}{4},\frac{1}{4},\frac{1}{2},\frac{1}{2};1-s^2\Big)\Big]^2\nonumber\\ 
&-\frac{3\sqrt{2}(1-s^2) }{16\pi^2 G_\mathrm{R}(1)}\Big[F\Big(-8,-\frac{29}{16}; \frac{3}{4},\frac{3}{4},\frac{3}{2},\frac{1}{2};1-s^2\Big)\Big]^2\nonumber\\
&-  \frac{3}{2\pi} (s^2-1)^{1/2} F\Big(-8,-\frac{1}{16}; \frac{1}{4},\frac{1}{4},\frac{1}{2},\frac{1}{2};1-s^2\Big)\nonumber\\
&\times F\Big(-8,-\frac{29}{16}; \frac{3}{4},\frac{3}{4},\frac{3}{2},\frac{1}{2};1-s^2\Big),
\end{align}
where $G_\mathrm{R}(1)=\frac{\pi}{2}\big[\Gamma\big(\frac{5}{8}\big) \Gamma\big(\frac{7}{8}\big) \big]^{-2} \approx 0.642882248294458$, cf.  \cite{joyce1973simple}.

Domain III. ($ 0 < s \le 1$)
\begin{align}
\label{eqn:green-sc-r3h}
G_\mathrm{R}(s)  &= \frac{2s}{\pi\sqrt{3}} F\Big(9,\frac{1}{8}; \frac{1}{4},\frac{1}{4},\frac{1}{2},\frac{1}{2};s^2\Big) F\Big(9,\frac{21}{8}; \frac{3}{4},\frac{3}{4},\frac{3}{2},\frac{1}{2};s^2\Big),\\
\label{eqn:green-sc-i3h}
 G_\mathrm{I}(s) &=G_\mathrm{I}(0)\Big[F\Big(9,\frac{1}{8}; \frac{1}{4},\frac{1}{4},\frac{1}{2},\frac{1}{2};s^2\Big)\Big]^2 -\frac{s^2}{3\pi^2 G_\mathrm{I}(0)} \Big[F\Big(9,\frac{21}{8}; \frac{3}{4},\frac{3}{4},\frac{3}{2},\frac{1}{2};s^2\Big)\Big]^2,
\end{align}
where $G_\mathrm{I}(0)=3\big[\Gamma\big(\frac{1}{3}\big)\big]^6/(2^{11/3}\pi^4) \approx 0.896440788776763$, cf. \cite{joyce1973simple}.

\begin{figure}
\begin{center}
 \includegraphics[width=0.50\textwidth]{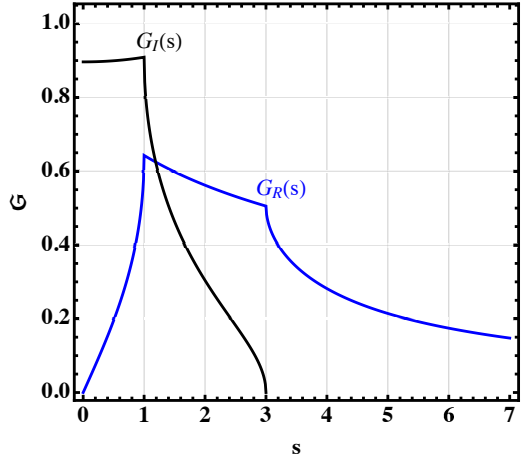}\\
\caption{The simple cubic lattice Green function  $G(s)=G_\mathrm{R}(s)+i G_\mathrm{I}(s) $  as a function of the energy parameter $s$.}
 \label{fig:gsc-heun}
 \end{center}
\end{figure}

 Figure \ref{fig:gsc-heun} displays $G_\mathrm{R}(s)$ and  $G_\mathrm{I}(s)$  in the range $ 0 \le  s  \le 7$ using the  package  \cite{Mathematica};  cf.  \cite{morita1971lattice} . For the record, we have also tabulated some values of these Green functions in that range in table \ref{tab:sc-Gfunctions}, cf. \cite{joyce1973simple}.  Both the elliptic numerical integrations and the general Heun function in \texttt{Mathematica} yield identical results. In the table, we have also included rescaled functions $\tilde{h}(s)=G_\mathrm{R}(3) h_0(\frac{3-s}{2})$  and $\tilde{g}(s)=\frac{\pi}{2}g_0(\frac{3-s}{2})$; cf. Eqs. \eqref{eqn:f-integral-real} and \eqref{eqn:dos-scc-real}.
\begin{table}[!htb]
  \begin{center}
    \caption{Green functions for simple cubic lattice and the corresponding rescaled functions $\tilde{h}(s)$ and $\tilde{g}(s)$.}
    \begin{tabular}[l]{lcccc}
      \hline
  $s$  &  $G_\mathrm{R}(s)$ &	 $G_\mathrm{I}(s)$  & $\tilde{h}(s)$ & $\tilde{g}(s)$\\
  \hline
 0. & 0. & 0.896440788776763 & 0. & 0.896101529858 \\
 0.5 & 0.195322880928494 & 0.899508458513519  & 0.197626844299 & 0.899074844722 \\
 1. & 0.642882248294458 & 0.909172794546930  & 0.622661438677 & 0.882179155256 \\
 1.5 & 0.598433718602123 & 0.463544765191000  & 0.604628664612 & 0.463674449721 \\
 2. & 0.562116099272940 & 0.303993825678427   & 0.568050117827 & 0.303992903951 \\
 2.5 & 0.531612102622276 & 0.182284855335886   & 0.537219933977 & 0.182207061358 \\
 3. & 0.505462019717332 & 0.   & 0.501864167279 & 0.008960324478 \\
 4. & 0.281862976225440 & 0.  & 0.284941093557 & 0.000028773349 \\
 5. & 0.214294082764824 & 0.  & 0.216629175152 & 0.000005092943\\
 6. & 0.174459564044919 & 0.  & 0.176362265226 & -0.000000479982\\
 7. & 0.147605297340978 & 0. & 0.14921700196 & -0.000001421573\\  \hline
 \end{tabular}
  \label{tab:sc-Gfunctions}
  \end{center}
\end{table}
Finally, in a more general setting the simple cubic lattice Green function at an arbitrary point $(l,m,n)$ is commonly expressed as \cite{abe1973lattice}
\begin{equation}
\label{eqn:green-lattice-cube-lmn}
G(E; l,m,n;\gamma) =   \iiint\limits_\mathrm{BZ} \frac{dk_1 dk_2 dk_3}{(2\pi)^3}   \frac{\cos lk_1\cos mk_2\cos nk_3}{E- (\gamma\cos k_1+\cos k_2+\cos k_3)}.
\end{equation}
\noindent
Here,  for the simple cubic lattice at the origin $(l = m = n = 0, \gamma = 1)$, for the square lattice $(l = 0, \gamma = 0)$ and for the linear lattice $(m=n = 0, E\to \gamma E, \gamma \to \infty)$. The integral in \eqref{eqn:green-lattice-cube-lmn} can be transformed into \cite{abe1973lattice}
\begin{equation}
\label{eqn:green-lattice-cube-bessel}
G(E; l,m,n;\gamma) =   i^{l+m+n+1}\int_0^\infty e^{-iEt} J_l(\gamma t)J_m(t)J_n(t)  dt,
\end{equation}
\noindent
where $J$'s are Bessel functions of integral order; cf. Eqs. \eqref{eqn:green-spd}.  The lattice Green function  $G(E; l,m,n;\gamma)$ can also be represented in terms of the Kamp\'e de F\'eriet function by analytic continuation, which is a generalized hypergeometric function of two variables, as shown in \cite{abe1973lattice}.
\section{Calculation of magnon decay rate in the long-wave limit}
\label{sec:appF}
\setcounter{figure}{0}    
\setcounter{table}{0}    
We consider the case of zero external field ($B=0$) and write Eqs. \eqref{eqn:case-gen-re1}-\eqref{eqn:case-gen-im1} combined with Eqs. \eqref{eqn:case-gen-phi}-\eqref{eqn:case-gen-psi} as 
\begin{align}
\label{eqn:case-gen-re1-0}
\bar{\Sigma}^{\prime}(\mathbf{q},z) &= 2c(1-c)^2\sigma q^2\big[\mathrm{Re} f_1(z)-\mathrm{Re} f_2(z)\big],\\
\label{eqn:case-gen-im1-0}
\bar{\Gamma}(\mathbf{q},z) &=   2c(1-c)^2\sigma q^2 \big[\mathrm{Im} f_1(z)-\mathrm{Im} f_2(z)\big].
\end{align}
\noindent
where we put $\mu_\mathbf{q} \sim q^2$  in the long-wavelength limit. The functions $\mathrm{Re} f_i(z)$ and $\mathrm{Im} f_i(z)$ are given by Eqs. \eqref{eqn:Re-fi}-\eqref{eqn:Im-fi}, which in turn are expressed in terms of $\Re \mathfrak{g}^0_i (z)= \mathrm{Re}\mathfrak{g}^0_i(z)$ and $\Im \mathfrak{g}^0_i(z) = \mathrm{Im}\mathfrak{g}^0_i(z)$, Eqs.\eqref {eqn:Re-base-green0-int-q}-\eqref{eqn:Im-base-green0-int-q}. For the case of the simple cubic lattice these relations  are 
\begin{align}
\label{eqn:Re-base-green0-int-q-0}
\Re \mathfrak{g}^0_1(z) &=  \int\frac{d^3p}{(2\pi)^3}\frac{p^2}{p^2-z},\\
\label{eqn:Im-base-green0-int-q-0}
\Im \mathfrak{g}^0_1(z) &=  \pi \int \frac{d^3p}{(2\pi)^3} p^2 \delta(p^2-z),
\end{align}
\noindent
\begin{equation}
\label{eqn:G10toG200}
\text{and} \quad \mathfrak{g}^0_2(\omega) =  -\frac{c}{1-c}\mathfrak{g}^0_1(\omega),
\end{equation}
\noindent
Let's evaluate first the integral \eqref{eqn:Im-base-green0-int-q-0}. We use spherical symmetry to write
\begin{equation}
\label{eqn:Im-base-green0-int-eval}
\Im \mathfrak{g}^0_1(z)\Bigg|_{z \to q^2} =  \pi \int \frac{dp}{(2\pi)^3} p^4 \delta(p^2-z)=\frac{q^3}{4\pi}.
\end{equation}
Evaluating next Eq. \eqref{eqn:Re-base-green0-int-q-0}, we take the shape of the Brillouin zone as a sphere of radius $\Lambda\approx \pi/a$
\begin{align}
\label{eqn:Re-base-green0-int-eval}
\Re \mathfrak{g}^0_1(z) &= 4\pi \int_0^\Lambda \frac{dp}{(2\pi)^3}\frac{p^4}{p^2-z},\nonumber\\
&= \frac{2}{(2\pi)^2} \Bigg[\frac{\Lambda^3}{3} + \Lambda z -z^{3/2} \tanh^{-1} \Big(\frac{\Lambda}{\sqrt{z}}\Big)\Bigg].
\end{align}
Hence
\begin{equation}
\label{eqn:Re-base-green0-int-eval2}
\Re \mathfrak{g}^0_1(z)\Bigg|_{z \to q^2} = \frac{2}{(2\pi)^2} \Big[\frac{\Lambda^3}{3} + \Lambda q^2 -O(q^3)  \Big].
\end{equation}
Furthermore, using relations \eqref{eqn:Im-base-green0-int-eval}-\eqref{eqn:Re-base-green0-int-eval2} in the $q\ll 1$ limit,  Eqs. \eqref{eqn:Re-fi}-\eqref{eqn:Im-fi} become
\begin{align}
\label{eqn:Re-fi-0}
\mathrm{Re} f_1 &= C_0+C_1 q^2 + O(q^4),\\
\mathrm{Im} f_1  &= \frac{q^3}{4\pi} + O(q^5),
\label{eqn:Im-fi-0}
\end{align}
where $C_0$ and $C_1$ are constants, and $f_2=-c f_1(1-c+f_1)^{-1}$. Finally, Eqs. \eqref{eqn:case-gen-re1-0}-\eqref{eqn:case-gen-im1-0}  with $z^\ast=q^2$ become
\begin{align}
\label{eqn:case-gen-re1-f}
\bar{\Sigma}^{\prime}(\mathbf{q},z^\ast) &= A c(1-c)^2\sigma q^2 + O(q^4),\\
\label{eqn:case-gen-im1-f}
\bar{\Gamma}(\mathbf{q},z^\ast) &=   Bc(1-c)^2\sigma q^5 + O(q^7).
\end{align}
Here, $A$ and $B$ are some $c$-dependent parameters.The decay rate of magnon is  $\lambda_q \sim \bar{\Gamma}(\mathbf{q},z^\ast)$.

\biboptions{sort&compress}
\bibliographystyle{model1-num-names}
\bibliography{spin}

\begin{thebibliography}{144}
\expandafter\ifx\csname natexlab\endcsname\relax\def\natexlab#1{#1}\fi
\providecommand{\bibinfo}[2]{#2}
\ifx\xfnm\relax \def\xfnm[#1]{\unskip,\space#1}\fi
\bibitem[{Bohrdt et~al.(2018)Bohrdt, J{\"a}gering, Eggert, and
  Schneider}]{bohrdt2018dynamic}
\bibinfo{author}{A.~Bohrdt}, \bibinfo{author}{K.~J{\"a}gering},
  \bibinfo{author}{S.~Eggert}, \bibinfo{author}{I.~Schneider},
\newblock \bibinfo{title}{Dynamic structure factor in impurity-doped spin
  chains},
\newblock \bibinfo{journal}{Phys. Rev. B} \bibinfo{volume}{98}
  (\bibinfo{year}{2018}) \bibinfo{pages}{020402}.
\bibitem[{Ohno(1998)}]{ohno1998making}
\bibinfo{author}{H.~Ohno},
\newblock \bibinfo{title}{Making nonmagnetic semiconductors ferromagnetic},
\newblock \bibinfo{journal}{Science} \bibinfo{volume}{281}
  (\bibinfo{year}{1998}) \bibinfo{pages}{951--956}.
\bibitem[{{\v{Z}}uti{\'c} et~al.(2004){\v{Z}}uti{\'c}, Fabian, and
  Sarma}]{vzutic2004spintronics}
\bibinfo{author}{I.~{\v{Z}}uti{\'c}}, \bibinfo{author}{J.~Fabian},
  \bibinfo{author}{S.~D. Sarma},
\newblock \bibinfo{title}{Spintronics: Fundamentals and applications},
\newblock \bibinfo{journal}{Rev. Mod. Phys.} \bibinfo{volume}{76}
  (\bibinfo{year}{2004}) \bibinfo{pages}{323--410}.
\bibitem[{MacDonald et~al.(2005)MacDonald, Schiffer, and
  Samarth}]{macdonald2005ferromagnetic}
\bibinfo{author}{A.~H. MacDonald}, \bibinfo{author}{R.~P. Schiffer},
  \bibinfo{author}{H.~N. Samarth},
\newblock \bibinfo{title}{Ferromagnetic semiconductors: moving beyond
  \mbox{(Ga, Mn)As}},
\newblock \bibinfo{journal}{Nature Materials} \bibinfo{volume}{4}
  (\bibinfo{year}{2005}) \bibinfo{pages}{195--202}.
\bibitem[{Barman et~al.(2021)Barman, Gubbiotti, Ladak et~al.}]{barman2021}
\bibinfo{author}{A.~Barman}, \bibinfo{author}{G.~Gubbiotti},
  \bibinfo{author}{S.~Ladak}, et~al.,
\newblock \bibinfo{title}{The 2021 magnonics roadmap},
\newblock \bibinfo{journal}{J. Phys.: Condensed Matter} \bibinfo{volume}{33}
  (\bibinfo{year}{2021}) \bibinfo{pages}{413001}.
\bibitem[{Chumak et~al.(2022)Chumak, Kabos, Wu, Abert
  et~al.}]{chumak2022roadmap}
\bibinfo{author}{A.~V. Chumak}, \bibinfo{author}{P.~Kabos},
  \bibinfo{author}{M.~Wu}, \bibinfo{author}{C.~Abert}, et~al.,
\newblock \bibinfo{title}{Roadmap on spin-wave computing},
\newblock \bibinfo{journal}{IEEE Transactions on Magnetics}
  \bibinfo{volume}{58} (\bibinfo{year}{2022}) \bibinfo{pages}{0800172}.
  \bibinfo{note}{Pp. 1--72}.
\bibitem[{Wolfram and Callaway(1963)}]{Wolfram_Callaway_1963}
\bibinfo{author}{T.~Wolfram}, \bibinfo{author}{J.~Callaway},
\newblock \bibinfo{title}{Spin-wave impurity states in ferromagnets},
\newblock \bibinfo{journal}{Phys. Rev.} \bibinfo{volume}{130}
  (\bibinfo{year}{1963}) \bibinfo{pages}{2207--2217}.
\bibitem[{Takeno(1963)}]{takeno1963spin}
\bibinfo{author}{S.~Takeno},
\newblock \bibinfo{title}{Spin--wave impurity levels in a \mbox{H}eisenberg
  ferromagnet},
\newblock \bibinfo{journal}{Prog. Theor. Phys.} \bibinfo{volume}{30}
  (\bibinfo{year}{1963}) \bibinfo{pages}{731--742}.
\bibitem[{Izyumov(1966{\natexlab{a}})}]{izyumov1966spin}
\bibinfo{author}{Y.~A. Izyumov},
\newblock \bibinfo{title}{Spin-wave theory of ferromagnetic crystals containing
  impurities},
\newblock \bibinfo{journal}{Proc. Phys. Soc.} \bibinfo{volume}{87}
  (\bibinfo{year}{1966}{\natexlab{a}}) \bibinfo{pages}{505--519}.
\bibitem[{Izyumov(1966{\natexlab{b}})}]{izyumov1966theory}
\bibinfo{author}{Y.~Izyumov},
\newblock \bibinfo{title}{The theory of inelastic scattering of neutrons in
  ferromagnetic crystals containing impurities},
\newblock \bibinfo{journal}{Proc. Phys. Soc.} \bibinfo{volume}{87}
  (\bibinfo{year}{1966}{\natexlab{b}}) \bibinfo{pages}{521--534}.
\bibitem[{Murray(1966{\natexlab{a}})}]{murray1966low}
\bibinfo{author}{G.~A. Murray},
\newblock \bibinfo{title}{Low-energy spin waves for impure \mbox{H}eisenberg
  ferromagnets},
\newblock \bibinfo{journal}{Proc. Phys. Soc.} \bibinfo{volume}{89}
  (\bibinfo{year}{1966}{\natexlab{a}}) \bibinfo{pages}{87--110}.
\bibitem[{Murray(1966{\natexlab{b}})}]{murray1966determination}
\bibinfo{author}{G.~A. Murray},
\newblock \bibinfo{title}{The determination of the critical concentration for a
  dilute \mbox{H}eisenberg ferromagnet from the low-energy spin waves},
\newblock \bibinfo{journal}{Proc. Phys. Soc.} \bibinfo{volume}{89}
  (\bibinfo{year}{1966}{\natexlab{b}}) \bibinfo{pages}{111--118}.
\bibitem[{Hone et~al.(1966)Hone, Callen, and Walker}]{Hone_et_al_1966}
\bibinfo{author}{D.~Hone}, \bibinfo{author}{H.~Callen}, \bibinfo{author}{L.~R.
  Walker},
\newblock \bibinfo{title}{Thermodynamic properties of spin impurities in
  \mbox{H}eisenberg ferromagnets},
\newblock \bibinfo{journal}{Phys. Rev.} \bibinfo{volume}{144}
  (\bibinfo{year}{1966}) \bibinfo{pages}{283--295}.
\bibitem[{Takeno and Homma(1968)}]{takeno1968low}
\bibinfo{author}{S.~Takeno}, \bibinfo{author}{S.~Homma},
\newblock \bibinfo{title}{A low-lying magnon resonant mode in an impure
  \mbox{H}eisenberg ferromagnet},
\newblock \bibinfo{journal}{Prog. Theor. Phys.} \bibinfo{volume}{40}
  (\bibinfo{year}{1968}) \bibinfo{pages}{452--461}.
\bibitem[{Kaneyoshi(1969)}]{kaneyoshi1969contribution}
\bibinfo{author}{T.~Kaneyoshi},
\newblock \bibinfo{title}{Contribution to the spin-wave theory of a dilute
  \mbox{H}eisenberg ferromagnet},
\newblock \bibinfo{journal}{Prog. Theor. Phys.} \bibinfo{volume}{42}
  (\bibinfo{year}{1969}) \bibinfo{pages}{477--493}.
\bibitem[{Edwards and Jones(1971)}]{edwards1971green}
\bibinfo{author}{S.~F. Edwards}, \bibinfo{author}{R.~C. Jones},
\newblock \bibinfo{title}{A \mbox{G}reen function theory of spin waves in
  randomly disordered magnetic systems. \mbox{I}. \mbox{T}he ferromagnet},
\newblock \bibinfo{journal}{J. Phys. C: Solid State Phys.} \bibinfo{volume}{4}
  (\bibinfo{year}{1971}) \bibinfo{pages}{2109--2126}.
\bibitem[{Foo and Wu(1972)}]{foo1972coherent}
\bibinfo{author}{E.-N. Foo}, \bibinfo{author}{D.-H. Wu},
\newblock \bibinfo{title}{Coherent-potential approximation for disordered
  ferromagnetic binary alloys},
\newblock \bibinfo{journal}{Phys. Rev. B} \bibinfo{volume}{5}
  (\bibinfo{year}{1972}) \bibinfo{pages}{98}.
\bibitem[{Tahir-Kheli(1972)}]{tahir1972spatially}
\bibinfo{author}{R.~A. Tahir-Kheli},
\newblock \bibinfo{title}{Spatially random \mbox{H}eisenberg spins at very low
  temperatures. \mbox{I. D}ilute ferromagnet},
\newblock \bibinfo{journal}{Phys. Rev. B} \bibinfo{volume}{6}
  (\bibinfo{year}{1972}) \bibinfo{pages}{2808--2825}.
\bibitem[{Matsubara(1973)}]{matsubara1973application}
\bibinfo{author}{T.~Matsubara},
\newblock \bibinfo{title}{An application of \mbox{CPA} to a random spin
  system},
\newblock \bibinfo{journal}{Prog. Theor. Phys. Suppl.} \bibinfo{volume}{53}
  (\bibinfo{year}{1973}) \bibinfo{pages}{202--221}.
\bibitem[{Tahir-Kheli(1977)}]{tahir1977effects}
\bibinfo{author}{R.~A. Tahir-Kheli},
\newblock \bibinfo{title}{Effects of randomness on three-dimensional magnetic
  ordering of quasi-low-dimensional spin systems},
\newblock \bibinfo{journal}{Phys. Rev. B} \bibinfo{volume}{16}
  (\bibinfo{year}{1977}) \bibinfo{pages}{2239--2253}.
\bibitem[{Dvey-Aharon and Fibich(1978)}]{dvey1978magnetic}
\bibinfo{author}{H.~Dvey-Aharon}, \bibinfo{author}{M.~Fibich},
\newblock \bibinfo{title}{Magnetic properties of a disordered substitutional
  alloy--\mbox{H}eisenberg model},
\newblock \bibinfo{journal}{Phys. Rev. B} \bibinfo{volume}{18}
  (\bibinfo{year}{1978}) \bibinfo{pages}{3491--3506}.
\bibitem[{Cowley and Buyers(1972)}]{cowley1972properties}
\bibinfo{author}{R.~A. Cowley}, \bibinfo{author}{W.~J.~L. Buyers},
\newblock \bibinfo{title}{The properties of defects in magnetic insulators},
\newblock \bibinfo{journal}{Rev. Mod. Phys.} \bibinfo{volume}{44}
  (\bibinfo{year}{1972}) \bibinfo{pages}{406--450}.
\bibitem[{Coombs and Cowley(1975)}]{coombs1975magnetic}
\bibinfo{author}{G.~J. Coombs}, \bibinfo{author}{R.~A. Cowley},
\newblock \bibinfo{title}{Magnetic excitations in two-dimensional random
  antiferromagnets},
\newblock \bibinfo{journal}{J. Phys. C: Solid State Physics}
  \bibinfo{volume}{8} (\bibinfo{year}{1975}) \bibinfo{pages}{1889--1900}.
\bibitem[{Korenblit and Shender(1978)}]{korenblit1978ferromagnetism}
\bibinfo{author}{I.~Y. Korenblit}, \bibinfo{author}{E.~F. Shender},
\newblock \bibinfo{title}{Ferromagnetism of disordered systems},
\newblock \bibinfo{journal}{Sov. Phys. Uspekhi} \bibinfo{volume}{21}
  (\bibinfo{year}{1978}) \bibinfo{pages}{832--851}.
\bibitem[{Wagner and Krey(1978)}]{wagner1978spin}
\bibinfo{author}{V.~Wagner}, \bibinfo{author}{U.~Krey},
\newblock \bibinfo{title}{Spin wave excitations in the dilute two-dimensional
  ferromagnet \mbox{K$_2$Cu$_{1-x}$Z$_{nx}$F$_4$}},
\newblock \bibinfo{journal}{Zeitschrift f{\"u}r Physik B Cond. Matter}
  \bibinfo{volume}{30} (\bibinfo{year}{1978}) \bibinfo{pages}{367--373}.
\bibitem[{Wagner and Krey(1979)}]{wagner1979spin}
\bibinfo{author}{V.~Wagner}, \bibinfo{author}{U.~Krey},
\newblock \bibinfo{title}{Spin correlations of a dilute two-dimensional
  \mbox{H}eisenberg ferromagnet near the percolation limit},
\newblock \bibinfo{journal}{J. Magnetism \& Magnetic Mater.}
  \bibinfo{volume}{14} (\bibinfo{year}{1979}) \bibinfo{pages}{169--171}.
\bibitem[{Da~Silva et~al.(2010)Da~Silva, Nakagomi, Silva, Franco~Jr, Garg,
  Oliveira, and Morais}]{daSilva2010effect}
\bibinfo{author}{S.~W. Da~Silva}, \bibinfo{author}{F.~Nakagomi},
  \bibinfo{author}{M.~S. Silva}, \bibinfo{author}{A.~Franco~Jr},
  \bibinfo{author}{V.~K. Garg}, \bibinfo{author}{A.~C. Oliveira},
  \bibinfo{author}{P.~C. Morais},
\newblock \bibinfo{title}{Effect of the \mbox{Zn} content in the structural and
  magnetic properties of \mbox{Mg$_{1-x}$Zn$_x$Fe$_2$O$_4$} mixed ferrites
  monitored by \mbox{R}aman and \mbox{M}{\"o}ssbauer spectroscopies},
\newblock \bibinfo{journal}{J. Appl. Phys.} \bibinfo{volume}{107}
  (\bibinfo{year}{2010}) \bibinfo{pages}{09B503}.
\bibitem[{Saqib et~al.(2019)Saqib, Rahman, Susilo, Chen, and
  Dai}]{saqib2019structural}
\bibinfo{author}{H.~Saqib}, \bibinfo{author}{S.~Rahman},
  \bibinfo{author}{R.~Susilo}, \bibinfo{author}{B.~Chen},
  \bibinfo{author}{N.~Dai},
\newblock \bibinfo{title}{Structural, vibrational, electrical, and magnetic
  properties of mixed spinel ferrites \mbox{Mg$_{1-x}$Zn$_x$Fe$_2$O$_4$}
  nanoparticles prepared by coprecipitation},
\newblock \bibinfo{journal}{AIP Advances} \bibinfo{volume}{9}
  (\bibinfo{year}{2019}) \bibinfo{pages}{055306}.
\bibitem[{Matsubara(1955)}]{Matsubara_1955}
\bibinfo{author}{T.~Matsubara},
\newblock \bibinfo{title}{A new approach to quantum statistical mechanics},
\newblock \bibinfo{journal}{Prog. Theor. Phys.} \bibinfo{volume}{14}
  (\bibinfo{year}{1955}) \bibinfo{pages}{351--378}.
\bibitem[{Abrikosov et~al.(1963)Abrikosov, Gorkov, and
  Dzyaloshinski}]{AGD_1963}
\bibinfo{author}{A.~A. Abrikosov}, \bibinfo{author}{L.~P. Gorkov},
  \bibinfo{author}{I.~E. Dzyaloshinski}, \bibinfo{title}{Methods of Quantum
  Field Theory in Statistical Physics}, \bibinfo{publisher}{Prentice-Hall,
  Inc.}, \bibinfo{address}{Englewood Cliffs, New Jersey}, \bibinfo{year}{1963}.
  \bibinfo{note}{See section 15}.
\bibitem[{Coleman(2015)}]{Coleman_2015}
\bibinfo{author}{P.~Coleman}, \bibinfo{title}{Introduction to Many-Body
  Physics}, \bibinfo{publisher}{Cambridge University Press},
  \bibinfo{address}{Cambridge}, \bibinfo{year}{2015}. \bibinfo{note}{Chap. 7}.
\bibitem[{Bogolyubov and Tyablikov(1959)}]{Bogolyubov_1959}
\bibinfo{author}{N.~N. Bogolyubov}, \bibinfo{author}{S.~V. Tyablikov},
\newblock \bibinfo{title}{Retarded and advanced \mbox{G}reen functions in
  statistical physics},
\newblock in: \bibinfo{booktitle}{Soviet Physics Doklady},
  volume~\bibinfo{volume}{4}, pp. \bibinfo{pages}{589--604}.
\bibitem[{Bonch-Bruevich and Tyablikov(1962)}]{Bonch-Bruevich_1962}
\bibinfo{author}{V.~L. Bonch-Bruevich}, \bibinfo{author}{S.~V. Tyablikov},
  \bibinfo{title}{The Green Function Method in Statistical Mechanics},
  \bibinfo{publisher}{North-Holland}, \bibinfo{address}{Amsterdam},
  \bibinfo{year}{1962}. \bibinfo{note}{Dover edition 2015}.
\bibitem[{Zubarev(1960)}]{Zubarev_1960}
\bibinfo{author}{D.~N. Zubarev},
\newblock \bibinfo{title}{Double-time green functions in statistical physics},
\newblock \bibinfo{journal}{Sov. Phys. Uspekhi} \bibinfo{volume}{3}
  (\bibinfo{year}{1960}) \bibinfo{pages}{320--345}.
\bibitem[{Rudoy(2011)}]{rudoy2011bogoliubov}
\bibinfo{author}{Y.~G. Rudoy},
\newblock \bibinfo{title}{The \mbox{B}ogoliubov-\mbox{T}yablikov \mbox{G}reen's
  function method in the quantum theory of magnetism},
\newblock \bibinfo{journal}{Theor. Math. Phys.} \bibinfo{volume}{168}
  (\bibinfo{year}{2011}) \bibinfo{pages}{1318--1329}.
\bibitem[{Kubo(1957)}]{Kubo_1957}
\bibinfo{author}{R.~Kubo},
\newblock \bibinfo{title}{Statistical-mechanical theory of irreversible
  processes. \mbox{I}. general theory and simple applications to magnetic and
  conduction problems},
\newblock \bibinfo{journal}{J. Phys. Soc. Japan} \bibinfo{volume}{12}
  (\bibinfo{year}{1957}) \bibinfo{pages}{570--586}.
\bibitem[{Oguchi and Honma(1963)}]{Oguchi_Honma_1963}
\bibinfo{author}{T.~Oguchi}, \bibinfo{author}{A.~Honma},
\newblock \bibinfo{title}{Theory of ferro- and antiferromagnetism by the method
  of \mbox{G}reen functions},
\newblock \bibinfo{journal}{J. Appl. Phys.} \bibinfo{volume}{34}
  (\bibinfo{year}{1963}) \bibinfo{pages}{1153--1160}.
\bibitem[{Callen(1963)}]{Callen_1963}
\bibinfo{author}{H.~B. Callen},
\newblock \bibinfo{title}{Green function theory of ferromagnetism},
\newblock \bibinfo{journal}{Phys. Rev.} \bibinfo{volume}{130}
  (\bibinfo{year}{1963}) \bibinfo{pages}{890--898}.
\bibitem[{Tahir-Kheli and Callen(1964)}]{Tahir-Kheli_1964}
\bibinfo{author}{R.~A. Tahir-Kheli}, \bibinfo{author}{H.~B. Callen},
\newblock \bibinfo{title}{Remarks on the \mbox{O}guchi transformation},
\newblock \bibinfo{journal}{J. Appl. Phys.} \bibinfo{volume}{35}
  (\bibinfo{year}{1964}) \bibinfo{pages}{948--949}.
\bibitem[{Tyablikov(1967)}]{Tyablikov_1967}
\bibinfo{author}{S.~V. Tyablikov}, \bibinfo{title}{Methods in the Quantum
  Theory of Magnetism}, \bibinfo{publisher}{Plenum Press},
  \bibinfo{address}{New York}, \bibinfo{year}{1967}.
\bibitem[{Nolting and Ramakanth(2009)}]{nolting2009quantum}
\bibinfo{author}{W.~Nolting}, \bibinfo{author}{A.~Ramakanth},
  \bibinfo{title}{Quantum Theory of Magnetism}, \bibinfo{publisher}{Springer
  Science \& Business Media}, \bibinfo{year}{2009}.
\bibitem[{Brout(1959)}]{brout1959statistical}
\bibinfo{author}{R.~Brout},
\newblock \bibinfo{title}{Statistical mechanical theory of a random
  ferromagnetic system},
\newblock \bibinfo{journal}{Phys. Rev.} \bibinfo{volume}{115}
  (\bibinfo{year}{1959}) \bibinfo{pages}{824}.
\bibitem[{Elliott(1960)}]{elliott1960some}
\bibinfo{author}{R.~J. Elliott},
\newblock \bibinfo{title}{Some properties of concentrated and dilute
  \mbox{H}eisenberg magnets with general spin},
\newblock \bibinfo{journal}{J. Phys. Chem. Solids} \bibinfo{volume}{16}
  (\bibinfo{year}{1960}) \bibinfo{pages}{165--168}.
\bibitem[{Smart(1960)}]{smart1960behavior}
\bibinfo{author}{J.~S. Smart},
\newblock \bibinfo{title}{The behavior of magnetic systems with dilution},
\newblock \bibinfo{journal}{J. Phys. Chem. Solids} \bibinfo{volume}{16}
  (\bibinfo{year}{1960}) \bibinfo{pages}{169--173}.
\bibitem[{Charap(1962)}]{charap1962critical}
\bibinfo{author}{S.~H. Charap},
\newblock \bibinfo{title}{Critical concentrations in magnetism},
\newblock \bibinfo{journal}{Phys. Rev.} \bibinfo{volume}{126}
  (\bibinfo{year}{1962}) \bibinfo{pages}{1393--1394}.
\bibitem[{Elliott and Heap(1962)}]{elliott1962theory}
\bibinfo{author}{R.~J. Elliott}, \bibinfo{author}{B.~R. Heap},
\newblock \bibinfo{title}{Theory of random dilute magnets with application to
  \mbox{MnZnF$_2$}},
\newblock \bibinfo{journal}{Proc. Roy. Soc. A} \bibinfo{volume}{265}
  (\bibinfo{year}{1962}) \bibinfo{pages}{264--283}.
\bibitem[{Izyumov and Medvedev(1965{\natexlab{a}})}]{izyumov1965impurity}
\bibinfo{author}{Y.~A. Izyumov}, \bibinfo{author}{M.~V. Medvedev},
\newblock \bibinfo{title}{Impurity atoms in a ferromagnetic crystal},
\newblock \bibinfo{journal}{Sov. Phys. JETP} \bibinfo{volume}{21}
  (\bibinfo{year}{1965}{\natexlab{a}}) \bibinfo{pages}{381--388}.
\bibitem[{Izyumov and Medvedev(1965{\natexlab{b}})}]{izyumov1965properties}
\bibinfo{author}{Y.~A. Izyumov}, \bibinfo{author}{M.~V. Medvedev},
\newblock \bibinfo{title}{Some properties of a ferromagnetic crystal containing
  magnetic impurity atoms},
\newblock \bibinfo{journal}{Sov. Phys. JETP} \bibinfo{volume}{21}
  (\bibinfo{year}{1965}{\natexlab{b}}) \bibinfo{pages}{1155--1166}.
\bibitem[{Izyumov and Medvedev(1966)}]{izyumov1966peculiarities}
\bibinfo{author}{Y.~A. Izyumov}, \bibinfo{author}{M.~V. Medvedev},
\newblock \bibinfo{title}{Peculiarities of the spin wave spectrum of a
  ferromagnet containing impurities and the temperature dependence of
  spontaneous magnetization},
\newblock \bibinfo{journal}{Sov. Phys. JETP} \bibinfo{volume}{22}
  (\bibinfo{year}{1966}) \bibinfo{pages}{1289--1293}.
\bibitem[{Izyumov and Medvedev(1967)}]{izyumov1967incoherent}
\bibinfo{author}{Y.~A. Izyumov}, \bibinfo{author}{M.~V. Medvedev},
\newblock \bibinfo{title}{Incoherent scattering of neutrons and the problem of
  reconstructing the magnon spectrum},
\newblock \bibinfo{journal}{Sov. Phys. JETP} \bibinfo{volume}{24}
  (\bibinfo{year}{1967}) \bibinfo{pages}{960--964}.
\bibitem[{Izyumov and Medvedev(1973)}]{izyumov1973magnetically}
\bibinfo{author}{Y.~A. Izyumov}, \bibinfo{author}{M.~V. Medvedev},
  \bibinfo{title}{Magnetically Ordered Crystals Containing Impurities},
  \bibinfo{publisher}{Consultants Bureau New York/London},
  \bibinfo{year}{1973}. \bibinfo{note}{A Special Research Report}.
\bibitem[{Last(1972)}]{last1972percolation}
\bibinfo{author}{B.~J. Last},
\newblock \bibinfo{title}{Percolation theory and the critical concentration of
  a dilute \mbox{H}eisenberg ferromagnet},
\newblock \bibinfo{journal}{J. Phys. C: Solid State Phys.} \bibinfo{volume}{5}
  (\bibinfo{year}{1972}) \bibinfo{pages}{2805--2812}.
\bibitem[{Kumar and Harris(1973)}]{kumar1973spin}
\bibinfo{author}{D.~Kumar}, \bibinfo{author}{A.~B. Harris},
\newblock \bibinfo{title}{Spin waves and long-range order in the diluted
  \mbox{H}eisenberg ferromagnet at zero temperature},
\newblock \bibinfo{journal}{Phys. Rev. B} \bibinfo{volume}{8}
  (\bibinfo{year}{1973}) \bibinfo{pages}{2166--2184}.
\bibitem[{Kaneyoshi(1970)}]{kaneyoshi1970contribution}
\bibinfo{author}{T.~Kaneyoshi},
\newblock \bibinfo{title}{Contribution to the theory of a dilute
  \mbox{H}eisenberg ferromagnet},
\newblock \bibinfo{journal}{Prog. Theor. Phys.} \bibinfo{volume}{44}
  (\bibinfo{year}{1970}) \bibinfo{pages}{328--338}.
\bibitem[{Sykes and Essam(1964)}]{sykes1964critical}
\bibinfo{author}{M.~F. Sykes}, \bibinfo{author}{J.~W. Essam},
\newblock \bibinfo{title}{Critical percolation probabilities by series
  methods},
\newblock \bibinfo{journal}{Phys. Rev.} \bibinfo{volume}{133}
  (\bibinfo{year}{1964}) \bibinfo{pages}{A310--A315}.
\bibitem[{Kirkpatrick(1973{\natexlab{a}})}]{Kirkpatrick1973nature}
\bibinfo{author}{S.~Kirkpatrick},
\newblock \bibinfo{title}{The nature of percolation channels},
\newblock \bibinfo{journal}{Solid State Commun.} \bibinfo{volume}{12}
  (\bibinfo{year}{1973}{\natexlab{a}}) \bibinfo{pages}{1279--1283}.
\bibitem[{Kirkpatrick(1973{\natexlab{b}})}]{kirkpatrick1973percolation}
\bibinfo{author}{S.~Kirkpatrick},
\newblock \bibinfo{title}{Percolation and conduction},
\newblock \bibinfo{journal}{Rev. Mod. Phys.} \bibinfo{volume}{45}
  (\bibinfo{year}{1973}{\natexlab{b}}) \bibinfo{pages}{574--588}.
\bibitem[{Shante and Kirkpatrick(1971)}]{shante1971introduction}
\bibinfo{author}{V.~K.~S. Shante}, \bibinfo{author}{S.~Kirkpatrick},
\newblock \bibinfo{title}{An introduction to percolation theory},
\newblock \bibinfo{journal}{Advances in Physics} \bibinfo{volume}{20}
  (\bibinfo{year}{1971}) \bibinfo{pages}{325--357}.
\bibitem[{Takeno(1968)}]{takeno1968solution}
\bibinfo{author}{S.~Takeno},
\newblock \bibinfo{title}{A solution of a \mbox{D}yson equation in
  many-impurity problems in solids},
\newblock \bibinfo{journal}{Prog. Theor. Phys.} \bibinfo{volume}{40}
  (\bibinfo{year}{1968}) \bibinfo{pages}{942--957}.
\bibitem[{Elliott et~al.(1974)Elliott, Krumhansl, and
  Leath}]{elliott1974theory}
\bibinfo{author}{R.~J. Elliott}, \bibinfo{author}{J.~A. Krumhansl},
  \bibinfo{author}{P.~L. Leath},
\newblock \bibinfo{title}{The theory and properties of randomly disordered
  crystals and related physical systems},
\newblock \bibinfo{journal}{Rev. Mod. Phys.} \bibinfo{volume}{46}
  (\bibinfo{year}{1974}) \bibinfo{pages}{465--543}.
\bibitem[{Rickayzen(2013)}]{Rickayzen_1984}
\bibinfo{author}{G.~Rickayzen}, \bibinfo{title}{Green's Functions and Condensed
  Matter}, \bibinfo{publisher}{Dover Publications}, \bibinfo{address}{New
  York}, \bibinfo{year}{2013}. \bibinfo{note}{Chap. 11}.
\bibitem[{Yonezawa and Morigaki(1973)}]{yonezawa1973coherent}
\bibinfo{author}{F.~Yonezawa}, \bibinfo{author}{K.~Morigaki},
\newblock \bibinfo{title}{Coherent potential approximation. \mbox{B}asic
  concepts and applications},
\newblock \bibinfo{journal}{Prog. Theor. Phys. Suppl.} \bibinfo{volume}{53}
  (\bibinfo{year}{1973}) \bibinfo{pages}{1--76}.
\bibitem[{Nolting(2009)}]{nolting2008fundamentals}
\bibinfo{author}{W.~Nolting}, \bibinfo{title}{Fundamentals of Many-body
  Physics}, \bibinfo{publisher}{Springer}, \bibinfo{address}{Berlin},
  \bibinfo{year}{2009}. \bibinfo{note}{Translator: W. D. Brewer}.
\bibitem[{Elliott and Pepper(1973)}]{elliott1973excitations}
\bibinfo{author}{R.~J. Elliott}, \bibinfo{author}{D.~E. Pepper},
\newblock \bibinfo{title}{Excitations in dilute magnets using the
  coherent-potential approximation},
\newblock \bibinfo{journal}{Phys. Rev. B} \bibinfo{volume}{8}
  (\bibinfo{year}{1973}) \bibinfo{pages}{2374--2378}.
\bibitem[{Pepper(1972)}]{pepper1972}
\bibinfo{author}{D.~E. Pepper}, \bibinfo{title}{Excitations in impure magnetic
  crystals}, Ph.D. thesis, Oxford University, \bibinfo{address}{Oxford, UK},
  \bibinfo{year}{1972}.
\bibitem[{Harris et~al.(1974)Harris, Leath, Nickel, and
  Elliott}]{harris1974excitations}
\bibinfo{author}{A.~B. Harris}, \bibinfo{author}{P.~L. Leath},
  \bibinfo{author}{B.~G. Nickel}, \bibinfo{author}{R.~J. Elliott},
\newblock \bibinfo{title}{Excitations in the dilute \mbox{H}eisenberg
  ferromagnet using the coherent potential approximation},
\newblock \bibinfo{journal}{J. Phys. C: Solid State Physics}
  \bibinfo{volume}{7} (\bibinfo{year}{1974}) \bibinfo{pages}{1693--1718}.
\bibitem[{Nickel(1974)}]{nickel1974method}
\bibinfo{author}{B.~G. Nickel},
\newblock \bibinfo{title}{A method of moments applied to the diluted
  ferromagnet},
\newblock \bibinfo{journal}{J. Phys. C: Solid State Physics}
  \bibinfo{volume}{7} (\bibinfo{year}{1974}) \bibinfo{pages}{1719--1734}.
\bibitem[{Theumann and Tahir-Kheli(1975)}]{theumann1975excitations}
\bibinfo{author}{A.~Theumann}, \bibinfo{author}{R.~A. Tahir-Kheli},
\newblock \bibinfo{title}{Excitations in randomly diluted ferromagnets},
\newblock \bibinfo{journal}{Phys. Rev. B} \bibinfo{volume}{12}
  (\bibinfo{year}{1975}) \bibinfo{pages}{1796--1818}.
\bibitem[{Brouers and Ducastelle(1975)}]{brouers1975theories}
\bibinfo{author}{F.~Brouers}, \bibinfo{author}{F.~Ducastelle},
\newblock \bibinfo{title}{On the theories of local environment effect in
  disordered alloys},
\newblock \bibinfo{journal}{J. Phys. F: Metal Physics} \bibinfo{volume}{5}
  (\bibinfo{year}{1975}) \bibinfo{pages}{45--55}.
\bibitem[{Alben et~al.(1977)Alben, Kirkpatrick, and Beeman}]{alben1977spin}
\bibinfo{author}{R.~Alben}, \bibinfo{author}{S.~Kirkpatrick},
  \bibinfo{author}{D.~Beeman},
\newblock \bibinfo{title}{Spin waves in random ferromagnets},
\newblock \bibinfo{journal}{Phys. Rev. B} \bibinfo{volume}{15}
  (\bibinfo{year}{1977}) \bibinfo{pages}{346--357}.
\bibitem[{Salzberg et~al.(1976)Salzberg, Da~Silva, and
  Falicov}]{salzberg1976spin}
\bibinfo{author}{J.~B. Salzberg}, \bibinfo{author}{C.~E. T.~G. Da~Silva},
  \bibinfo{author}{L.~M. Falicov},
\newblock \bibinfo{title}{Spin waves in dilute ferromagnets:
  Cluster-\mbox{B}ethe-lattice approach},
\newblock \bibinfo{journal}{Phys. Rev. B} \bibinfo{volume}{14}
  (\bibinfo{year}{1976}) \bibinfo{pages}{1314--1322}.
\bibitem[{Katsura and Takizawa(1974)}]{katsura1974bethe}
\bibinfo{author}{S.~Katsura}, \bibinfo{author}{M.~Takizawa},
\newblock \bibinfo{title}{Bethe lattice and the \mbox{B}ethe approximation},
\newblock \bibinfo{journal}{Prog. Theor. Phys.} \bibinfo{volume}{51}
  (\bibinfo{year}{1974}) \bibinfo{pages}{82--98}. \bibinfo{note}{And references
  therein}.
\bibitem[{Blackman et~al.(1971)Blackman, Esterling, and
  Berk}]{blackman1971generalized}
\bibinfo{author}{J.~A. Blackman}, \bibinfo{author}{D.~M. Esterling},
  \bibinfo{author}{N.~F. Berk},
\newblock \bibinfo{title}{Generalized locator-coherent-potential approach to
  binary alloys},
\newblock \bibinfo{journal}{Phys. Rev. B} \bibinfo{volume}{4}
  (\bibinfo{year}{1971}) \bibinfo{pages}{2412--2428}.
\bibitem[{Theumann(1974)}]{theumann1974generalized}
\bibinfo{author}{A.~Theumann},
\newblock \bibinfo{title}{Generalized \mbox{CPA} approach to the disordered
  \mbox{H}eisenberg ferromagnet},
\newblock \bibinfo{journal}{J. Phys. C: Solid State Phys.} \bibinfo{volume}{7}
  (\bibinfo{year}{1974}) \bibinfo{pages}{2328--2346}.
\bibitem[{Lage and Stinchcombe(1977)}]{lage1977generalized}
\bibinfo{author}{E.~J.~S. Lage}, \bibinfo{author}{R.~B. Stinchcombe},
\newblock \bibinfo{title}{A generalized coherent-potential approximation for
  site-disordered spin systems},
\newblock \bibinfo{journal}{J. Phys. C: Solid State Phys.} \bibinfo{volume}{10}
  (\bibinfo{year}{1977}) \bibinfo{pages}{295--312}.
\bibitem[{Stinchcombe(1979)}]{stinchcombe1979critical}
\bibinfo{author}{R.~B. Stinchcombe},
\newblock \bibinfo{title}{Critical properties of dilute \mbox{H}eisenberg and
  \mbox{I}sing magnets},
\newblock \bibinfo{journal}{J. Phys. C: Solid State Phys.} \bibinfo{volume}{12}
  (\bibinfo{year}{1979}) \bibinfo{pages}{4533--4552}.
\bibitem[{McGurn and Thorpe(1983)}]{mcgurn1983spin}
\bibinfo{author}{A.~R. McGurn}, \bibinfo{author}{M.~F. Thorpe},
\newblock \bibinfo{title}{Spin wave theory of dilute one-dimensional magnets},
\newblock \bibinfo{journal}{J. Phys. C: Solid State Physics}
  \bibinfo{volume}{16} (\bibinfo{year}{1983}) \bibinfo{pages}{1255--1269}.
\bibitem[{Rettori and Pini(1988)}]{rettori1988effect}
\bibinfo{author}{A.~Rettori}, \bibinfo{author}{M.~G. Pini},
\newblock \bibinfo{title}{Effect of a field on the thermodynamics of the
  randomly dilute \mbox{1D H}eisenberg ferromagnet},
\newblock \bibinfo{journal}{Le Journal de Physique Colloques}
  \bibinfo{volume}{49} (\bibinfo{year}{1988})
  \bibinfo{pages}{C8--1417--C8--1418}.
\bibitem[{Hilbert and Nolting(2004)}]{hilbert2004disorder}
\bibinfo{author}{S.~Hilbert}, \bibinfo{author}{W.~Nolting},
\newblock \bibinfo{title}{Disorder in diluted spin systems},
\newblock \bibinfo{journal}{Phys. Rev. B} \bibinfo{volume}{70}
  (\bibinfo{year}{2004}) \bibinfo{pages}{165203}.
\bibitem[{Reed et~al.(2001)Reed, El-Masry, Stadelmaier, Ritums
  et~al.}]{reed2001room}
\bibinfo{author}{M.~L. Reed}, \bibinfo{author}{N.~A. El-Masry},
  \bibinfo{author}{H.~H. Stadelmaier}, \bibinfo{author}{M.~K. Ritums}, et~al.,
\newblock \bibinfo{title}{Room temperature ferromagnetic properties of
  (\mbox{Ga, Mn)N}},
\newblock \bibinfo{journal}{Appl. Phys. Lett.} \bibinfo{volume}{79}
  (\bibinfo{year}{2001}) \bibinfo{pages}{3473--3475}.
\bibitem[{Park et~al.(2002)Park, Hanbicki, Erwin et~al.}]{park2002group}
\bibinfo{author}{Y.~D. Park}, \bibinfo{author}{A.~T. Hanbicki},
  \bibinfo{author}{S.~C. Erwin}, et~al.,
\newblock \bibinfo{title}{A group-\mbox{IV} ferromagnetic semiconductor:
  \mbox{Mn$_x$Ge$_{1- x}$}},
\newblock \bibinfo{journal}{Science} \bibinfo{volume}{295}
  (\bibinfo{year}{2002}) \bibinfo{pages}{651--654}.
\bibitem[{Tang and Nolting(2007)}]{tang2007effects}
\bibinfo{author}{G.~Tang}, \bibinfo{author}{W.~Nolting},
\newblock \bibinfo{title}{Effects of dilution and disorder on magnetism in
  diluted spin systems},
\newblock \bibinfo{journal}{physica status solidi (b)} \bibinfo{volume}{244}
  (\bibinfo{year}{2007}) \bibinfo{pages}{735--747}.
\bibitem[{Martin et~al.(2016)Martin, Reining, and
  Ceperley}]{martin2016interacting}
\bibinfo{author}{R.~M. Martin}, \bibinfo{author}{L.~Reining},
  \bibinfo{author}{D.~M. Ceperley}, \bibinfo{title}{Interacting electrons},
  \bibinfo{publisher}{Cambridge University Press}, \bibinfo{year}{2016}.
\bibitem[{Mookerjee(1973{\natexlab{a}})}]{mookerjee1973averaged}
\bibinfo{author}{A.~Mookerjee},
\newblock \bibinfo{title}{Averaged density of states in disordered systems},
\newblock \bibinfo{journal}{J. Phys. C: Solid State Phys.} \bibinfo{volume}{6}
  (\bibinfo{year}{1973}{\natexlab{a}}) \bibinfo{pages}{1340--1349}.
\bibitem[{Mookerjee(1973{\natexlab{b}})}]{mookerjee1973new}
\bibinfo{author}{A.~Mookerjee},
\newblock \bibinfo{title}{A new formalism for the study of
  configuration-averaged properties of disordered systems},
\newblock \bibinfo{journal}{J. Phys. C: Solid State Phys.} \bibinfo{volume}{6}
  (\bibinfo{year}{1973}{\natexlab{b}}) \bibinfo{pages}{L205--L208}.
\bibitem[{Bouzerar and Bruno(2002)}]{bouzerar2002rpa}
\bibinfo{author}{G.~Bouzerar}, \bibinfo{author}{P.~Bruno},
\newblock \bibinfo{title}{\mbox{RPA-CPA} theory for magnetism in disordered
  \mbox{H}eisenberg binary systems with long-range exchange integrals},
\newblock \bibinfo{journal}{Phys. Rev. B} \bibinfo{volume}{66}
  (\bibinfo{year}{2002}) \bibinfo{pages}{014410}.
\bibitem[{Whitelaw(1981)}]{whitelaw1981single}
\bibinfo{author}{D.~J. Whitelaw},
\newblock \bibinfo{title}{A single-site theory for binary alloys with
  environmental and off-diagonal disorder},
\newblock \bibinfo{journal}{J. Phys. C: Solid State Phys.} \bibinfo{volume}{14}
  (\bibinfo{year}{1981}) \bibinfo{pages}{2871--2886}.
\bibitem[{Yonezawa(1968)}]{Yonezawa_1968}
\bibinfo{author}{F.~Yonezawa},
\newblock \bibinfo{title}{A systematic approach to the problems of random
  lattices: \mbox{I}},
\newblock \bibinfo{journal}{Prog. Theor. Phys.} \bibinfo{volume}{40}
  (\bibinfo{year}{1968}) \bibinfo{pages}{734--757}.
\bibitem[{Theumann(1974)}]{theumann1973dilute}
\bibinfo{author}{A.~Theumann},
\newblock \bibinfo{title}{Dilute \mbox{H}eisenberg ferromagnet within a
  single-site formulation},
\newblock \bibinfo{journal}{J. Phys. C: Solid State Phys.} \bibinfo{volume}{6}
  (\bibinfo{year}{1974}) \bibinfo{pages}{2822--2832}.
\bibitem[{Tang and Nolting(2006)}]{tang2006magnetic}
\bibinfo{author}{G.~X. Tang}, \bibinfo{author}{W.~Nolting},
\newblock \bibinfo{title}{Magnetic properties of disordered heisenberg binary
  spin system with long-range exchange},
\newblock \bibinfo{journal}{Phys. Rev. B} \bibinfo{volume}{73}
  (\bibinfo{year}{2006}) \bibinfo{pages}{024415}.
\bibitem[{Buczek et~al.(2016)Buczek, Sandratskii, Buczek, Thomas, Vignale, and
  Ernst}]{buczek2016magnons}
\bibinfo{author}{P.~Buczek}, \bibinfo{author}{L.~M. Sandratskii},
  \bibinfo{author}{N.~Buczek}, \bibinfo{author}{S.~Thomas},
  \bibinfo{author}{G.~Vignale}, \bibinfo{author}{A.~Ernst},
\newblock \bibinfo{title}{Magnons in disordered nonstoichiometric
  low-dimensional magnets},
\newblock \bibinfo{journal}{Phys. Rev. B} \bibinfo{volume}{94}
  (\bibinfo{year}{2016}) \bibinfo{pages}{054407}.
\bibitem[{Buczek et~al.(2018)Buczek, Thomas, Marmodoro, Buczek, Zubizarreta,
  Hoffmann, Balashov, Wulfhekel, Zakeri, and Ernst}]{buczek2018spin}
\bibinfo{author}{P.~Buczek}, \bibinfo{author}{S.~Thomas},
  \bibinfo{author}{A.~Marmodoro}, \bibinfo{author}{N.~Buczek},
  \bibinfo{author}{X.~Zubizarreta}, \bibinfo{author}{M.~Hoffmann},
  \bibinfo{author}{T.~Balashov}, \bibinfo{author}{W.~Wulfhekel},
  \bibinfo{author}{K.~Zakeri}, \bibinfo{author}{A.~Ernst},
\newblock \bibinfo{title}{Spin waves in disordered materials},
\newblock \bibinfo{journal}{J. Phys.: Condens. Matter} \bibinfo{volume}{30}
  (\bibinfo{year}{2018}) \bibinfo{pages}{423001}.
\bibitem[{Paischer et~al.(2021)Paischer, Buczek, Buczek, Eilmsteiner, and
  Ernst}]{paischer2021spin}
\bibinfo{author}{S.~Paischer}, \bibinfo{author}{P.~A. Buczek},
  \bibinfo{author}{N.~Buczek}, \bibinfo{author}{D.~Eilmsteiner},
  \bibinfo{author}{A.~Ernst},
\newblock \bibinfo{title}{Spin waves in alloys at finite temperatures:
  Application to the \mbox{FeCo} magnonic crystal},
\newblock \bibinfo{journal}{Phys. Rev. B} \bibinfo{volume}{104}
  (\bibinfo{year}{2021}) \bibinfo{pages}{024403}.
\bibitem[{Kittel(1987)}]{kittel1987quantum}
\bibinfo{author}{C.~Kittel}, \bibinfo{title}{Quantum Theory of Solids},
  \bibinfo{publisher}{Wiley}, \bibinfo{address}{New York},
  \bibinfo{year}{1987}. \bibinfo{note}{Chaps. 4, 10}.
\bibitem[{Dyson(1956)}]{Dyson_1956a}
\bibinfo{author}{F.~J. Dyson},
\newblock \bibinfo{title}{General theory of spin-wave interactions},
\newblock \bibinfo{journal}{Phys. Rev.} \bibinfo{volume}{102}
  (\bibinfo{year}{1956}) \bibinfo{pages}{1217--1230}.
\bibitem[{Maleev(1958)}]{Maleev_1958}
\bibinfo{author}{S.~V. Maleev},
\newblock \bibinfo{title}{Scattering of slow neutrons in ferromagnets},
\newblock \bibinfo{journal}{Sov. Phys. JEPT} \bibinfo{volume}{33}
  (\bibinfo{year}{1958}) \bibinfo{pages}{1010--1021}.
\bibitem[{Weiss(1964)}]{Weiss_1964a}
\bibinfo{author}{Z.~Weiss}, \bibinfo{title}{Spin-wave theory in a nonideal
  lattice}, \bibinfo{year}{1964}. \bibinfo{note}{Parts I \& II; unpublished
  notes}.
\bibitem[{Bloch(1932)}]{Bloch_1932}
\bibinfo{author}{F.~Bloch},
\newblock \bibinfo{title}{Zur \mbox{T}heorie des \mbox{A}ustauschproblems und
  der \mbox{R}emanenzerscheinung der \mbox{F}erromagnetika.},
\newblock \bibinfo{journal}{Z. Physik} \bibinfo{volume}{74}
  (\bibinfo{year}{1932}) \bibinfo{pages}{295--335}.
\bibitem[{Dyson(1956)}]{Dyson_1956b}
\bibinfo{author}{F.~J. Dyson},
\newblock \bibinfo{title}{Thermodynamic behavior of an ideal ferromagnet},
\newblock \bibinfo{journal}{Phys. Rev.} \bibinfo{volume}{102}
  (\bibinfo{year}{1956}) \bibinfo{pages}{1230--1244}.
\bibitem[{Vaks et~al.(1968)Vaks, Larkin, and Pikin}]{Vaks1968thermodynamics}
\bibinfo{author}{V.~G. Vaks}, \bibinfo{author}{A.~I. Larkin},
  \bibinfo{author}{S.~A. Pikin},
\newblock \bibinfo{title}{Thermodynamics of an ideal ferromagnetic substance},
\newblock \bibinfo{journal}{Sov. Phys. JEPT} \bibinfo{volume}{26}
  (\bibinfo{year}{1968}) \bibinfo{pages}{188--199}.
\bibitem[{Jones(1971)}]{jones1971impurity}
\bibinfo{author}{R.~C. Jones},
\newblock \bibinfo{title}{Impurity spin wave models in a simple cubic
  \mbox{H}eisenberg ferromagnet},
\newblock \bibinfo{journal}{J. Phys. C: Solid State Phys.}
  (\bibinfo{year}{1971}) \bibinfo{pages}{2903--2918}.
\bibitem[{Edwards(1958)}]{edwards1958new}
\bibinfo{author}{S.~F. Edwards},
\newblock \bibinfo{title}{A new method for the evaluation of electric
  conductivity in metals},
\newblock \bibinfo{journal}{Philos. Mag.} \bibinfo{volume}{3}
  (\bibinfo{year}{1958}) \bibinfo{pages}{1020--1031}.
\bibitem[{Ashcroft and Mermin(1976)}]{Ashcroft_Mermin_1976}
\bibinfo{author}{N.~W. Ashcroft}, \bibinfo{author}{N.~D. Mermin},
  \bibinfo{title}{Solid State Physics}, \bibinfo{publisher}{Harcourt Inc.},
  \bibinfo{address}{Orlando}, \bibinfo{year}{1976}. \bibinfo{note}{Chapter 8}.
\bibitem[{Landau and Lifshitz(1980)}]{Landau_Lifshitz_1980}
\bibinfo{author}{L.~D. Landau}, \bibinfo{author}{E.~M. Lifshitz},
  \bibinfo{title}{Statistical Physics}, \bibinfo{publisher}{Pergamon},
  \bibinfo{address}{Oxford}, \bibinfo{edition}{third} edition,
  \bibinfo{year}{1980}. \bibinfo{note}{Part 1, Section 71}.
\bibitem[{Jelitto(1969)}]{Jelitto_1969}
\bibinfo{author}{R.~J. Jelitto},
\newblock \bibinfo{title}{The density of states of some simple excitations in
  solids},
\newblock \bibinfo{journal}{J. Phys. Chem. Solids} \bibinfo{volume}{30}
  (\bibinfo{year}{1969}) \bibinfo{pages}{609--626}.
\bibitem[{Morita and Horiguchi(1971)}]{Morita_Horiguchi_1971}
\bibinfo{author}{T.~Morita}, \bibinfo{author}{T.~Horiguchi},
\newblock \bibinfo{title}{Formulas for the lattice green's functions for the
  cubic lattices in terms of the complete elliptic integrals},
\newblock \bibinfo{journal}{J. Phys. Soc. Japan} \bibinfo{volume}{30}
  (\bibinfo{year}{1971}) \bibinfo{pages}{957--964}.
\bibitem[{Joyce(1973)}]{joyce1973simple}
\bibinfo{author}{G.~S. Joyce},
\newblock \bibinfo{title}{On the simple cubic lattice green function},
\newblock \bibinfo{journal}{Philos.Trans. Royal Soc. Lond. A}
  \bibinfo{volume}{273} (\bibinfo{year}{1973}) \bibinfo{pages}{583--610}.
\bibitem[{\mbox{Wolfram} Research{,}~Inc.(2019)}]{Mathematica}
\bibinfo{author}{\mbox{Wolfram} Research{,}~Inc.}, \bibinfo{title}{Mathematica,
  {V}ersion 12.0}, \bibinfo{year}{2019}. \bibinfo{note}{Champaign, IL, USA}.
\bibitem[{Auerbach(1994)}]{Auerbach_1994}
\bibinfo{author}{A.~Auerbach}, \bibinfo{title}{Interacting Electrons and
  Quantum Magnetism}, \bibinfo{publisher}{Springer}, \bibinfo{address}{New
  York}, \bibinfo{year}{1994}.
\bibitem[{Mano(1982)}]{mano1982discrepancy}
\bibinfo{author}{H.~Mano},
\newblock \bibinfo{title}{Discrepancy between spin-wave dispersion and
  temperature dependence of magnetization in amorphous ferromagnets},
\newblock \bibinfo{journal}{J. Phys. Soc. Japn.} \bibinfo{volume}{51}
  (\bibinfo{year}{1982}) \bibinfo{pages}{3157--3165}.
\bibitem[{Chakraborty and Bouzerar(2015)}]{chakraborty2015long}
\bibinfo{author}{A.~Chakraborty}, \bibinfo{author}{G.~Bouzerar},
\newblock \bibinfo{title}{Long wavelength spin dynamics in diluted magnetic
  systems: Scaling of magnon lifetime},
\newblock \bibinfo{journal}{J. Magn. Magn. Mater.} \bibinfo{volume}{381}
  (\bibinfo{year}{2015}) \bibinfo{pages}{50--55}.
\bibitem[{Whittaker and Watson(1927)}]{Whittaker_Watson_1927}
\bibinfo{author}{E.~T. Whittaker}, \bibinfo{author}{G.~N. Watson},
  \bibinfo{title}{A Course of Modern Analysis}, \bibinfo{publisher}{Cambridge
  University Press}, \bibinfo{address}{Cambridge}, \bibinfo{year}{1927}.
  \bibinfo{note}{Chap. 13}.
\bibitem[{Olver et~al.(2010)Olver, Lozier, Boisvert, and
  Clark}]{NIST_Math_Handbook}
\bibinfo{author}{F.~W.~J. Olver}, \bibinfo{author}{D.~W. Lozier},
  \bibinfo{author}{R.~F. Boisvert}, \bibinfo{author}{C.~W. Clark},
  \bibinfo{title}{NIST Handbook of Mathematical Functions},
  \bibinfo{publisher}{Cambridge University Press},
  \bibinfo{address}{Cambridge}, \bibinfo{year}{2010}.
\bibitem[{Vaks et~al.(1968)Vaks, Larkin, and Pikin}]{Vaks1968spinwaves}
\bibinfo{author}{V.~G. Vaks}, \bibinfo{author}{A.~I. Larkin},
  \bibinfo{author}{S.~A. Pikin},
\newblock \bibinfo{title}{Spin waves and correlation functions in a
  ferromagnetic},
\newblock \bibinfo{journal}{Sov. Phys. JEPT} \bibinfo{volume}{26}
  (\bibinfo{year}{1968}) \bibinfo{pages}{647--655}.
\bibitem[{Mattis(1985)}]{Mattis_ii_1985}
\bibinfo{author}{D.~C. Mattis}, \bibinfo{title}{The Theory of Magnetism: II
  Thermodynamics and Statistical Mechanics}, volume~\bibinfo{volume}{2},
  \bibinfo{publisher}{Springer-Verlag}, \bibinfo{address}{Berlin},
  \bibinfo{year}{1985}. \bibinfo{note}{Chap. 2.12}.
\bibitem[{Dresselhaus et~al.(2007)Dresselhaus, Dresselhaus, and
  Jorio}]{dresselhaus2007group}
\bibinfo{author}{M.~S. Dresselhaus}, \bibinfo{author}{G.~Dresselhaus},
  \bibinfo{author}{A.~Jorio}, \bibinfo{title}{Group Theory: Application to the
  Physics of Condensed Matter}, \bibinfo{publisher}{Springer Science \&
  Business Media}, \bibinfo{address}{Berlin}, \bibinfo{year}{2007}.
  \bibinfo{note}{Chaps. 5, 10}.
\bibitem[{Mattuck(1964)}]{mattuck1964lifetime}
\bibinfo{author}{R.~D. Mattuck},
\newblock \bibinfo{title}{Lifetime of quasi particles in fermi system},
\newblock \bibinfo{journal}{Physics Letters} \bibinfo{volume}{11}
  (\bibinfo{year}{1964}) \bibinfo{pages}{29--31}.
\bibitem[{Mattuck(1992)}]{mattuck1992guide}
\bibinfo{author}{R.~D. Mattuck}, \bibinfo{title}{A Guide to Feynman Diagrams in
  the Many-body Problem}, \bibinfo{publisher}{Dover Publications Inc.},
  \bibinfo{address}{New York}, \bibinfo{edition}{second} edition,
  \bibinfo{year}{1992}. \bibinfo{note}{Chap. 11}.
\bibitem[{Oguchi(1960)}]{Oguchi_1960}
\bibinfo{author}{T.~Oguchi},
\newblock \bibinfo{title}{Theory of spin-wave interactions in ferro-
  antiferromagnetism},
\newblock \bibinfo{journal}{Phys. Rev.} \bibinfo{volume}{117}
  (\bibinfo{year}{1960}) \bibinfo{pages}{117--123}.
\bibitem[{Bloch(1962)}]{bloch1962magnon}
\bibinfo{author}{M.~Bloch},
\newblock \bibinfo{title}{Magnon renormalization in ferromagnets near the
  \mbox{C}urie point},
\newblock \bibinfo{journal}{Phys. Rev. Lett.} \bibinfo{volume}{9}
  (\bibinfo{year}{1962}) \bibinfo{pages}{286--287}.
\bibitem[{Kaganov and Chubukov(1987)}]{kaganov1987interacting}
\bibinfo{author}{M.~I. Kaganov}, \bibinfo{author}{A.~V. Chubukov},
\newblock \bibinfo{title}{Interacting magnons},
\newblock \bibinfo{journal}{Soviet Phys. Uspekhi} \bibinfo{volume}{30}
  (\bibinfo{year}{1987}) \bibinfo{pages}{1015--1040}.
\bibitem[{Zhitomirsky and Chernyshev(2013)}]{zhitomirsky2013colloquium}
\bibinfo{author}{M.~E. Zhitomirsky}, \bibinfo{author}{A.~L. Chernyshev},
\newblock \bibinfo{title}{Colloquium: \mbox{S}pontaneous magnon decays},
\newblock \bibinfo{journal}{Rev. Mod. Phys.} \bibinfo{volume}{85}
  (\bibinfo{year}{2013}) \bibinfo{pages}{219--243}.
\bibitem[{Mubayi and Lange(1969)}]{mubayi1969phase}
\bibinfo{author}{V.~Mubayi}, \bibinfo{author}{R.~V. Lange},
\newblock \bibinfo{title}{Phase transition in the two-dimensional
  \mbox{H}eisenberg ferromagnet},
\newblock \bibinfo{journal}{Phys. Rev.} \bibinfo{volume}{178}
  (\bibinfo{year}{1969}) \bibinfo{pages}{882--894}.
\bibitem[{Oguchi(1971)}]{oguchi1971phase}
\bibinfo{author}{A.~Oguchi},
\newblock \bibinfo{title}{Phase transition of the \mbox{H}eisenberg
  ferromagnet},
\newblock \bibinfo{journal}{Prog. Theor. Phys.} \bibinfo{volume}{46}
  (\bibinfo{year}{1971}) \bibinfo{pages}{63--76}.
\bibitem[{Takahashi(1986)}]{takahashi1986quantum}
\bibinfo{author}{M.~Takahashi},
\newblock \bibinfo{title}{Quantum \mbox{H}eisenberg ferromagnets in one and two
  dimensions at low temperature},
\newblock \bibinfo{journal}{Prog. Theor. Phys. Suppl.} \bibinfo{volume}{87}
  (\bibinfo{year}{1986}) \bibinfo{pages}{233--246}.
\bibitem[{Takahashi(1987{\natexlab{a}})}]{takahashi1987few}
\bibinfo{author}{M.~Takahashi},
\newblock \bibinfo{title}{Few-dimensional \mbox{H}eisenberg ferromagnets at low
  temperature},
\newblock \bibinfo{journal}{Phys. Rev. Lett.} \bibinfo{volume}{58}
  (\bibinfo{year}{1987}{\natexlab{a}}) \bibinfo{pages}{168--170}.
\bibitem[{Takahashi(1987{\natexlab{b}})}]{takahashi1987classical}
\bibinfo{author}{M.~Takahashi},
\newblock \bibinfo{title}{Classical \mbox{H}eisenberg ferromagnet in two
  dimensions},
\newblock \bibinfo{journal}{Phys. Rev. B} \bibinfo{volume}{36}
  (\bibinfo{year}{1987}{\natexlab{b}}) \bibinfo{pages}{3791--3797}.
\bibitem[{Takahashi(1987{\natexlab{c}})}]{takahashi1987two}
\bibinfo{author}{M.~Takahashi},
\newblock \bibinfo{title}{Two-dimensional \mbox{H}eisenberg ferromagnet},
\newblock \bibinfo{journal}{Jpn. J. Appl. Phys.} \bibinfo{volume}{26}
  (\bibinfo{year}{1987}{\natexlab{c}}) \bibinfo{pages}{869--870}.
\bibitem[{Takahashi(1990)}]{takahashi1990dynamics}
\bibinfo{author}{M.~Takahashi},
\newblock \bibinfo{title}{Dynamics of \mbox{H}eisenberg ferromagnets at low
  temperature},
\newblock \bibinfo{journal}{Phys. Rev. B} \bibinfo{volume}{42}
  (\bibinfo{year}{1990}) \bibinfo{pages}{766--770}.
\bibitem[{Mermin and Wagner(1966)}]{mermin1966absence}
\bibinfo{author}{N.~D. Mermin}, \bibinfo{author}{H.~Wagner},
\newblock \bibinfo{title}{Absence of ferromagnetism or antiferromagnetism in
  one-or two-dimensional isotropic \mbox{H}eisenberg models},
\newblock \bibinfo{journal}{Phys. Rev. Lett.} \bibinfo{volume}{17}
  (\bibinfo{year}{1966}) \bibinfo{pages}{1133--1136}.
\bibitem[{Zakeri(2014)}]{zakeri2014elementary}
\bibinfo{author}{K.~Zakeri},
\newblock \bibinfo{title}{Elementary spin excitations in ultrathin itinerant
  magnets},
\newblock \bibinfo{journal}{Phys. Rep.} \bibinfo{volume}{545}
  (\bibinfo{year}{2014}) \bibinfo{pages}{47--93}.
\bibitem[{Burch et~al.(2018)Burch, Mandrus, and Park}]{burch2018magnetism}
\bibinfo{author}{K.~S. Burch}, \bibinfo{author}{D.~Mandrus},
  \bibinfo{author}{J.~Park},
\newblock \bibinfo{title}{Magnetism in two-dimensional van der \mbox{W}aals
  materials},
\newblock \bibinfo{journal}{Nature} \bibinfo{volume}{563}
  (\bibinfo{year}{2018}) \bibinfo{pages}{47--52}.
\bibitem[{Ortmanns et~al.(2021)Ortmanns, Bauer, and
  Blanter}]{ortmanns2021magnon}
\bibinfo{author}{L.~C. Ortmanns}, \bibinfo{author}{G.~E.~W. Bauer},
  \bibinfo{author}{Y.~M. Blanter},
\newblock \bibinfo{title}{Magnon dispersion in bilayers of two-dimensional
  ferromagnets},
\newblock \bibinfo{journal}{Phys. Rev. B} \bibinfo{volume}{103}
  (\bibinfo{year}{2021}) \bibinfo{pages}{155430}.
\bibitem[{Evers et~al.(2015)Evers, M{\"u}ller, and Nowak}]{evers2015spin}
\bibinfo{author}{M.~Evers}, \bibinfo{author}{C.~A. M{\"u}ller},
  \bibinfo{author}{U.~Nowak},
\newblock \bibinfo{title}{Spin-wave localization in disordered magnets},
\newblock \bibinfo{journal}{Phys. Rev. B} \bibinfo{volume}{92}
  (\bibinfo{year}{2015}) \bibinfo{pages}{014411}.
\bibitem[{W{\"o}lfle and Vollhardt(2010)}]{wolfle2010self}
\bibinfo{author}{P.~W{\"o}lfle}, \bibinfo{author}{D.~Vollhardt},
\newblock \bibinfo{title}{Self-consistent theory of anderson localization:
  General formalism and applications},
\newblock \bibinfo{journal}{International Journal of Modern Physics B}
  \bibinfo{volume}{24} (\bibinfo{year}{2010}) \bibinfo{pages}{1526--1554}.
\bibitem[{Girvin and Yang(2019)}]{girvin2019modern}
\bibinfo{author}{S.~M. Girvin}, \bibinfo{author}{K.~Yang},
  \bibinfo{title}{Modern Condensed Matter Physics},
  \bibinfo{publisher}{Cambridge University Press}, \bibinfo{year}{2019}.
  \bibinfo{note}{Chap. 11}.
\bibitem[{Matsubara and Yonezawa(1965)}]{Matsubara_Yonezawa_1965}
\bibinfo{author}{T.~Matsubara}, \bibinfo{author}{F.~Yonezawa},
\newblock \bibinfo{title}{Note on cumulant average used in random lattice
  problem},
\newblock \bibinfo{journal}{Prog. Theor. Phys.} \bibinfo{volume}{34}
  (\bibinfo{year}{1965}) \bibinfo{pages}{871--872}.
\bibitem[{Bruus and Flensberg(2004)}]{Bruus_Flensberg_2004}
\bibinfo{author}{H.~Bruus}, \bibinfo{author}{K.~Flensberg},
  \bibinfo{title}{Many-Body Quantum Theory in Condensed Matter Physics},
  \bibinfo{publisher}{Oxford University Press}, \bibinfo{address}{Oxford},
  \bibinfo{year}{2004}. \bibinfo{note}{Chap. 8}.
\bibitem[{Callaway(1963)}]{Callaway_1963}
\bibinfo{author}{J.~Callaway},
\newblock \bibinfo{title}{Scattering of spin waves by magnetic defects},
\newblock \bibinfo{journal}{Phys. Rev.} \bibinfo{volume}{132}
  (\bibinfo{year}{1963}) \bibinfo{pages}{2003--2009}.
\bibitem[{Van~Hove(1953)}]{van1953occurrence}
\bibinfo{author}{L.~Van~Hove},
\newblock \bibinfo{title}{The occurrence of singularities in the elastic
  frequency distribution of a crystal},
\newblock \bibinfo{journal}{Phys. Rev.} \bibinfo{volume}{89}
  (\bibinfo{year}{1953}) \bibinfo{pages}{1189--1193}. \bibinfo{note}{And
  references therein}.
\bibitem[{Shi and Eyink(2016)}]{shi2016resonance}
\bibinfo{author}{Y.-K. Shi}, \bibinfo{author}{G.~L. Eyink},
\newblock \bibinfo{title}{Resonance \mbox{Van Hove} singularities in wave
  kinetics},
\newblock \bibinfo{journal}{Physica D: Nonlinear Phenomena}
  \bibinfo{volume}{332} (\bibinfo{year}{2016}) \bibinfo{pages}{55--72}.
\bibitem[{Morita and Horiguchi(1972)}]{morita1972analytic}
\bibinfo{author}{T.~Morita}, \bibinfo{author}{T.~Horiguchi},
\newblock \bibinfo{title}{Analytic properties of the lattice green function},
\newblock \bibinfo{journal}{J. Phys. A: General Physics} \bibinfo{volume}{5}
  (\bibinfo{year}{1972}) \bibinfo{pages}{67--77}.
\bibitem[{Morita and Horiguchi(1971)}]{morita1971lattice}
\bibinfo{author}{T.~Morita}, \bibinfo{author}{T.~Horiguchi},
\newblock \bibinfo{title}{Lattice green's functions for the cubic lattices in
  terms of the complete elliptic integral},
\newblock \bibinfo{journal}{J. Math. Phys.} \bibinfo{volume}{12}
  (\bibinfo{year}{1971}) \bibinfo{pages}{981--986}.
\bibitem[{Abe and Katsura(1973)}]{abe1973lattice}
\bibinfo{author}{Y.~Abe}, \bibinfo{author}{S.~Katsura},
\newblock \bibinfo{title}{Lattice green's function for the simple cubic and
  tetragonal lattices at arbitrary points},
\newblock \bibinfo{journal}{Ann. Phys.} \bibinfo{volume}{75}
  (\bibinfo{year}{1973}) \bibinfo{pages}{348--380}.

\end{thebibliography}

\end{document}